\documentclass[11pt, fleqn]{article}

\usepackage{amsmath}
\usepackage{amssymb}
\usepackage{amsfonts}
\usepackage{amsthm}
\usepackage{dsfont}
\usepackage{mathrsfs}
\usepackage{ulem}
\usepackage{graphicx}

\usepackage{mdframed}

\usepackage{enumerate}

\usepackage[arrowdel]{physics}
\usepackage{tensor}

\usepackage{simplewick}
\usepackage{feynmf}

\usepackage{tabu}
\usepackage{physics}
\usepackage{mathdots}

\newcommand{\operator}[1]{\hat{#1}}

\newcommand{\R}{\mathbb{R}}

\newcommand{\Z}{\mathbb{Z}}

\newcommand{\T}{\mathbb{T}}

\newcommand{\pauli}[1]{
    \ifnum#1=1
        \operator{\sigma}_{x}
    \else
        \ifnum#1=2
           \operator{\sigma}_{y}
        \else
            \ifnum#1=3
                \operator{\sigma}_{z}
            \else
                \errmessage{Incorrect number given to pauli}
            \fi
        \fi
    \fi
}

\usepackage{geometry}
\usepackage{comment}

\usepackage{soul}
\usepackage{subcaption}

\usepackage{jheppub}
\usepackage{amsmath,amssymb,amsthm}
\usepackage{bm}
\usepackage{slashed}
\usepackage{graphicx}
\usepackage[T1]{fontenc}
\usepackage{fix-cm}

\usepackage{tabu}
\usepackage{physics}

\makeatletter
\newcommand*\bigcdot{\mathpalette\bigcdot@{.5}}
\newcommand*\bigcdot@[2]{\mathbin{\vcenter{\hbox{\scalebox{#2}{$\m@th#1\bullet$}}}}}
\makeatother

\newlength{\dummysp}
\newcommand{\ba}{\begin{eqnarray}}
\newcommand{\ea}{\end{eqnarray}}
\newcommand{\ca}{{\cal{A}}}

\usepackage[font=scriptsize]{caption}
\usepackage{mwe}

\def\R{{\mathbb R}}
\def\S{{\mathbb S}}
\def\Z{{\mathbb Z}}
\def\T{{\mathbb T}}

\def\tr{\,{\rm tr}\,}

\title{The Nahm transform of multi-fractional instantons}

 \author[a]{Mohamed M. Anber,}\author[b]{Erich Poppitz} 
  
\affiliation[a]{Centre for Particle Theory, Department of Mathematical Sciences, Durham University, South Road, Durham DH1 3LE, UK}
\affiliation[b]{Department of Physics,   University of Toronto, 60 St George St., 
Toronto, ON M5S 1A7, Canada}
\emailAdd{mohamed.anber@durham.ac.uk}\emailAdd{poppitz@physics.utoronto.ca}   
   
\abstract{

{\flushleft{We}} embed the multi-fractional instantons of  $SU(N)$ gauge theories on  $\T^4$ with 't Hooft twisted boundary conditions into $U(N)$ bundles and use the Nahm transform to study the corresponding configurations on the dual $\hat \T^4$. We first show that $SU(N)$  fractional instantons of    topological charge $Q={r \over N}$,  $r \in \{1, 2,...,N-1\}$, are mapped to fractional instantons of $SU(\hat N)$ of charge $\hat Q = {r \over \hat N}$, where $\hat N = N q_1 q_3 - r q_3 + q_1$ and  $q_{1,3}$ are  integer-quantized $U(1)$ fluxes. We then explicitly construct the Nahm transform of constant field strength fractional instantons of $SU(N)$ and find the  $SU(\hat N)$ configurations they map to. Both the $\T^4$ instantons and their $\hat \T^4$ images are self-dual for appropriately tuned  torus periods. The Nahm duality  can be extended to  tori with detuned periods, with detuning parameter $\Delta$,  mapping solutions with $\Delta >0$ on $\T^4$ to ones with $\hat\Delta <0$ on $\hat \T^4$.  We also recall that  fractional instantons appear in string theory precisely via the $U(N)$ embedding,  suggesting that studying  the end point of tachyon condensation for $\Delta \ne 0$ is needed---and is perhaps feasible in a small-$\Delta$ expansion, as in field theory studies---in order to understand the appearance and role of fractional instantons in $D$-brane constructions.    }

\begin{document}

\maketitle

\section{Introduction and summary}

Recently, there has been  renewed interest in  investigating the role Euclidean configurations with fractional topological charge  \cite{tHooft:1979rtg,tHooft:1981sps,tHooft:1981nnx,vanBaal:1982ag} play in gauge dynamics, motivated by several developments. The most recent  is  the discovery of generalized anomalies  involving $1$-form (center) symmetry \cite{Gaiotto:2014kfa,Gaiotto:2017yup}, where the introduction of fractional charge backgrounds  reveals various   't Hooft anomalies involving center symmetry (these  lead to, for example, exact spectral degeneracies \cite{Cox:2021vsa}). The other is the realization that objects with fractional topological charge are responsible---as shown in many cases  using reliable semiclassical tools \cite{RTN:1993ilw,Gonzalez-Arroyo:1995ynx,Unsal:2007jx,Unsal:2008ch,Tanizaki:2022ngt,Tanizaki:2022plm}---for nonperturbative gauge theory  phenomena such as confinement and chiral symmetry breaking, a development recently reviewed in \cite{Poppitz:2021cxe,Gonzalez-Arroyo:2023kqv}.

Instantons of fractional topological charge $Q={r \over N}$ in $SU(N)$ gauge theories on $\T^4$ were discovered by 't Hooft \cite{tHooft:1979rtg,tHooft:1981sps}, who  found explicit solutions with commuting constant field strength \cite{tHooft:1981nnx} but nonabelian transition functions (see \cite{Gonzalez-Arroyo:1997ugn} for an introduction and review). Further studies were slowed by the lack of explicit solutions for more general---nonabelian and space-time-dependent---instantons, which had to be studied using numerical techniques, as in \cite{GarciaPerez:1992fj,Gonzalez-Arroyo:1998hjb}. An important advance was the introduction \cite{GarciaPerez:2000aiw} of an analytic technique for finding self-dual, or minimum action,  fractional instantons via an expansion in a small parameter $\Delta$, the asymmetry parameter of $\T^4$. This technique was further developed in \cite{Gonzalez-Arroyo:2004fmv,Gonzalez-Arroyo:2019wpu,Anber:2023sjn}. 

The  (multi-)fractional instantons  found in these studies were  used in the semiclassical  calculations of the gaugino condensates in super-Yang-Mills theory on a small twisted-$\T^4$ \cite{Anber:2022qsz,Anber:2024mco}, finding agreement with the weak-coupling semiclassical $\R^4$ result (see \cite{Shifman:1999kf,Shifman:1999mv} and the review \cite{Dorey:2002ik}) and the recent independent lattice determination \cite{Bonanno:2024bqg}. This  completed a calculation first attempted 40 years ago \cite{Cohen:1983fd}---which, we stress,  could neither be performed nor its result   interpreted  before understanding both the relevant generalized anomalies and  the somewhat intricate multi-fractional instanton moduli space.

Another  recent observation involving fractional instantons is  that seemingly different objects of fractional topological charge, such as center vortices and monopole-instantons, shown to be responsible for semiclassical confinement in various geometries with small compact spaces, $\R \times \T^3$, $\R^3 \times \S^1$, or $\R^2 \times \T^2$  \cite{RTN:1993ilw,Gonzalez-Arroyo:1995ynx,Unsal:2007jx,Unsal:2008ch,Tanizaki:2022ngt,Tanizaki:2022plm}, are in fact smoothly related to each other by taking different limits of twists and $\T^4$ periods. This was argued or shown in \cite{GarciaPerez:1999hs,Unsal:2020yeh,Poppitz:2022rxv,Hayashi:2024yjc,Guvendik:2024umd, Wandler:2024hsq,Hayashi:2024psa} (some of these works use analytic  and continuity arguments, while others rely on numerical tools). The above developments  underscore the need for finding more diverse analytic approaches to fractional instantons.

In this paper, we study  the Nahm transform\footnote{The  Nahm transform  in its original form  \cite{Nahm:1979yw}  relates BPS monopoles of charge $N$ on $\R^3$ to the solutions of a set of equations for $N\times N$ matrices on an interval; see \cite{Manton:2004tk} for a review and more references. The common thread with the $\T^4$ construction  \cite{Schenk:1986xe,Braam:1988qk} is the use of  families of fundamental Dirac operators in the construction of the Nahm dual. We also recall that the Nahm transform in the non-compact $\R^3 \times \S^1$ was crucial in finding  the well-studied caloron solution  \cite{Kraan:1998pm}.}  of fractional   instantons on $\T^4$,  motivated by a desire to better understand the space of fractional instantons solutions and their moduli. As described in  \cite{Schenk:1986xe,Braam:1988qk} and reviewed further below, starting from a topologically nontrivial classical background in a $U(N)$ gauge theory   on $\T^4$, of periods $L_\mu$, $\mu=1,2,3,4$, the Nahm transform constructs a classical  background for a $U(\hat N)$ gauge theory on the dual torus $\hat \T^4$. The dual torus is one of inverted periods $\hat L_\mu = {1\over L_\mu}$.\footnote{As this  transformation relates classical gauge  backgrounds, the scale needed to make up the dimensions can be taken arbitrary (it equals $4 \pi^2 \alpha'$ in string theory).} The rank  $\hat N$ of the dual gauge theory on $\hat \T^4$   equals the second Chern character (the topological charge or Dirac index) of the $U(N$) background  on  $\T^4$, while the topological charge of the $U(\hat N)$ background equals the rank $N$ of the $\T^4$ gauge theory. Thus,  Nahm duality exchanges rank with topological charge, acting akin to $T$-duality in string theory, see e.g.~\cite{Polchinski:1998rr,Tong:2005un}. The relation between Nahm transform and $T$-duality, using the families index theorem \cite{Atiyah:1970ws}, has been explored in a context closest to ours in \cite{Hori:1999me} (see also \cite{Douglas:1996sw,Douglas:1996uz}).

Our approach to finding the Nahm transform of fractional instantons can be briefly described as follows. We first embed the $\T^4$ $SU(N)$ fractional instantons of charge $Q={r \over N}$, $1 \le r < N$, into $U(N)$ bundles of integer topological charge, by adding appropriate fractional and integer $U(1)$ fluxes (or first Chern characters). The Nahm transformation is then applied to this integer-charge $U(N)$ background.
The topological properties of the 
 Nahm dual on $\hat \T^4$ can be easily found via the families index theorem: one first finds the dual $U(1)$ fluxes and uses them to infer the $SU(\hat N)$ topological charge. The rank of the dual group, $\hat N = N q_1 q_3 - r q_3 + q_1$, is fixed by the integer quantized fluxes $q_{1,3}$ of $U(1)$, which are arbitrary but chosen so that $\hat N>1$. The   topological charge of the Nahm dual fractional instanton of  $SU(\hat N)$ on $\hat \T^4$ is then found to be $\hat Q = {r \over \hat N}$. Thus,  the Nahm dual of a  minimal charge  fractional instanton is also  a minimal-charge one.

Finding the explicit form of the Nahm transform of a given fractional instanton background, however, requires a calculation. 
 We explicitly perform the Nahm transformation of the constant field strength solutions of \cite{tHooft:1981nnx}. We find that constant field strength solutions of $SU(N$) of charge $Q$ are mapped to constant field strength configurations of charge $\hat Q$ in the $SU(\hat N)$ Nahm dual and  determine the precise relation between the two. Both the $\T^4$ solution and its Nahm dual are self-dual when the torus periods are appropriately tuned: the asymmetry parameter of $\T^4$ vanishes, $\Delta =0$, as well as that of $\hat \T^4$, $\hat \Delta =0$.
 
The explicit Nahm transform can be  extended---with extra future effort---to  tori with detuned periods in the framework of the $\Delta$-expansion  \cite{GarciaPerez:2000aiw}. Then, a small-$\Delta$ nonabelian fractional instanton on $\T^4$ with $\Delta >0$ is mapped to a small-$\hat \Delta$ fractional instanton on $\hat \T^4$ with $\hat \Delta <0$. Thus, ultimately, it could be used to relate fractional instantons on $\R \times \T^3$ to fractional instantons on the dual $\S^1 \times \R^3$, i.e. the monopole-instantons.
 
  The Nahm transform of fractional instantons has been investigated before, albeit following a different route than ours \cite{Gonzalez-Arroyo:1998nku,GarciaPerez:1999bc}. There, staying within $SU(N)$, the twisted $\T^4$ was replicated a sufficient number of times to make the $SU(N)$ topological charge integer. The Nahm transform was performed on the replicated torus and a symmetry was found which allows to extract a fractional topological charge instanton on an appropriately reduced dual torus. Despite the different setup of the construction, our results are broadly consistent with those of  \cite{Gonzalez-Arroyo:1998nku,GarciaPerez:1999bc}.\footnote{Yet another  construction in \cite{Gonzalez-Arroyo:1998nku} introduced additional flavors to  cancel the twist.  The possibility to add $U(1)$ fluxes to cancel the twist via $U(N)$ embedding was also mentioned there, but  was not pursued.} 
 
A strong  motivation to pursue our construction of the Nahm transform via the $U(N)$-embedding  is that this is precisely how fractional instantons  appear in string theory $D$-brane constructions. These use  stacks of intersecting $D$-branes at angles wrapped on $\T^4$, see \cite{Guralnik:1997sy,Hashimoto:1997gm}. These constructions give explicit realizations of the constant field strength solutions of 't Hooft  in string theory and, as in \cite{tHooft:1981nnx}, are stable, or BPS, only for appropriately tuned $\T^4$ periods such that  the asymmetry parameter $\Delta =0$ (otherwise, there is a tachyon in the spectrum of strings connecting the different stacks of branes). Given the utility of $D$-brane constructions in elucidating the ADHM construction of instantons on $\R^4$, see \cite{Tong:2005un} and references therein, it is natural to wonder if any insight into fractional instantons can be obtained from their string theory embedding. In particular, there may be an analogue of the $\Delta$-expansion \cite{GarciaPerez:2000aiw} allowing a study of the end point of tachyon condensation.
   
{\flushleft\bf{The main body of this paper:}}   
The main results of this paper are presented in as concise a manner as possible. 
We begin in section \ref{sec:review}, with a review of the Nahm transform for $U(N)$ topologically nontrivial backgrounds on $\T^4$. 
The embedding of $SU(N)$ fractional instantons into $U(N)$ bundles is reviewed in section \ref{sec:embedding}. In section \ref{sec:transitionfunctions}, we discuss the $SU(N) \times U(1)_B/\Z_N$ transition functions   and, in section \ref{sec:propertiesviafamilies}, derive the topological properties of the Nahm dual from the families index theorem. 

The steps needed to perform an explicit calculation of the Nahm dual are outlined in section \ref{explicit:constant1}. In section \ref{sec:nahmdual}, we give the results and discuss the properties of the Nahm dual of 't Hooft's constant-field strength fractional instantons, their self-duality, and moduli spaces. In section \ref{sec:towards} we outline the steps needed to generalize the explicit calculation to fractional instantons with non-constant field strength and stress the connection to $D$-brane constructions. 

{\flushleft\bf{The appendices:}} The details of our calculations are relegated to several voluminous Appendices.
The form of the Nahm transform is motivated in a pedestrian way in Appendix \ref{appx:notationDefining} and the families index theorem applied to it is reviewed in Appendix \ref{appx:notationFamilies}. 
Appendix \ref{appx:calculating}  is devoted to the details of the calculation of the Nahm dual of 't Hooft's constant field strength fractional instantons. In Appendices \ref{appx:solution}--\ref{Embeddings}, we find the normalizable zero modes of the fundamental Dirac equation in the fractional instanton background on $\T^4$. In appendix \ref{sec:nahmdualgaugefield}, we use them to calculate   the Nahm dual gauge background on $\hat\T^4$, determine the transition functions on the dual torus  in Appendix \ref{appx:transitiondual}, and cast them in canonical (moduli-independent) form in Appendix \ref{appx:transitiondualcanonical}. The final Appendix \ref{appx:moduli} is devoted to a study of the Nahm dual moduli space.
 
 \section{The Nahm transform of $SU(N)$ fractional instantons via  $U(N)$-embedding }
 
 \subsection{Brief review of the Nahm transform for $U(N)$ gauge backgrounds on $\T^4$}
 
 \label{sec:review}
 
We work on  $\T^4$ of periods $L_\mu$, with transition functions $\Sigma_\mu \in U(N)$. For brevity we denote $x+L_\mu \equiv x + \hat e_\mu L_\mu$ with $\hat e_\mu$ a unit vector in the $\mu$ direction. The boundary conditions on the $U(N)$ gauge fields are 
\begin{eqnarray}\label{gauge1}
{\cal A}(x+L_\mu) &=& \Sigma_\mu(x) ({\cal{A}}(x) - i d) \Sigma^{-1}_\mu(x), ~~\Sigma_\mu(x + L_\nu) \Sigma_\nu(x) = \Sigma_\nu (x + L_\mu) \Sigma_\mu(x)~,
 \end{eqnarray}
 where the second equation gives the cocycle condition for the $U(N)$ transition functions on the torus.
We next consider  a $U(N)$ background (\ref{gauge1}) that has nonzero second Chern character:
 \begin{equation}\label{chern21}
 {ch}_2[{\cal A}] = {1 \over 8 \pi^2} \int\limits_{\T^4} \tr {\cal F} \wedge {\cal F} = \hat N~,
 \end{equation}
 assuming $\hat N > 1$.
To define the Nahm transform, we also need to consider a family of $U(N)$ fundamental representation Dirac operators $D$ and $\bar D$, parameterized by the flat $U(1)$ connections $z_\mu$ and defined as\footnote{For our spinor notation, see Appendix \ref{appx:notation12}.} 
\begin{eqnarray}\label{fundDirac}
D &=& \sigma_\mu (\partial_\mu + i {\cal A}_\mu + 2 \pi i z_\mu) \equiv \sigma_\mu D_\mu,   \\
\bar D &=& \bar\sigma_\mu (\partial_\mu + i {\cal A}_\mu+ 2 \pi i z_\mu).
\end{eqnarray}
The periodic gauge transformation $e^{i  2 \pi  {x^\mu\over L_\mu}}$ shifts $z_\mu$ by   $\hat L_\mu = {1\over L_\mu}$, thus $z_\mu$ takes values in the dual torus $\hat\T^4$.
The background connection ${\cal A}_\mu$ is further  assumed to be ``irreducible,'' i.e. such that the operator $\bar D D$ has no zero modes, but the operator $D \bar D$ has $\hat N$ zero modes, as per the index theorem relating the Dirac index to the Chern character $ch_2[{\cal A}]$ of (\ref{chern21}).\footnote{Irreducibility is  often  guaranteed if the gauge background ${\cal A}$ is self-dual, owing to the fact that ${\bar{D}} D = D_\mu D^\mu$ for such backgrounds (usually, a negative definite operator), but there are exceptions: some self-dual constant field strength backgrounds are not irreducible (see discussion in section \ref{explicit:constant1}).} The equations,  boundary conditions, and normalization obeyed by the zero modes  are
\begin{eqnarray}\label{zeromodes1}
\bar D \psi^a (x,z) &=& \bar\sigma_\mu(\partial_\mu + i {\cal A}_\mu(x) + 2 \pi i z_\mu) \psi^a(x,z)= 0, ~~ a, b= 1,...,\hat{N},\nonumber \\ 
\psi^a(x + L_\mu,z) &=& \Sigma_\mu(x) \psi^a(x,z),~~\int_{\T^4} (\psi^a)^\dagger \cdot \psi^b  =\delta^{ab}~.
\end{eqnarray}
The boundary condition on the fundamental zero modes $\psi^a$, from the last line of  (\ref{zeromodes1}) is consistent due to the non-twist cocycle condition (\ref{gauge1}). 
In the work below, we shall often need to display the whole set of indices (in addition to the index $a$ labeling the $\hat N$ zero modes, also including the spinor $\alpha$ and $U(N)$-fundamental index $A$) of the zero modes as shown below:
\begin{equation} \label{zeromodes3}
\psi^{a}_{A \alpha}(x,z)~, ~ a=1,... \hat{N}, ~ A=1,... N, ~ \alpha = 1,2~.
\end{equation}
Any $z$-dependent linear combination of the $\bar D$ zero modes $\psi^a$ is also a zero mode. One can also consider linear combinations that preserve the normalization condition (\ref{zeromodes1}): \begin{eqnarray}\label{UQdef}
g_{\hat{N}}(z): ~\psi^a(x,z) \rightarrow\psi'^{a}(x,z) =  \psi^b(x,z) (g_{\hat{N}}^{-1}(z))^{ba} = (g_{\hat{N}}^{ab})^* \; \psi^b~, ~~ g_{\hat{N}}(z) \in U(\hat N)~,
\end{eqnarray}
provided the matrix $g_{\hat N}(z)$ is  unitary $\hat N \times \hat N$.\footnote{We stress that $g_{\hat{N}}(z)$ here denotes a general gauge transformation and not a transition function $\hat\Omega_\mu(z)$ on the dual torus (these are given in (\ref{transition matrix in the first group for SUM P}, \ref{transition matrix in the second group for SUM P})).}

 It is said that the zero modes of $\bar D$, $\psi^{a}_{A \alpha}(x,z)$, define a  bundle over the dual $\hat \T^4$ where the natural inner product is an integral over $\T^4$. The $\psi^a_{A \alpha}(x,z)$ define the basis in the space of zero modes and the connection on the dual bundle, the Nahm dual $U(\hat N)$ connection  $\hat {\cal A}_\mu^{ab}(z)$ on $\hat \T^4$ is:\footnote{In Appendix \ref{appx:notationDefining}, we motivate the form of the Nahm dual connection in a pedestrian way.}
\begin{eqnarray}\label{defofdualA2}
\hat {\cal A}(z)_\mu^{ab} = - i \int_{\T^4}  (\psi^{a}_{A \alpha}(x,z))^* \partial_{z_\mu} \psi^b_{A \alpha}(x,z)~,~ a,b=1,... \hat{N}, 
\end{eqnarray}
with the spinor and $U(N)$ indices summed over.
That this is a $U(\hat N)$ connection is  seen from the fact that applying (\ref{UQdef}), we find that under $g_{\hat N}(z)$:
 $ \hat {\cal A}(z) \rightarrow g_{\hat N}(z) (\hat {\cal A}(z) - i d) g^{-1}_{\hat N}(z)$. 
 
The topological properties of the Nahm dual as a $U(\hat N)$ gauge field on $\hat \T^4$ of field strength 
\ba\label{defofdualA3}
  \hat {\cal F}_{\mu\nu}^{ab}(z)= \partial_{z_\mu} \hat {\cal A}_{\nu}^{ab} - \partial_{z_\nu}\hat {\cal A}_{\mu}^{ab}+ i \sum_c (\hat {\cal A}_\mu^{ac} \hat {\cal A}_\nu^{cb} - \hat {\cal A}_\nu^{ac} \hat {\cal A}_\mu^{ca}) 
  \ea
  follow from the families index theorem (see Appendix \ref{appx:notationFamilies})\footnote{We use ${\cal A}$, ${\cal F}$ for $U(N)$ quantities and, in further sections, $A, F$ for $SU(N)$ ones (${\cal A} = A + {\mathbf 1}_N a$ with $\tr A =0$ and $\tr {\cal A}= N a$). Here $a$ is the $U(1) \equiv U(1)_B$ connection. Similar notation with hats is used for $U(\hat N)$ quantities.}
\ba \label{familyindex22}
 \tr {\hat {\cal F} \wedge \hat {\cal F} \over 8 \pi^2}  &=& N \int_{\T^4} {1 \over 4!} (dz_\mu\wedge dx_\mu)^4,  
 \ea
 \ba
rk(\hat {\cal F}) &=&  \int_{\T^4}  \tr {  {\cal F} \wedge  {\cal F} \over 8 \pi^2},\nonumber\\
  \tr {\hat {\cal F}\over 2\pi} &=&- {1 \over 2} dz_\mu \wedge dz_\nu \int_{\T^4} dx_\mu\wedge dx_\nu \wedge \tr { {\cal F} \over 2\pi}, \nonumber
\ea
where these relations should be understood as relations between topological classes, i.e. ones determining $ch_2[\hat {\cal F}]$ (defined similar to (\ref{chern21})) and $ch_1[\hat {\cal F}] = - \oint \tr {\hat{\cal F} \over 2 \pi} $.

The first two relations in (\ref{familyindex22}), for example, establish that the topological charge $ch_2[\hat A]$ and rank of the $U(\hat N)$ bundle over $\hat\T^4$ are related to the rank and topological charge of the $U(N)$ bundle on $\T^4$, respectively. The last relation gives the relations between the corresponding $U(1)$ fluxes over the various $2$-planes in $\T^4$ and $\hat\T^4$ and will be important to our embedding of fractional instantons.

In  \cite{Schenk:1986xe,Braam:1988qk} it is further shown   that   if the $\T^4$ $U(N)$ connection $\cal A$ is   self-dual  then $\hat{\cal A}$ is a self-dual $U(\hat N)$ connection on $\hat\T^4$, that  the Nahm transform is  an involution, and that it gives an isometry between the moduli spaces of the $\T^4$ and $\hat \T^4$ instantons.

 \subsection{Embedding the fractional instanton of $SU(N)$ into $U(N)$}
 \label{sec:embedding}
 
 Here, we discuss the embedding of the $SU(N)$ fractional instanton into $U(N)$ (section \ref{sec:transitionfunctions})  and find the properties of the Nahm dual from the families index theorem (section \ref{sec:propertiesviafamilies}).
 
\subsubsection{The $SU(N)\times U(1)_B$ transition functions on $\mathbb T^4$}
\label{sec:transitionfunctions}
Now we consider Yang-Mills theory on $\mathbb T^4$ with a $SU(N)\times U(1)_B/\mathbb Z_N$ bundle. To study the Nahm transform, we need to solve the Dirac equation of the fundamental fermions in the background of an instanton of this bundle. We take $\Omega_{\mu}$ and $\omega_\mu$ to be the $SU(N)$ and $U(1)_B$ transition functions. The boundary conditions obeyed by the $SU(N)$ connection $A$ and $U(1)_B$ connection $a$, respectively, are as in (\ref{gauge1}) with $\Sigma$ replaced by the corresponding transition function, $\Omega$ for $A$ and $\omega$ for $a$. We recall that the $U(N)$ connection is ${\cal A} = A + {\mathbf{1}}_N a$.
  These $SU(N)$ and $U(1)$ transition functions satisfy the cocycle conditions
 \begin{eqnarray}\label{cocycle conditions for both}
 \nonumber
 \Omega_{\mu}(x+  L_\nu)\Omega_\nu(x)&=&e^{i\frac{2\pi n_{\mu\nu}}{N}} \Omega_\nu(x+  L_\mu)\Omega_\mu (x)\,,\\
  \omega_{\mu}(x+  L_\nu)\omega_\nu(x)&=&e^{-i\frac{2\pi n_{\mu\nu}}{N}} \omega_\nu(x+ L_\mu)\omega_\mu (x)\,,
 \end{eqnarray}
upon traversing $\mathbb T^4$ in any direction. The integers $n_{\mu\nu}$ satisfy  $n_{\mu\nu}=-n_{\nu\mu}$ and are taken (mod $N$). Notice the negative sign difference in the $\mathbb Z_N$ phases of the two equations in (\ref{cocycle conditions for both}), which ensures that the fundamental fermions are well defined on $\mathbb T^4$. To see this, we let $\psi$ be a left-handed Weyl fermion in the defining representation of $SU(N)$ with unit $U(1)$ charge.  Upon traversing $\mathbb T^4$ in the $L_\mu$ direction, $\psi$ satisfies the boundary conditions:
\begin{eqnarray}\label{BCS for F}
\psi(x+ L_\mu)=\Sigma_\mu (x)\psi (x)\equiv\Omega_\mu(x)\omega_\mu (x)\psi(x)\,, \quad \mu=1,2,3,4\,,
\end{eqnarray}
such that the combined transition functions $\Sigma_\mu (x) \equiv \Omega_\mu(x)\omega_\mu (x)$ satisfy the $U(N)$ cocycle condition from (\ref{gauge1}).

We will build upon our earlier research  \cite{Anber:2023sjn}, where we extensively examined $SU(N)$ instantons characterized by fractional topological charges $Q_{SU(N)}=r/N$. As a quick summary, we employed 't Hooft's idea of embedding $SU(k)\times SU(\ell)\times U(1)$ within $SU(N)$, where $\ell+k=N$. 
To this end, we use 
\begin{equation}\label{omega}
\omega=2\pi\mbox{diag}\left[\underbrace{\ell, \ell,...,\ell}_{k\, \mbox{times}},\underbrace{ -k,-k,...,-k}_{\ell\,\mbox{times}}\right].
\end{equation} The matrices $P_\ell$ and $Q_\ell$  are the $\ell\times \ell$ shift and clock matrices:
\begin{eqnarray}\label{pandq}
P_\ell=\gamma_\ell\left[\begin{array}{cccc} 0& 1&0&...\\ 0&0&1&...\\... \\ ...&  &0&1 \\1&0&...&0\end{array}\right]\,,\quad Q_\ell=\gamma_\ell\; \mbox{diag}\left[1, e^{\frac{i 2\pi}{\ell}}, e^{2\frac{i 2\pi}{\ell}},...\right]\,,
\end{eqnarray}
which satisfy the relation $P_\ell Q_\ell=e^{i\frac{2\pi}{\ell}}Q_\ell P_\ell$. The factor $\gamma_\ell \equiv e^{\frac{i\pi (1-\ell)}{\ell}}$ ensures that $\det Q_\ell=1$ and $\det P_\ell=1$ ($P_k$, $Q_k$, and $\gamma_k$ are defined similarly).

  We recall the cocycle conditions (\ref{cocycle conditions for both}), taking the nonzero values of $n_{\mu\nu}$ to be
\begin{eqnarray}\label{twists1}
n_{12}=-r\,,\quad n_{34}=1\,.
\end{eqnarray}
The nonabelian transition functions, obeying the cocycle condition in the first line of (\ref{cocycle conditions for both}) (with twists (\ref{twists1})),  are
 \begin{eqnarray}
\nonumber
\Omega_1&=&P_k^{-r}\oplus {\mathbf 1}_\ell e^{i \omega \frac{r x_2}{Nk L_2}} = \left[\begin{array}{cc}P_k^{-r}e^{i2\pi \ell r \frac{x_2}{Nk L_2}}&0\\0& e^{-i 2\pi r\frac{x_2}{NL_2}}{\mathbf 1}_\ell\end{array}\right],\\
\nonumber
\Omega_2&=&Q_k\oplus {\mathbf 1}_\ell = \left[\begin{array}{cc}Q_k&0\\0& {\mathbf 1}_\ell\end{array}\right],\\
\nonumber
\Omega_3&=&{\mathbf 1}_k\oplus P_\ell e^{i \omega \frac{x_4}{N\ell L_4}} = \left[\begin{array}{cc} e^{i2\pi  \frac{x_4}{N L_4}} {\mathbf 1}_k&0\\0& e^{-i 2\pi k\frac{x_4}{N \ell L_4}}P_\ell\end{array}\right],\\
\Omega_4&=&{\mathbf 1}_k\oplus Q_\ell = \left[\begin{array}{cc}{\mathbf 1}_k&0\\0& Q_\ell\end{array}\right].
\label{the set of transition functions for Q equal r over N, general solution}
\end{eqnarray}
The abelian transition functions, obeying the cocycle condition in the second line in (\ref{cocycle conditions for both}),  are taken to be
\begin{eqnarray}\label{u1transition}
\omega_1=e^{i\frac{2\pi(r-q_1 N)x_2}{N L_2}}\,,\quad \omega_3=e^{-i\frac{2\pi(1+q_3 N)x_4}{N L_4}}\,, \quad \omega_2=\omega_4=1\,,
\end{eqnarray}
where $q_1$  and $q_3$ are (for now) unrestricted arbitrary integer-quantized $U(1)$-fluxes.  The topological charges\footnote{The topological charges  only depend on the transition functions. In practice, it is easier to obtain them   by considering a particular background obeying boundary conditions with transition functions (\ref{cocycle conditions for both}), such as the one in section \ref{explicit:constant1}.} of the nonabelian ($A \in su(N)$) and abelian  ($a \in U(1)_B$) sectors are, recalling that ${\cal F} = F + \mathbf{1}_N d a$:
\begin{eqnarray}\label{charges 2}
Q_{SU(N)}&=& {1\over 8 \pi^2}\int\limits_{\T^4} \tr F \wedge F = \frac{r}{N} \; (\rm{mod} 1)\\
 Q_{U(1)_B}&=& {1\over 8 \pi^2}\int\limits_{\T^4} da \wedge da  =-\left(\frac{r}{N}-q_1\right)\left(\frac{1}{N}+q_3\right) \nonumber
\end{eqnarray}
and the corresponding Dirac index, or second Chern character, is
\begin{eqnarray}\label{dirac index 1}
{\rm ch}_2({\cal F}) &=&{1 \over 8 \pi^2} \int\limits_{\T^4} \tr {\cal F} \wedge {\cal F} =  Q_{SU(N)} + N Q_{U(1)_B} =  N q_1 q_3 - r q_3 + q_1 = \hat N\,.
\end{eqnarray}
In what follows, we shall consider   values of $q_{1,3}$ such that $\hat N > 1$.

\subsubsection{Properties of the Nahm dual of the fractional instanton via the families index}
\label{sec:propertiesviafamilies}
Given an $SU(N) \times U(1)_B/\Z_N$ irreducible background ${\cal A}$ of topological charge $\hat N$ given in  (\ref{dirac index 1}), the Dirac operator $\hat D$ (\ref{fundDirac}) has $\hat N$ zero modes and we can define, as in (\ref{defofdualA2}, \ref{defofdualA3}),  the Nahm dual $\hat{\cal A} \equiv \hat A + {\mathbf 1}_{\hat N} \hat a$.  The $U(\hat N)$ Nahm dual is, similarly,  in a $SU(\hat N) \times U(1)_{\hat B}/\Z_{\hat N}$ bundle.

The nonzero $U(1)_B$ fluxes (times $N$) associated with the transition functions  (\ref{u1transition}) are in the $12$ and $34$ planes only and are given by integrals over the corresponding two-planes in $\T^4$:
\ba
\label{u1fluxesonT4}
\oint_{\T^4 } {\tr {\cal F}_{12} \over 2 \pi} &=& N \; {q_1 N - r \over N}  = q_1 N - r,\\
\oint_{\T^4} {\tr {\cal F}_{34} \over 2 \pi} &=& N \; {q_3 N + 1 \over N} = q_3 N + 1,\nonumber
\ea
where we kept the factor of $N$ in the first equality in each line to emphasize the role played by the $U(N)$ embedding.

Without explicit calculation of $\hat{\cal A}$, we now use the last of equation in 
(\ref{familyindex22})
to find the $U(1)_{\hat B}$ fluxes in the Nahm dual, given by integrals over the corresponding two-planes in $\hat\T^4$:
\ba
\label{u1fluxesonhatT4}
\oint_{\hat\T^4} {\tr \hat{\cal F}_{12} \over 2 \pi} &=& -  \oint_{\T^4} {\tr {\cal F}_{34} \over 2 \pi}  = -  (q_3 N +1) ,\\
\oint_{\hat\T^4} {\tr \hat{\cal F}_{34} \over 2 \pi} &=&   -  \oint_{\T^4} {\tr {\cal F}_{12} \over 2 \pi} = - (q_1 N - r),\nonumber
\ea
These imply that $\hat f_{34} = - {2 \pi \over \hat L_3 \hat L_4} {q_1 N -r \over \hat N}$ and $\hat f_{12} = - {2 \pi \over \hat L_1 \hat L_2} {q_3 N + 1 \over \hat N}$. Thus, similar to (\ref{charges 2}), we have for the topological charge of the $U(1)_{\hat B}$ on $\hat\T^4$:
\ba \label{U1dualcharge}
\hat Q_{U(1)_{\hat B}} =  {1\over 8 \pi^2}\int\limits_{\hat\T^4} d\hat a \wedge d\hat a = {1 \over \hat{N}^2} (N q_3 +1) (N q_1 - r)~.
\ea

Next, from the  first equation in families index relations (\ref{familyindex22}), we know that $ch_2 (\hat{\cal F}) = \oint_{\hat\T^4} \tr {\hat{\cal F} \wedge \hat{\cal F} \over 8 \pi^2} = N$. Thus, using the analogue of (\ref{dirac index 1}) for $U(\hat N)$ and the value of $\hat Q_{U(1)_{\hat B}}$ from (\ref{U1dualcharge}), we find for the topological charge of the $SU(\hat N)$ part of the Nahm dual background, recalling that $\hat N = N q_1 q_3 - rq_3 + q_1$:
 \ba\label{fractionaldual1}
 \hat{Q}_{SU(\hat{N})} = ch_2 (\hat{\cal{F}}) - \hat N \hat Q_{U(1)_{\hat B}} = N -  {1 \over \hat{N}} (N q_3 +1) (N q_1 - r) = {r \over \hat{N}}~.
 \ea
Thus, an $SU(N)$ fractional instanton of charge $r/N$ on $\T^4$ is mapped, on the dual $\hat\T^4$, to a $SU(\hat{N})$ fractional instanton of charge $r/\hat N$, where $\hat N = N q_1 q_3 - r q_3 + q_1$.

From these equations, we observe that one can achieve self-duality of the fractional instanton under the Nahm transform: for example, requiring $\hat N=N$ for $r=1$, can be achieved if one takes  the $U(1)_B$ fluxes on $\T^4$ to be either $q_1=q_3=1$ or $q_3=0, q_1=N$. Recall, however, that irreducibility of the background $\cal A$ is necessary for the Nahm dual construction. As we shall see in our study of the Nahm duals of  constant field strength fractional instantons, irreducibility fails to hold for these particular choices and using the Nahm transform requires deforming the background, by including nonabelian components as in \cite{GarciaPerez:2000aiw,Gonzalez-Arroyo:2019wpu,Anber:2023sjn}. However, we stress that the relation (\ref{fractionaldual1}) is general, i.e.~it holds for the Nahm dual $\hat{\cal{A}}$ of any topologically nontrivial irreducible background $\cal{A}$  on $\T^4$.
 
\section{Explicitly finding the Nahm dual of constant-flux fractional instantons}

\label{explicit:constant1}

The $U(1)_B$ and $SU(N)$ constant flux backgrounds on the $\T^4$ with the above transition functions (and, therefore, topological charges as computed in the previous section) are, not including moduli, recalling the definition of $\omega$ from (\ref{omega}),
\begin{eqnarray}\label{the full abelian bck general r}
\nonumber
A_2&=&-\omega\left(\frac{r x_1}{Nk L_1 L_2}\right)\,,\quad  A_4=-\omega\left(\frac{x_3}{N\ell L_3 L_4}\right)\,, \quad A_1=A_3=0\,,\\
a_2&=&- \frac{2\pi (r-q_1 N)x^1}{N L_1L_2}\,,\quad a_4= \frac{2\pi (1+q_3 N)x_3}{N L_3L_4}\,, \quad a_1=a_3=0\,.
\end{eqnarray} 
In what follows, we shall be interested in studying the Dirac equation for fermions in the fundamental representation of $SU(N)$ and unit charge under $U(1)_B$, thus coupling to the linear combination 
\begin{equation}
\ca_\mu \equiv A_\mu +{\mathbf 1}_N a_\mu. \label{calA}
\end{equation}

To continue, we switch to the index notation where $C', D',... = 1,..., k$ and $C,D,... = 1,... \ell$. Then,  including also the general moduli, we rewrite the background (\ref{the full abelian bck general r}), in terms of ${\cal{A}}_\mu$.   The $k\times k$ components of the background (\ref{calA}) are:
\begin{eqnarray}
\label{Ukbackground}
\ca_{1 \; C'D'} &=& \delta_{C'D'} \; 2 \pi (z_1 - \ell \; \phi_{1 \; C'}) \nonumber \\
\ca_{2 \; C'D'} &=&\delta_{C'D'} \; 2 \pi (z_2 - \ell \; \phi_{2 \; C'} + {k q_1 - {r  } \over  k L_1 L_2} x_1)\\
\ca_{3 \; C'D'} &=&  \delta_{C'D'} \; 2 \pi (z_3 - \ell \;\phi_{3 \; C'}) \nonumber\\
\ca_{4 \; C'D'} &=& \delta_{C'D'} \; 2 \pi ( z_4 -  \ell \;  \phi_{4 \; C'} + { q_3 \over L_3 L_4}  x_3) \nonumber
\end{eqnarray}
where the $SU(k) \times U(1) \subset SU(N)$ moduli allowed are labeled $\phi_{\mu\; C'}$ and the $U(1)$ moduli are labeled $z_\mu$. The $SU(N)$ moduli are not all independent, but, in order to be consistent with the transition functions, are subject to the identifications\footnote{Thus, there are $4 \times$gcd$(k,r)$ independent $SU(N)$ holonomies $\phi_{\mu C'}$. For $r=1, k$, this is exactly the number required by the index theorem for a charge-$r/N$ $SU(N)$ instanton; for further discussion, see section \ref{sec:nahmdual}. Also, we use the notation $[x]_k \equiv x (\text{mod} k)$ throughout.}
\begin{eqnarray} \label{SUNmoduli}
\phi_{\mu \; C'} &=& \phi_{\mu \; C'-r (\text{mod} \; k)} \equiv \phi_{\mu\; [C'-r]_k},~\text{and we define } \; \tilde\phi_{\mu}  \equiv  {1 \over k} \sum\limits_{C'=1}^k \phi_{\mu \; C'}~.
\end{eqnarray}
The $ \ell \times \ell$ components of (\ref{calA}) are: 
\begin{eqnarray}
\label{Ulbackground}
\ca_{1 \; CD} &=&   \delta_{CD} \; 2 \pi(z_1 +  k \;  \tilde\phi_{1}) \nonumber \\ 
\ca_{2 \; CD} &=& \delta_{CD} \; 2 \pi (z_2 +  k \;  \tilde\phi_{2} + {q_1 \over   L_1 L_2} x_1 )\\
\ca_{3 \; CD} &=&  \delta_{CD} \;  2 \pi(z_3 + k\;  \tilde\phi_{3}) \nonumber\\
\ca_{4 \; CD} &=&   \delta_{CD}\; 2 \pi (z_4 + k \;   \tilde\phi_{4} + {\ell q_3+1 \over   \ell L_3 L_4} x_3)~.\nonumber
\end{eqnarray}
 In terms of the $U(N)$ field strength $\cal{F}$ of (\ref{Ulbackground}, \ref{Ukbackground}), the Dirac index (\ref{dirac index 1}) is seen to be 
 \begin{eqnarray} \label{dirac index 2}
 {\rm{ch}}_2({\cal{F}}) = {1 \over 8 \pi^2} \int\limits_{\T^4} \tr {\cal{F}} \wedge {\cal{F}}  = q_3 (k q_1 - r) + q_1 (\ell q_3 + 1) = N q_1 q_3 - r q_3 + q_1=\hat N\,, 
 \end{eqnarray}
while the projections to the $U(1)_B$ and $SU(N)$ topological charges (\ref{charges 2}) are also easily computable from (\ref{the full abelian bck general r}). For further convenience, we now introduce the notation:
\ba \label{hatkldef}
\hat N = \hat k + \hat \ell, ~ \hat k \equiv q_3 (k q_1 -r),~ \hat \ell \equiv q_1 (\ell q_3 +1)~,
\ea
where both $\hat k$ and $\hat \ell$ are assumed to be positive and greater than zero. As we show below for such values of $r, q_1, q_3$, the background (\ref{Ukbackground}, \ref{Ulbackground}) is irreducible. 

\subsection{The zero modes of the fundamental Dirac operator and the Nahm dual of the constant flux instanton}
\label{sec:nahmdual}

Since the $U(N = k+\ell)$ background (\ref{Ukbackground}, \ref{Ulbackground}) splits into $U(k)$ and $U(\ell)$ parts, the 
fundamental Dirac operator $\bar D=\bar\sigma_\mu D_\mu$ also factorizes.
 Explicitly, in the index notation of the previous section (recalling $B',C'=1,...k$, $B,C=1,...\ell$), we have the following equations for the normalizable $\bar D$ zero modes:
\ba
\label{DiracconstantF}
\bar\sigma_\mu (\delta_{C'B'}\partial_\mu + i {\cal A}_{\mu \; C'B'}) \Psi_{B' \; i}^{t_2, t_4} &=& 0, \nonumber \\
\text{where}~&t_2&= 0, ..., {kq_1 - r \over {\rm{gcd}}(k,r)} -1; ~i=0,...{\rm{gcd}}(k,r)-1;~ t_4 = 0,..., q_3-1; \nonumber \\
\bar\sigma_\mu(\delta_{CB}\partial_\mu + i {\cal A}_{\mu \; CB}) \Psi_{B}^{p_2,p_4} &=& 0, \nonumber \\
\text{where}~&p_2&= 0,..., q_1-1; ~ p_4 = 0,..., \ell q_3~.
\ea
In the above equation, we have labelled the zero modes by a convenient multi-index notation instead of the single $\hat N$-valued index $a$ form eqn.~(\ref{defofdualA2}). This is prompted by the fact that $\hat N = \hat k + \hat \ell$ and the parameterization of the normalizable zero modes found in Appendices \ref{appx:B5} and \ref{Embeddings} by explicitly solving for the zero modes. The $U(k)$ part of the $U(N)$ fundamental Dirac operator has $\hat k = q_3 (k q_1 -r)$ zero modes, $\Psi_{B' \; i}^{t_2, t_4}$, whose explicit expressions are given in (\ref{The final expressions of PSI in suk and sul}) and (\ref{final expression of Sigma moduli}). The $U(\ell)$ part has $\hat \ell = q_1 (\ell q_3 +1)$ zero modes, $\Psi_{B}^{p_2,p_4}$  given in (\ref{The final expressions of PSI in suk and sul}) and (\ref{final expression of Phi moduli}). This is exactly as per the index theorem (\ref{dirac index 2}, \ref{hatkldef}). Thus, for $\hat k, \hat \ell >0$ the background from the previous section is irreducible.\footnote{As already noted, a background with, say, $r=1$, $q_3=0$ and $q_1 = N$ is ``Nahm self-dual,'' i.e. leading to $\hat N = N$. However, the constant flux background ${\cal A}$ from (\ref{Ukbackground}, \ref{Ulbackground})  is now not irreducible: one can explicitly show that the $\bar D$ operator has more than $\hat N$ zero modes ($2N-1$, for $k=r=1$), implying that $D$ also has zero modes.}

The explicit expressions for the zero modes are given in the Appendices  \ref{appx:B5} and \ref{Embeddings}. They are used to explicitly find the $U(\hat N)$ Nahm dual (\ref{defofdualA2}) on $\T^4$. Since $\Psi_{B}^{p_2,p_4}$ and $\Psi_{B' \; i}^{t_2,t_4}$ are orthogonal, as are their derivatives (since they occupy different subgroups of $U(N)$), the Nahm dual  to the background  (\ref{Ukbackground}, \ref{Ulbackground}) automatically splits into a $U(\hat k)$ and a $U(\hat \ell)$ part. Explicitly, in terms of the normalized solutions of (\ref{DiracconstantF}), these components are:
\ba \label{explicitnahm1}\nonumber
\hat{\cal{A}}_\mu^{(p_2, p_4; p_2', p_4')} &=& - i \int\limits_{\T^4}\sum\limits_{\alpha, B}(\Psi_{\alpha \; B}^{p_2, p_4})^* \; { \partial_{z_\mu}} \Psi_{\alpha \; B}^{p_2', p_4'} \in u(\hat{k}), \nonumber \\\nonumber
&&~ \; p_2, p_2'= 0,...q_1 -1;\quad p_4, p_4' = 0,..., \ell q_3, \nonumber \\\nonumber
\hat{\cal{A}}_\mu^{(t_2, i, t_4; t_2', i', t_4')} &=&- i \int\limits_{\T^4} \sum\limits_{\alpha, B'}(\Psi_{\alpha \; B' \; i}^{t_2, t_4})^*  \;{ \partial_{z_\mu}} \Psi_{\alpha \; B' \; i'}^{t_2', t_4'} \in u(\hat{\ell}), \\\nonumber
&& t_2, t_2'= 0,..., {k q_1 -r \over \text{gcd}(k,r)}-1;\quad i, i' = 0,...\text{gcd}(k,r)-1; \quad t_4, t_4' = 0,...,  q_3, \nonumber\\
\ea
where we do not show the  $x, z$, and moduli dependence. The explicit expressions for the Nahm dual gauge field (\ref{explicitnahm1}) are given in the Appendix, see eqns.~(\ref{Uhatkfield}, \ref{Uhatellfield})   in the index notation employed above, while in a $\hat{k} \times \hat{\ell}$ matrix notation, they are given   in eqns.~(\ref{dual1}, \ref{dual2}, \ref{dual3}). We warn the reader that, depending on the gauge used, some of the moduli dependence is in the dual transition functions. The transition functions are given in Appendix \ref{appx:transitiondual}, in a gauge where they contain some moduli dependence, and in Appendix \ref{appx:transitiondualcanonical}, in a canonical form, where the moduli dependence is entirely in the gauge background.
 
 In terms of the field strength, 
the net result of the Nahm transformation is as follows.
The instanton on $\T^4$, given by eqns.~(\ref{Ukbackground}, \ref{Ulbackground}),  embedded into $U(N)$, with $N=k+\ell$ and $Q(SU(N))= {r\over N}$ has field strength given here in an explicit matrix notation:
\begin{eqnarray}\label{fieldstrengthoriginal}
\T^4:~  {\cal F}_{12} &=&{2\pi \over   L_1 L_2} \left(\begin{array}{cc}  { k q_1 - r \over k } \mathbf{1}_{  k} & 0 \cr 0 &  q_1 \mathbf{1}_{  \ell} \end{array}\right), \nonumber \\
  {\cal F}_{34} &=& {2\pi \over   L_3 L_4}\left(\begin{array}{cc} q_3 \mathbf{1}_{
 k} & 0 \cr 0 &   {  \ell q_3+1\over \ell }  \mathbf{1}_{  \ell} \end{array}\right).
\end{eqnarray}
The $U(N)$ background on $\T^4$ of eqn.~(\ref{fieldstrengthoriginal}) is self-dual provided:
\ba \label{t4selfduality}
{L_1 L_2 \over L_3 L_4} = {k q_1-r \over k q_3} = {q_1 \ell \over \ell q_3 +1}~.
\ea
Its $U(\hat N)$  Nahm dual   on $\hat\T^4$, with $\hat L_\mu=1/L_\mu$ and $\hat N = \hat k + \hat \ell$ and $\hat Q(SU(\hat N)) = {r \over \hat N}$, has also constant field strength:
 \ba \label{fieldstrengthdualU}
 \hat\T^4:~ \hat {\cal F}_{12} &=&{2\pi \over \hat L_1\hat L_2} \left(\begin{array}{cc} - {  k \over kq_1-r} \mathbf{1}_{\hat k} & 0 \cr 0 & - {1 \over q_1}  \mathbf{1}_{\hat \ell} \end{array}\right), ~ \hat k \equiv q_3 (k q_1 - r), ~\hat \ell \equiv q_1 (\ell q_3 + 1)~, \\
 \hat {\cal F}_{34} &=& {2\pi \over \hat L_3\hat L_4}\left(\begin{array}{cc} - {1  \over q_3} \mathbf{1}_{
\hat k} & 0 \cr 0 & - { \ell \over \ell q_3+1}  \mathbf{1}_{\hat \ell} \end{array}\right), \nonumber
 \ea
and, as is easily seen from the above equation, is self-dual provided ${\hat L_3 \hat L_4/(\hat L_1 \hat L_2)}$ equals the same expressions as the ones appearing on the r.h.s. in (\ref{t4selfduality})---in accordance with the general fact  \cite{Schenk:1986xe,Braam:1988qk}  that $U(N)$ self-duality implies self-duality of the $U(\hat N)$ Nahm dual.

Writing the field strength of the $U(\hat N)$ Nahm-dual $\hat {\cal F}_{\mu\nu}$ as $\hat {\cal F}_{\mu\nu}=\hat F_{\mu\nu}+\hat f_{\mu\nu}{\bm 1}_{\hat N}$, the field strength of the $SU(\hat N)$ part can be found by projecting the above equations to the traceless part:%
\begin{eqnarray}\nonumber
\hat\T^4: \hat F_{12}(z)&=&-{2\pi L_1L_2 \over \hat N} \left[\begin{array}{cc} r {\ell q_3 + 1 \over k q_1 - r} \; {\bf 1}_{\hat{k}} &0\\ 0 & - { r q_3 \over q_1}\;{\bf 1}_{\hat{\ell}} \end{array}\right]\,,\\
\hat F_{34}(z)&=&-{2\pi L_3 L_4 \over \hat N} \left[\begin{array}{cc} {q_1 \over q_3} \; {\bf 1}_{\hat{k}} &0\\ 0 &- {k q_1 - r\over \ell q_3+1}\;{\bf 1}_{\hat{\ell}} \end{array}\right]\,, \label{dualsuhatN}
\end{eqnarray}
while  the field strength of the dual $U(1)$ gauge field $\hat f$ is
\begin{eqnarray}\label{dula U1 fields}
\hat f_{12}(z)=-2\pi L_1 L_2 \frac{(q_3N+1)}{\hat N}\,,\quad \hat f_{34}(z)=-2\pi L_3 L_4 \frac{(Nq_1-r)}{\hat N}\,.
\end{eqnarray}
One can use these expressions to explicitly calculate the $SU(\hat N)$ topological charge (this is done in the Appendix, see eqns.~(\ref{indexappendix1}, \ref{indexappendix2}, \ref{indexappendix3})), finding agreement of the explicit calculation with the  families index theorem determination of Section \ref{sec:propertiesviafamilies}.

{\flushleft{S}}everal comments are now due, regarding this pair of constant field strength Nahm duals:
\begin{enumerate}
\item The self duality conditions for the $U(N)$ background on $\T^4$ and its $U(\hat{N})$ Nahm dual on $\hat \T^4$ are  identical, in agreement with the general properties of the Nahm transform that a $U(N)$ self dual background has a self dual $U(\hat N)$ background \cite{Schenk:1986xe,Braam:1988qk}.  

To see what self-duality entails in our case, we note that demanding $U(N)$ self duality (\ref{t4selfduality}) (and hence automatically $U(\hat N)$ self duality) can be assured by taking the $q_1, q_3$ fluxes obey:
\ba\label{conditions1}
r=1:&&~ (\ell q_3 +1) (k q_1 -1) = kq_1 \ell q_3  \; \implies \;  \ell q_3 + 1= k q_1, \nonumber \\
r=k:&&~(\ell q_3 +1) ( q_1 -1) =  q_1 \ell q_3  \; \implies \;  \ell q_3 + 1= q_1.
\ea

\item  We stress that the $r=1$ and $r=k$ cases shown are of most interest: only in these  cases does  the number of independent holonomies $\phi_C$ (the ones commuting with the transition functions, as per (\ref{SUNmoduli})) equal  the number of moduli expected by the index theorem for a solution of $r$ times the minimal charge.\footnote{It was seen in \cite{Anber:2023sjn} that this mismatch of the number of independent holonomies and number of moduli creates problems in the leading-order $\Delta$ expansion. These could be fixed at higher order, but this has not been yet understood. We thank A. Gonz\' alez-Arroyo for discussions on this.} We thus focus in our examples on $r=1,k$.

\item For studies of $SU(N)$ or $SU(\hat N)$ field theory dynamics one is interested in self-dual  $SU(N)/SU(\hat N)$ backgrounds. Imposing self duality for  $SU({N})$ only, as follows from the first line in (\ref{the full abelian bck general r}), implies that:
\ba \label{SUNduality}
 SU(N): \; {L_1 L_2 \over L_3 L_4} = {r \ell \over k}.
\ea
Clearly, this condition does not imply the $SU(\hat{N})$ self-duality condition, which, as per (\ref{dualsuhatN}), has the form:
\ba\label{SUhatNduality}
SU(\hat N): \; {\hat L_3 \hat L_4 \over \hat L_1 \hat L_2} = {L_1 L_2 \over L_3 L_4} = { q_1 (k q_1 -r) \over r q_3 (\ell q_3 + 1)}. \ea
Thus, 
 if we demand that both $SU(N)$ and $SU(\hat N)$ self-duality hold, we find:
\ba\label{conditions2}
r=1:&&~ (\ell q_3 +1) \ell q_3 = kq_1 (k q_1 -1), \; \implies \;  \ell q_3 + 1= k q_1, \nonumber \\
r=k:&&~ (\ell q_3 +1) \ell q_3 =  q_1 (q_1 -1), \; \implies \;  \ell q_3 + 1= q_1.
\ea
We note that while the conditions on the l.h.s. of (\ref{conditions1}) appear different from those on the l.h.s. of (\ref{conditions2}), their solutions given on the r.h.s.  in the $SU(N/\hat N)$ self duality condition are the same as the solutions of the $U(N)$ self-duality conditions on the r.h.s. in (\ref{conditions1}).\footnote{
That the constraints on the r.h.s. of (\ref{conditions1}, \ref{conditions2}) can be solved is seen by considering examples: e.g.~taking $N=2$ with $k = \ell = 1$, solving (\ref{conditions2}) by $q_1 = 2, q_3=1$, we find $\hat N=5$. Thus, a self-dual minimal charge solution for $SU(2)$ is mapped to a minimal charge solution for $SU(5)$ (recall $q_1 >1$ and $q_3 \ge 1$ for the constant flux background to be irreducible). In some cases, however, there are no solutions: notably, for $\ell$ and $k$ even, the $r=1$ condition can not be solved for integer $q_{1,3}$. }

\item In constructing the Nahm dual, we have kept track of the moduli dependence throughout, see eqns.~(\ref{Uhatkfield}, \ref{Uhatellfield}) for the Nahm dual gauge field $\hat{\cal A}$. Thus, we can also study the moduli space of both the $U(N)$ instanton and its $U(\hat N)$ Nahm dual. We define the $U(N)$ moduli space metric, without the conventional factor of $2/g^2$ used in instanton calculus,  as\footnote{For brevity, eqn.~(\ref{modulimetric}) is written for $r=1$. For $r=k$, the definition is the same, except that the metric has to now have a composite index, i.e. read $g_{\nu' i', \nu i}$ with $i=1,...,k,$ and the derivatives $\delta^{(\nu i)} {\cal A}_\mu$ should now be w.r.t. $\phi_{\mu  i} L_\mu$.}
\ba \label{modulimetric}
g_{\nu'\nu}^{U(N)} = \int\limits_{\T^4} \tr  \delta^{(\nu)} {\cal A}_\mu \; \delta^{(\nu')} {\cal A}_\mu, \;  \text{with} \; \delta^{(\nu)} {\cal A}_\mu = {\delta {\cal A}_\mu \over \delta (\phi_\mu L_\mu)}, \ea
where $\phi_\mu$ are the moduli (\ref{SUNmoduli}). We use  an  identical expression to define $\hat g_{\nu'\nu}^{U(\hat N)}$, the $U(\hat N)$ instanton moduli space metric,  by replacing ${\cal A}$ in (\ref{modulimetric}) with its Nahm dual ${\cal \hat A}$, and the integral over $\T^4$ by one over $\hat\T^4$. Then, by explicit calculation\footnote{We do not give the details, as the result follows from eqns.~(\ref{Ukbackground}, \ref{Ulbackground}) for $\cal A$ and  (\ref{Uhatkfield}, \ref{Uhatellfield}) for $\hat{\cal A}$.} we find, for both $r=1$ and $r=k$, after imposing the self-duality condition (\ref{t4selfduality}):
\begin{equation}\label{metricsdual2}
{1 \over \sqrt{V}} \; g_{\nu'\nu}^{U(N)} = {1 \over \sqrt{\hat V}}\; \hat g_{\nu'\nu}^{U(\hat N)}~.
\end{equation}
In other words, the two moduli-space metrics are identical up to a volume normalization factor, agreeing with the general statement of \cite{Gonzalez-Arroyo:1998nku,GarciaPerez:1999bc} of the Nahm transform giving an isometry of the moduli spaces.
\item A moduli space metric for the $SU(N)$ and $SU(\hat N)$ factors can be defined similarly, by replacing ${\cal A} \rightarrow A$, $\hat {\cal A} \rightarrow \hat A$ in the definition (\ref{modulimetric}). Now we note that $g_{\nu'\nu}^{U(N)} = g_{\nu'\nu}^{SU(N)}$, due to the fact that the moduli $\phi_\mu$ appear in the $SU(N)$ part  of the $U(N)$ background only. However, this is not true for the Nahm dual, for which $\hat g^{SU(\hat N)}_{\nu' \nu} \ne \hat g^{U(\hat N)}_{\nu' \nu}$, because the moduli $\phi_\mu$ also appear in the $U(1)$ part of the $U(\hat N)$ Nahm dual---as seen for example in eqn.~(\ref{expressions of dual abelian field for general r}) or in the form of the $U(1)$ Wilson lines on $\hat \T^4$ (\ref{dualU1wilson}).

However, one can study the moduli space metric of the $SU(\hat N)$ fractional instanton independently of the $U(1)$ part of the Nahm dual.  Briefly, as in \cite{Anber:2022qsz,Anber:2024mco}, in Appendix \ref{appx:moduli} we  study the behaviour of the fractional instanton under the $\Z_{\hat N}^{(1)}$ $1$-form center symmetry of the $SU(\hat N)$ theory on $\hat\T^4$. As in \cite{Anber:2022qsz,Anber:2024mco},  one thus determines the range of the moduli in the $SU(\hat N)$ theory, denoted by $\Gamma(\phi)$, by extending the range of the moduli such that all center symmetry images are included. The volume of the moduli space $\mu^{SU(\hat N)}$ then determines the normalization of  the gaugino condensate. 
In the $SU(\hat N)$ theory, one finds the same expression for the volume of the moduli space
as in the $SU(N)$ theory, with $N \rightarrow \hat N$ and the gauge coupling replaced by the gauge coupling of the $SU(\hat N)$ theory---as shown in eqn.~(\ref{final result sun using phi}), for $r=1$, reproduced below \begin{eqnarray}\label{final result sun using phi 1}
\mu^{SU(\hat N)}=\int_{\Gamma(\phi)}d\mu^{SU(\hat N)}=\frac{16\pi^2 \hat N^2}{g^4}\,.
\end{eqnarray}
As already noted, up to the replacement of coupling and rank, this is the same expression obtained in the gaugino condensate calculation on $\T^4$ \cite{Anber:2024mco}. \end{enumerate}

\subsection{Towards the Nahm dual of non-constant fractional instantons via the $\Delta$-expansion}

\label{sec:towards}

One  motivation to study the Nahm transform is that it could, in principle, allow one to construct new solutions, by taking a known fractional instanton solution on $\T^4$, adding appropriate $U(1)$ fluxes, and constructing its $\hat\T^4$ Nahm dual. As noted before,  the Nahm transform is well defined for generic self-dual solutions, as for the generic case self-duality guarantees irreducibility of the background.  

However, there are only a few  known self-dual fractional instanton solutions on $\T^4$. The only closed form solutions  to date are those of 't Hooft, the ones considered so far in this paper. For these solutions, as we showed, constant-$F$ (self-dual) backgrounds on $\T^4$ are mapped to constant-$\hat F$ (self-dual) backgrounds on $\hat\T^4$. These Nahm duals also belong to the general class of solutions constructed by 't Hooft, so, strictly speaking, we have not obtained any new fractional instanton solutions.

There are, however, also self-dual fractional instantons in $SU(N)$, known in a series expansion in a small parameter, the detuning parameter $\Delta$ of $\T^4$. The Nahm transform can be applied to these solutions. One can thus pose the question as to the nature of their Nahm dual.

 Here, we simply consider a few generic features, with any explicit calculations left outside the scope of this paper. To this end, it suffices to consider a particular case as an illustration.
 Consider $N=2$, $\ell = k =1$, $Q=1/2$, example, where the $\T^4$ asymmetry parameter is
\ba \label{delta1}
\Delta = {L_1 L_2 - L_3 L_4 \over \sqrt{L_1 L_2 L_3 L_4}}~ \implies {L_1 L_2 \over L_3 L_4} \simeq 1 + \Delta +  \ldots,
\ea
where we assumed $\Delta >0$ and, on the r.h.s., expanded for small $\Delta$. Taking $\Delta=0$, one obtains the self-dual constant-$F$ background. For $\Delta>0$, a nonabelian self-dual background is constructed as a series in the $\T^4$ asymmetry parameter $\Delta$, see \cite{GarciaPerez:2000aiw, Anber:2022qsz}.

The Nahm transform---after adding  appropriate $U(1)$ fluxes $q_1, q_3$, as we discuss below---can be applied to these self-dual $SU(2)$ solutions with nonabelian field strength, by a straightforward generalization of the procedure used to obtain (\ref{explicitnahm1}).
The nonabelian components are small, of order $\sqrt{\Delta}$ (and higher powers thereof). The fundamental Dirac zero modes can then also be found in terms of powers of $\sqrt\Delta$. Their leading $\Delta^0$ term is given by  the zero modes determined in this paper, but there will be order-$\sqrt\Delta$ terms mixing the $U(\hat k)$ and $U(\hat \ell)$ components in (\ref{DiracconstantF}). 
Then, the Nahm dual connection can also be determined in a $\Delta$ expansion, by generalizing (\ref{explicitnahm1}) to also include off-diagonal components.
Clearly, the result will be equal to the Nahm dual to the constant-$F$ background plus small nonabelian pieces---the off-diagonal $\hat k \times \hat \ell$ components, proportional to
 $\sqrt\Delta$  and higher powers. 
 
With the sides of $\T^4$ tuned as in (\ref{delta1}), we have for the asymmetry parameter of the dual $\hat\T^4$ the following expression:
\ba
\label{delta2}
 {\hat L_1 \hat L_2 \over \hat L_3 \hat L_4} = 1 - \Delta +... .
\ea

Now, while an $U(\hat N)$,  $\hat N = 2 q_1 q_3 - q_1 + q_3$ (recall that $N=2$, $r=1$ here) Nahm dual can be constructed as outlined above, it will not be self-dual unless we choose $q_1$ and $q_3$ so that the $U(2)$ background on $\T^4$ is self dual. As the $SU(2)$ part is self dual by construction, order by order in the $\Delta$ expansion, we have to impose $U(1)$ self duality.\footnote{We stress that the $U(1)$ self-duality requirement holds independent of what $SU(N)$ self-dual fractional instanton we start with, and, in particular, on whether we use the $\Delta$-expansion or not. Our $SU(2)$ example suffices to  illustrate this point.}  The second line in (\ref{the full abelian bck general r}) implies that we have to ensure that 
\ba\label{u1selfduality}
{L_1 L_2 \over L_3 L_4} = {2 q_1 -1 \over 2 q_3 +1}, \; \text{or}\;  q_1 = q_3 \; {L_1 L_2 \over L_3 L_4} + {1 \over 2}( {L_1 L_2 \over L_3 L_4} +1). \ea
Remembering that $q_1$ and $q_3$ are integer $U(1)_B$ fluxes, we conclude that for general values of the ratio $L_1L_2/(L_3 L_4)$, the $U(1)$ self-duality condition (\ref{u1selfduality})   restricts the values of $q_1, q_3$. In particular, finding a solution requires that the ratio $L_1 L_2/(L_3 L_4)$ be rational.
For the example when (\ref{delta1}) holds, defining $L_1 L_2/(L_3 L_4) = 1 + {2 s\over K}$, with integer $K,s$ and $K \gg s$, we find the condition $K q_1 = q_3 (K + 2 s) + s + K$ that should be obeyed by the integer $U(1)$ fluxes in order that the Nahm dual be self-dual. Clearly, this equation can be solved, but we shall not pursue this further here.

Thus, it appears that constructing Nahm duals in the framework of the $\Delta$-expansion that give rise to self-dual $SU(\hat N)$ fractional instantons on $\hat \T^4$ is constrained by the need to impose $U(1)$ self-duality on $\T^4$, which restricts the choice of $U(1)$ fluxes $q_1, q_3$ in the $12$ and $34$ planes, as we just illustrated.\footnote{One might also introduce further integer $U(1)$ fluxes on $\T^4$, e.g.~in the $13$ and $24$ planes in addition to $q_1, q_3$; these extra fluxes appear to make the $U(1)$ self-duality condition even more restrictive, although we have not studied this in detail.} It might thus be of interest to investigate how the methods of \cite{Gonzalez-Arroyo:1998nku,GarciaPerez:1999bc}---which do not use the addition of $U(1)$ fluxes---construct the Nahm dual of non-constant field strength solutions within the $\Delta$ expansion.

Finally, we note that the $U(N)$ embedding of fractional instantons of $SU(N)$ on $\T^4$  is precisely how 't Hoofts constant-$F$ fractional instanton backgrounds have been found to appear in string theory. It was argued there, see \cite{Guralnik:1997sy,Hashimoto:1997gm}, that the BPS condition for the $D$-brane configuration does not require $U(1)$ self-duality. In fact,  the  BPS conditions for the two stacks of $k$ and $\ell$ $D_2$-branes\footnote{Recall that $N=k+\ell$. The  stacks of $k$ and $\ell$ $D_2$ branes at angles are $T$-dual to fluxes in the $D_4$-brane gauge theory.} wrapped on $\T^4$ constrains the ratio of the sides of the torus to be precisely the $SU(N)$ self-duality condition, the one  in our eqn.~(\ref{SUNduality}). If the ratio $L_1 L_2/(L_3 L_4)$ does not obey  (\ref{SUNduality}), one finds a tachyon in the spectrum of open strings between the stacks of $D$-branes. It might be of interest to study if any further insight into fractional instantons with nonconstant-$F$ can be gained by studying the string theory embedding and the fate of the tachyonic instability as the sides of the torus are detuned from the BPS condition. 

{\bf {\flushleft{Acknowledgments:}}}  We would like to express our gratitude to Margarita García Pérez and Antonio González-Arroyo for the many insightful discussions, as well as to the Instituto de Física Teórica, UAM-CSIC, Madrid, for their warm hospitality. M.A. is supported by STFC through grant ST/T000708/1.   E.P. is supported by a Discovery Grant from NSERC.

 \appendix
 \section{Notation and some properties of the  Nahm transform}
 
\label{appx:notation1}
 
 \subsection{Spinor notation}
 
 \label{appx:notation12}
 
 We work in 4d Euclidean space, $\mu = 1,2,3,4$. The notation is as in \cite{Anber:2022qsz,Anber:2023sjn} (all as in \cite{Dorey:2002ik}, except we use Hermitean gauge fields):
\begin{eqnarray}\label{sigmadefinitions}
\sigma_\mu &=& (i \vec\tau, 1),~~\bar\sigma_\mu = (-i \vec\tau, 1),\nonumber\\
\sigma_{\mu\nu} &=& {1 \over 4}(\sigma_\mu \bar\sigma_\nu - \sigma_\nu\bar\sigma_\mu), ~~\bar\sigma_{\mu\nu} = {1 \over 4}(\bar\sigma_\mu  \sigma_\nu - \bar\sigma_\nu\sigma_\mu),\nonumber \\
\sigma_\mu \bar\sigma_\nu + \sigma_\nu\bar\sigma_\nu &=& 2 \delta_{\mu\nu},~~\bar\sigma_\mu\sigma_\nu +  \bar\sigma_\nu \sigma_\nu = 2 \delta_{\mu\nu}~,
\end{eqnarray}
where $\sigma_{\mu\nu} = {1 \over 2} \epsilon_{\mu\nu\lambda\sigma} \sigma_{\lambda\sigma}$ is selfdual (so $\sigma_{\mu\nu}\bar\sigma_{\mu\nu} =0$, with $\bar\sigma_{\mu\nu}$ antiselfdual). 
In some occasions we shall need to display the spinor indices, as shown below:
\begin{eqnarray}\label{abc1}
\sigma_{\mu \; \alpha \dot\alpha}, ~ \bar\sigma_{\mu}^{\dot\alpha \alpha}, ~(\sigma_{\mu\nu})_\alpha^{~~\beta}, ~ (\bar\sigma_{\mu\nu})^{\dot \alpha}_{~~\dot\beta}~.
\end{eqnarray}
The matrices shown in (\ref{sigmadefinitions}) are assumed to have their indices as displayed in (\ref{abc1}).

 \subsection{Motivating the Nahm dual connection on $\hat{\T}^4$} 
 \label{appx:notationDefining}

 To understand the construction of the Nahm dual connection in a pedestrian way, consider a $z$-dependent linear combination of the zero modes of the fundamental Dirac operator $\bar D$ (\ref{zeromodes1}, \ref{zeromodes3}). It is defined by its expansion coefficients,  $f^a(z)$, which live on $\hat{\T}^4$. Thus, define the $z$-dependent vector in space of the zero modes, $\{ \psi^a, a=1,...,\hat N\}$,  via its expansion as a sum over the $\hat{N}$ zero modes
 \begin{eqnarray}\label{vector1} 
\psi^a_{A\alpha}(x,z)  f^a(z). 
 \end{eqnarray}
Here, $f^a$ is taken to transform as fundamental under $U(\hat{N})$: since $\psi^a \rightarrow \psi^b (g_{\hat{N}}^{-1})^{ba}$ from (\ref{UQdef}), we have that $f^a \rightarrow g_{\hat{N}}^{ab} f^b$.
To transport (\ref{vector1}) over $z$, we attempt to  define a $z$-derivative
 \begin{eqnarray}\label{derivative}
 \partial_{z_\mu} (f^a(z) \psi^a_{A\alpha}(x,z)) =  \partial_{z_\mu} f^a(z) \psi^a_{A\alpha}(x,z)) +  f^a(z) \partial_{z_\mu} \psi^a_{A\alpha}(x,z),
 \end{eqnarray}
but find that we are taken out of the space with basis $\psi^a_{A \alpha}$, because the vectors themselves change, and, in particular, $\partial_{z_\mu}\psi$ is not in the zero mode space. But we can project back to the $\psi^a$ space using the projector on the space of zero modes of $D \bar D$:
\ba\label{project}
  \sum_a \psi^a_{A \; \alpha}(x,z) (\psi_{B \; \beta}^a(y,z))^*, \ea
where the projection action  (upon  vectors $\psi_{B \; \beta}^c(y,z)$) includes summing over $B \beta$ and integrating over $y$, as in the first line of eqn.~(\ref{nahmmotivate}) below.

Thus, we define the covariant derivative action on $f^a(z)$ from (\ref{vector1}) as obtained upon the action of the projection operator (\ref{project}) on the derivative (\ref{derivative}). This allows us to define the Nahm dual connection as follows:
\begin{eqnarray}\label{nahmmotivate}
\hat\nabla_\mu (f^b(z) \psi^b_{A\alpha}(x,z))&\equiv&  \underbrace{ \sum_{a, B, \beta}  \psi^a_{A \; \alpha}(x,z) \int d^4 y  (\psi_{B \; \beta}^a(y,z))^*}_\text{projector} \; \underbrace{\partial_{z_\mu} (f^b(z) \psi^b_{B\beta}(y,z))}_\text{derivative} \nonumber\\
&=& \partial_{z_\mu} f^a(z) \psi^a_{A\alpha}(x,z)  + \sum_{a, b} f^b(z)  \psi^a_{A \; \alpha}(x,z) \underbrace{ \sum_{B, \beta} \int d^4 y  (\psi_{B \; \beta}^a(y,z))^*\; \partial_{z_\mu}  \psi^b_{B\beta}(y,z)}_\text{connection = $i \hat A_\mu^{ab}(z)$}  \nonumber\\
&=& \psi^a_{A\alpha}(x,z) (\partial_{z_\mu} \delta^{ab}  + i \hat A_\mu^{ab}(z)) f^b(z)~.
\end{eqnarray}
We conclude that the connection $\hat A_\mu^{ab}(z)$, the Nahm dual connection, is 
\begin{eqnarray}\label{defofdualA1}
\hat{\cal{A}}(z)_\mu^{ab} = - i \int_{\T^4}  (\psi^{a}_{A \alpha}(x,z))^* \partial_{z_\mu} \psi^b_{A \alpha}(x,z)~,~ a,b=1,... \hat{N}, 
\end{eqnarray}
with the spinor and $U(N)$ indices summed over.
That this is a $U(\hat{N})$ connection is also seen from the fact that applying (\ref{UQdef}), $\psi^a(x,z) \rightarrow\psi'^{a}(x,z) =  \psi^b(x,z) (g_{\hat{N}}^{-1}(z))^{ba}$, we find that:
\begin{eqnarray}\label{dualgauge}
g_{\hat{N}}(z): ~   \hat{\cal{A}}(z) \rightarrow g_{\hat{N}}(z) (\hat{\cal{A}}(z) - i d) g^{-1}_{\hat{N}}(z).
\end{eqnarray}

\subsection{The families index and the topological properties of the Nahm dual background}
\label{appx:notationFamilies}

 These follow from the families index theorem, which 
relates  the total  Chern character $ch(\hat E)$ of the Nahm dual background on $\hat \T^4$ to the one of  $\T^4$ background, $ch(E)$.  Using  a notation explained below (see e.g. \cite{Schenk:1986xe,Braam:1988qk,Hori:1999me}), it takes the form 
\ba
ch(\hat E) =  \int_{\T^4} ch(E) ch(P)~. \label{familyindex11}
\ea
This is  a formal expression where  in each term on the r.h.s. one should pick up and integrate a four-form over $\T^4$, see the explicit expression in (\ref{usingrelation1}) below.
Here $ch(P)$ is the character of the Poincar\' e bundle  over $\T^4 \times \hat \T^4$, with curvature $dz_\mu \wedge dx_\mu$ (summed over $\mu$),\footnote{This is the curvature (over $2\pi$) of the flat $U(1)$ connection on $\T^4$, $a = 2 \pi z_\mu dx_\mu$, but now considered over $\T^4 \times \hat\T^4$, i.e. as a function of both $x$ and $z$.} explicitly:
\ba\label{poincarechern}
ch(P) = e^{ dz_\mu \wedge dx_\mu}\,.
\ea
On the other hand, writing for $E$ only (identical formulae apply for $\hat E$, with ${\cal F}\rightarrow \hat{\cal F}$), we have for the total Chern character:
\ba
ch E &=& ch_0({\cal F}) + ch_1({\cal F}) + ch_2({\cal F})~, \label{chern1}
\ea
which has an expansion in terms of the $k$-th Chern characters $ch_k(F)$. 
 The zeroth Chern character is  $ch_0(F)$ (the ``rank of the fiber,'' or simply the dimensionality of the group generators $T^a$, e.g. $rk({\cal F})=k$ for the fundamental of $U(k)$).
The Chern characters relevant for $\T^4$ are, then (recalling that we use Hermitean connections as opposed  to \cite{Nakahara:2003nw}):
\ba \label{chern3}
ch_0({\cal F}) &=& rk({\cal F}), \nonumber \\
ch_1({\cal F}) &=&  c_1({\cal F}) = - \tr {{\cal F} \over 2\pi},\nonumber \\
ch_2({\cal F}) &=&  {1\over 2} (c_1({\cal F}))^ 2- c_2({\cal F})= {1 \over 8\pi^2}  \tr{{\cal F} \wedge {\cal F} } ~,
\ea
where $c_1, c_2$ are the first and second Chern classes, whose explicit expression can be inferred from the above.
Thus,  plugging (\ref{chern3}) for ${\cal F}$ and $\hat {\cal F}$ and (\ref{poincarechern}) in (\ref{familyindex11}), we find the relation:
\ba\label{usingrelation1}
&&rk(\hat {\cal F}) -  \tr {\hat {\cal F} \over 2\pi} + \tr {\hat {\cal F} \wedge \hat {\cal F} \over 8 \pi^2}   \\
&=& \int_{\T^4} \left(1 + {1 \over 2} (dz_\mu\wedge dx_\mu)^2+ {1 \over 4!} (dz_\mu\wedge dx_\mu)^4\right) \wedge \left(N -      \tr { {\cal F} \over 2\pi}  + \tr {  {\cal F} \wedge  {\cal F} \over 8 \pi^2}   \right) \nonumber \\
&=&  N \int_{\T^4} {1 \over 4!} (dz_\mu\wedge dx_\mu)^4 + {1 \over 2} dz_\mu \wedge dz_\nu \int_{\T^4} dx_\mu\wedge dx_\nu \wedge \tr { {\cal F} \over 2\pi} + \int_{\T^4} \left(\tr {  {\cal F} \wedge  {\cal F} \over 8 \pi^2} \right)  \nonumber 
\ea
We conclude, after integration over $\T^4$, that:
\ba \label{familyindex1}
 \tr {\hat {\cal F} \wedge \hat {\cal F} \over 8 \pi^2}  &=& N \int_{\T^4} {1 \over 4!} (dz_\mu\wedge dx_\mu)^4\,, \nonumber \\
rk(\hat {\cal F}) &=&  \int_{\T^4}  \tr {  {\cal F} \wedge  {\cal F} \over 8 \pi^2}\,,\\
  \tr {\hat {\cal F} \over 2\pi} &=&- {1 \over 2} dz_\mu \wedge dz_\nu \int_{\T^4} dx_\mu\wedge dx_\nu \wedge \tr { {\cal F} \over 2\pi}\,. \nonumber
\ea
The first two relations in (\ref{familyindex1}) establish that the topological charge and rank of the $U(\hat N)$ bundle over $\hat\T^4$ are related to the rank and topological charge of the $U(N)$ bundle on $\T^4$, respectively. The third equation above   relates  the corresponding $U(1)$ fluxes over the various $2$-planes in $\T^4$ and $\hat\T^4$.

In  \cite{Schenk:1986xe,Braam:1988qk} it is further shown  that $\hat{\cal A}$ is a self dual $U(\hat N)$ connection on $\hat\T^4$  if the $\T^4$ $U(N)$ connection ${\cal A}$ is   self dual of charge $\hat N$ (we will not show details, but only note that these statements follow by explicit calculations from the definition (\ref{defofdualA1}) of $\hat{\cal{A}}^{ab}$ and the irreducibility of ${\cal A}$).

\section{Calculating the Nahm dual of the constant flux background}
\label{appx:calculating}
\subsection{Solution of the Weyl equation}
\label{appx:solution}
Here, we solve the Weyl equation $D_\mu \bar\sigma^\mu \psi = 0$, where $D_\mu = \partial_\mu + i A_\mu + i a_\mu + i 2\pi z_\mu$, and $A_\mu$, $a_\mu$ are defined in (\ref{the full abelian bck general r}).\footnote{Here we only include the $U(1)$ Wilson line $z_\mu$; the other moduli  appearing in (\ref{Ukbackground}) and (\ref{Ulbackground}), $\phi_\mu$, will be included at a later stage.}
To proceed, we express $\psi_\alpha$ as a combination of $k \times 1$ and $\ell \times 1$ column vectors, denoted by $\psi_{\alpha}^{(k)}$ and $\psi_{\alpha}^{(\ell)}$:
$
\psi_{\alpha} = \left[\begin{array}{c} \psi_\alpha^{(k)} \\ \psi_\alpha^{(\ell)} \end{array}\right]\,,
$
where $\alpha=1,2$ are the spinor indices. 
This transforms the Weyl equation into the following system of differential equations:
\begin{eqnarray}\label{set 3}
\nonumber
&&\left(\partial_4-i\partial_3+i\frac{2\pi q_3x_3}{L_3L_4}+2\pi(z_3+i z_4)\right)\psi_1^{(k)}+ \left(-i\partial_1-\partial_2+i\frac{2\pi(\frac{r}{k}-q_1)x_1}{L_1L_2}+2\pi(z_1-i z_2)\right)\psi_2^{(k)}=0\,,\\
\nonumber
&&\left(-i\partial_1+\partial_2-i\frac{2\pi (\frac{r}{k}-q_1)x_1}{L_1L_2}+2\pi(z_1+i z_2)\right)\psi_1^{(k)}+ \left(\partial_4+i\partial_3+i\frac{2\pi q_3 x_3}{L_3 L_4}+2\pi(-z_3+i z_4)\right)\psi_2^{(k)}=0\,, \\
\end{eqnarray}
and
\begin{eqnarray}\label{set 4}
\nonumber
&&\left(\partial_4-i\partial_3+ i\frac{2\pi (1+\ell q_3) x_3}{\ell L_3L_4}+2\pi(z_3+i z_4)\right)\psi_1^{(\ell)}+ \left(-i\partial_1-\partial_2-i\frac{2\pi q_1 x_1}{L_1L_2}+2\pi(z_1-i z_2)\right)\psi_2^{(\ell)}=0\,,\\
\nonumber
&&\left(-i\partial_1+\partial_2+i\frac{2\pi q_1x_1}{L_1L_2}+2\pi(z_1+i z_2)\right)\psi_1^{(\ell)} + \left(\partial_4+i\partial_3+i\frac{2\pi (1+\ell q_3) x_3}{\ell L_3L_4}+2\pi(-z_3+i z_4)\right)\psi_2^{(\ell)}=0\,.\\
\end{eqnarray}
These equations are supplemented with the boundary conditions:
\begin{eqnarray}
\nonumber
\left[\begin{array}{c} \psi_\alpha^{(k)}\\\psi_\alpha^{(\ell)} \end{array}\right](x+\hat e_1 L_1)=\left[\begin{array}{c} e^{i \frac{2\pi (r/k-q_1)x_2}{L_2}}P_k^{-r}\psi^{(k)}_\alpha\\ e^{-i\frac{2\pi q_1x_2}{L_2}}\psi_\alpha^{(\ell)} \end{array}\right](x)\,,\quad \left[\begin{array}{c} \psi_\alpha^{(k)}\\\psi_\alpha^{(\ell)} \end{array}\right](x+\hat e_2 L_2)=\left[\begin{array}{c} Q_k \psi^{(k)}_\alpha\\ \psi_\alpha^{(\ell)} \end{array}\right](x)\,,\\
\label{bc k}
\end{eqnarray}
and
\begin{eqnarray}
\nonumber
\left[\begin{array}{c} \psi_\alpha^{(k)}\\\psi_\alpha^{(\ell)} \end{array}\right](x+\hat e_3 L_3)=\left[\begin{array}{c} e^{-i\frac{2\pi q_3 x_4}{L_4}}\psi^{(k)}_\alpha\\ e^{-i\frac{2\pi (1+\ell q_3) x_4}{\ell L_4}} P_\ell \psi_\alpha^{(\ell)} \end{array}\right](x)\,,\quad \left[\begin{array}{c} \psi_\alpha^{(k)}\\\psi_\alpha^{(\ell)} \end{array}\right](x+\hat e_4 L_4)=\left[\begin{array}{c}  \psi^{(k)}_\alpha\\ Q_\ell \psi_\alpha^{(\ell)} \end{array}\right](x)\,,\\
\label{bc ell}
\end{eqnarray}
which can be derived by the use of (\ref{BCS for F}) and (\ref{the set of transition functions for Q equal r over N, general solution}).

Before we continue, we also recall the index notation used in (\ref{Ukbackground}, \ref{Ulbackground}). In this notation we can rename the $U(k)$ and $U(\ell)$ components of the fundamental fermions as  $\psi_\alpha^{(k)} \rightarrow \psi_{\alpha C'}$, $C'=1,...,k$, and $\psi_\alpha^{(\ell)} \rightarrow \psi_{\alpha C}$, $C=1,...,\ell$. This index notation is used throughout in what follows. 

\subsection{The second order equation and the normalizability condition}
\label{sec:normalizability}
Here, we use the second order equations for $\psi^{(k)}$ and $\psi^{(\ell)}$ derived from iterating the first order equations (\ref{set 3}, \ref{set 4})  to argue that a normalizable solution to the above system of equations is obtained only if we set $\psi^{(k)}_1=\psi^{(\ell)}_1=0$. 

The corresponding 2nd order equations for $\psi_{\alpha C'}$ are written in terms of the frequencies defined as:
\begin{eqnarray}
\omega'_{12} = {2 \pi (q_1 - r/k) \over L_1 L_2}, ~\omega'_{34}  = {2 \pi   q_3 \over  L_3 L_4} \label{omegaprime}~.
\end{eqnarray}
From (\ref{set 3}), we obtain a single equation for $\psi_{\alpha C'}$, $C'=1,...,k$:
\begin{eqnarray}\label{psiCprimeeqns11}
&&~\; \left[-{1\over 2}(\partial_1 + 2\pi i z_1 )^2 + {1 \over 2} \left(\omega'_{12} x_1 + 2 \pi  z_2   - i \partial_2\right)^2 \right] \psi_{\alpha C'} \nonumber \\
&&  + \left[-{1\over 2}(\partial_3 + 2\pi i  z_3  )^2 + {1 \over 2} \left(\omega'_{34} x_3 + 2 \pi  z_4    - i \partial_4\right)^2\right]  \psi_{\alpha C'} \nonumber \\ 
 &=&(-1)^\alpha \; {\omega'_{12} + \omega'_{34} \over 2} \; \psi_{\alpha C'}~.
\end{eqnarray}
Similarly, the  2nd order equations for $\psi_{\alpha C}$ are given in terms of the frequencies defined as:
\begin{eqnarray} 
\omega''_{12} =  {2 \pi q_1 \over L_1 L_2}, ~\omega''_{34}  = {2 \pi (1 + \ell q_3)    \over \ell L_3 L_4}. \label{omegaprimeprime}
\end{eqnarray}
We obtain, from (\ref{set 4}), a single equation for $\psi_{\alpha C}$, $C=1,...,k$:
\begin{eqnarray}\label{psiCeqns11}
&&~\; \left[-{1\over 2}(\partial_1 + 2\pi i  z_1   )^2 + {1 \over 2} \left(\omega''_{12} x_1 + 2 \pi  z_2    - i \partial_2\right)^2 \right] \psi_{\alpha C} \nonumber  \\
&& + \left[-{1\over 2}(\partial_3 + 2\pi i  z_3  )^2 + {1 \over 2} \left(\omega''_{34} x_3 + 2 \pi  z_4   - i \partial_4\right)^2\right]  \psi_{\alpha C} \nonumber \\ 
 &=&(-1)^\alpha \; {\omega''_{12} + \omega''_{34} \over 2} \; \psi_{\alpha C}~.
\end{eqnarray}
The  boundary conditions (\ref{bc k}, \ref{bc ell}) take the explicit form, for $\psi_{\alpha C}$: 
\begin{eqnarray}\label{the set of BCS 2}
\nonumber
&&  \psi_{\alpha C}(x+\hat e_1 L_1)=e^{-i\frac{2\pi q_1 x_2}{L_2}}\psi_{\alpha C}(x)\,,\\
\nonumber
&&  \psi_{\alpha C}(x+\hat e_2 L_2)=\psi_{\alpha C}(x)\,,\\
\nonumber
&&  \psi_{\alpha C}(x+\hat e_3 L_3)=\gamma_\ell e^{-i \frac{2\pi (1+\ell q_3) x_4}{\ell L_4}}\psi_{\alpha  (C+1)}(x)\,,\\
&& \psi_{\alpha C}(x+\hat e_4 L_4)=\gamma_\ell e^{i\frac{2\pi (C-1)}{\ell}}\psi_{\alpha C}(x)\,,
\end{eqnarray}
and for $\psi_{\alpha C'}$:
\begin{eqnarray}\label{the set of BCS 3}
\nonumber
&&\psi_{\alpha  C'}(x+\hat e_1 L_1)=\gamma_k^{-r}e^{i\frac{2\pi(r/k- q_1) x_2}{L_2}}\psi_{\alpha  (C'-r)}(x), \\
\nonumber
&&\psi_{\alpha C'}(x+\hat e_2 L_2)=\gamma_k e^{i\frac{2\pi(C'-1)}{k}}\psi_{\alpha C'}(x),\\
\nonumber
&&\psi_{\alpha C'}(x+\hat e_3 L_3)=e^{-i\frac{2\pi q_3 x_4}{L_4}}\psi_{\alpha C'}(x),\\
&&\psi_{\alpha C'}(x+\hat e_4 L_4)=\psi_{\alpha C'}(x)\, ,
\end{eqnarray}
and we take $\psi_{\alpha C}$ and  $\psi_{\alpha C'}$ to be cyclic in $C$ and $C'$, i.e., $\psi_{\alpha C}\equiv\psi_{\alpha (C+\ell)}$ and $\psi_{\alpha  C'}\equiv \psi_{\alpha C'+k }$.

We now observe that the second order equations for $\psi_{\alpha C}$ and $\psi_{\alpha C'}$ are remarkably similar. Hence, we will discuss in some detail only the normalizability of $\psi_{\alpha C}$, noting that the discussion for $\psi_{\alpha C'}$ proceeds similarly.
 The boundary conditions (\ref{the set of BCS 2}) allow us to write a Fourier expansion in $x_2$ and $x_4$:
 \begin{eqnarray}\label{fourier1}
\psi_{\alpha C }(x) = \sum\limits_{n_2, n_4} \psi_{\alpha C , n_2, n_4}(x_1, x_3) e^{ i  2\pi {x_2 \over L_2} n_2 + i 2 \pi {x_4 \over L_4}(n_4 + {2 C  - 1 -\ell \over 2 \ell}) },
 \end{eqnarray}
 where the Fourier modes $\psi_{\alpha C, n_2, n_4}(x_1, x_3)$ obey boundary conditions in $x_1$, $x_3$ that can be deduced from (\ref{the set of BCS 2}).  The Fourier modes  themselves obey the second order equations following from (\ref{psiCeqns11}): \begin{eqnarray}\label{psiCeqns22}
&&~\; \left[-{1\over 2}(\partial_1 + 2\pi i  z_1)^2 + {1 \over 2} \left(\omega''_{12} x_1 + 2 \pi z_2  + {2 \pi n_2 \over L_2}\right)^2 \right] \psi_{\alpha C, n_2,n_4}(x_1,x_3) \nonumber  \\
&& + \left[-{1\over 2}(\partial_3 + 2\pi i  z_3 )^2 + {1 \over 2} \left(\omega''_{34} x_3 + 2 \pi  z_4  + {2 \pi \over L_4}(n_4 + {2 C  - 1 -\ell \over 2 \ell})\right)^2\right]  \psi_{2 C, n_2,n_4}(x_1,x_3)  \nonumber \\ 
 &&~=  (-1)^\alpha \; {\omega''_{12} + \omega''_{34} \over 2} \; \psi_{\alpha C, n_2,n_4}(x_1,x_3) ~.
\end{eqnarray}

 The norm of the $U(\ell)$ part of the zero mode with components $\psi_{\alpha C}$, defined in the first equality below,  after substituting the Fourier expansion (\ref{fourier1}) and integrating over $x_2$ and $x_4$, is:
\ba \label{norm2}
|| \psi_{U(\ell)}||^2 &=& \int\limits_{\T^4} d^4 x \sum\limits_{\alpha, C} |\psi_{\alpha C}|^2 = L_2 L_4 \sum\limits_{n_2, n_4} \int\limits_{0}^{L_1} d x_1 \int\limits_{0}^{L_3}  d x_3 \sum_{\alpha, C} |\psi_{\alpha C, n_2, n_4}(x_1,x_3)|^2 \nonumber \\
&=&L_2 L_4  \sum_{\alpha} \sum_{p_2 = 0}^{q_1 -1}  \int\limits_{-\infty}^{\infty} d x_1 \int\limits_{0}^{L_3}  d x_3   \sum_{C=1}^{\ell }   \sum_{n_4 = -\infty}^\infty |\psi_{\alpha C, p_2, n_4}(x_1,x_3)|^2 ~.
\ea
In the last line, we used the $x_1$ boundary condition in (\ref{the set of BCS 2}) to extend the range of the $x_1$ integration over the entire line. This leaves, as indicated, a sum over $q_1$ values of $n_2$, from $0$ to $q_1 -1$ (labeled by $p_2$) since a shift of $x_1$ by $L_1$ is found to shift  $n_2$ by $q_1$. 

The more tedious part is to show that the $x_3$ boundary conditions in (\ref{the set of BCS 2}), imposed on the Fourier modes and iterated repeatedly,\footnote{For brevity, we skip the straightforward but tedious details.}  allow us to show that the norm of the solution can be written as an integral over the entire $x_1, x_3$ plane:
 \ba \label{norm4}
|| \psi_{U(\ell)}||^2 &=& L_2 L_4  \sum_{\alpha=1}^2 \sum_{p_2 = 0}^{q_1 -1}  \sum_{p_4=0}^{\ell q_3}  \int\limits_{-\infty}^{\infty} d x_1 \int\limits_{-\infty}^{\infty}  d x_3  |\psi_{\alpha \;1, p_2, p_4}(x_1,x_3)|^2 
\ea

The point of the somewhat lengthy discussion above is to argue that the extension of the integral determining the zero mode norm to the  infinite range of values of $x_1$ and $x_3$ implies that we are looking for solutions of the second order equation (\ref{psiCeqns22}) for the Fourier modes\footnote{We note, also without elaborating the details, that an equation identical to (\ref{norm4}) can be derived with $\psi_{\alpha \; 1, p_2, p_4}$ replaced by $\psi_{\alpha \; C, p_2, p_4}$, for any chosen value of $C$.} $\psi_{\alpha \; C, n_2, n_4}(x_1,x_3)$ which is normalizable on the $\R^2$-plane $x_1, x_3$, rather than on $\T^4$ only.

Thus, returning to eqn.~(\ref{psiCeqns22}), we observe that it has the form of the equation for two decoupled simple harmonic oscillators (SHOs): one in the $x_1$ direction with frequency $\omega_{12}^{''}$ and another in the $x_2$ direction with frequency $\omega_{34}^{''}$, given in (\ref{omegaprimeprime}). Then, we also observe that for $\alpha = 2$, eqn.~(\ref{psiCeqns22}),   is exactly the Schroedinger equation for the ground state of these two  SHOs, and, hence, for $\alpha=2$ there exists a normalizable solution in the $x_{1,3}$ plane---the product of the two SHOs ground state wave functions. In contrast, for $\alpha=1$, there is no normalizable solution. 

{\flushleft{We}} end this diversion into the second order equation with some comments:
\begin{enumerate}
\item The analysis for $\psi_{\alpha C'}$ proceeds similarly and leads to the conclusion that a normalizable zero mode has $\psi_{1 C'}=0$. 
\item One can start  with the normalizable solutions of the 2nd order equations (\ref{psiCeqns22}) as a product of ground state wave functions for the Fourier modes (\ref{fourier1}) and observe that with $\psi_{1 C}=\psi_{1 C'}=0$ these also obey the 1st order Dirac equation.  After imposing  the boundary conditions on these ground-state wave functions, one arrives, after a substantial amount of work, at an expression identical to the one obtained by solving the first order equations (it is the latter procedure that we chose to present in some detail in the subsequent sections).\footnote{
In fact, upon inspecting the final result, eqn.~(\ref{final  expression of Phi}), the reader may observe the product of ground state wave functions for the SHOs in $x_1$ and $x_3$ with frequencies (\ref{omegaprimeprime}).}
\end{enumerate}
We now continue our analysis by presenting the normalizable solutions to the first order equations (\ref{set 3}, \ref{set 4}) with $\psi_{1 C} = \psi_{1 C'}=0$.
\subsection{Solving the first order equation}
To simplify the analysis, we explicitly restore the color indices for the non-zero fermion fields, leading to the following system of equations:
\begin{eqnarray}\label{the set of eqs to solve 2}
\nonumber
\left(\partial_1-i\partial_2+\frac{2\pi(q_1-r/k)x_1}{L_1L_2}+2\pi(z_2+i z_1)\right)\psi_{2C'}&=&0\,,\\\nonumber
\left(\partial_3-i\partial_4+\frac{2\pi q_3 x_3}{L_3 L_4}+2\pi(z_4+i z_3)\right)\psi_{2C'}&=&0\,,\\
\nonumber
\left(\partial_1-i\partial_2+\frac{2\pi q_1x_1}{L_1L_2}+2\pi(z_2+i z_1)\right)\psi_{2C}&=&0\,,\\
\left(\partial_3-i\partial_4+\frac{2\pi(1+\ell q_3) x_3}{\ell L_3L_4}+2\pi(z_4+i z_3)\right)\psi_{2C}&=&0\,,
\end{eqnarray}

{\flushleft \underline{\bf Solution of $\psi_{2C'}(x)$}.} 
We begin by solving the first two equations in (\ref{the set of eqs to solve 2}), followed by the solutions to the remaining two equations. To do so, we introduce the function $V_{C'}(x)$ through the following change of variables:
\begin{eqnarray}\label{def of V}
\psi_{2C'}(x)=e^{-\frac{\pi(q_1-r/k)x_1^2}{L_1L_2}}e^{-\frac{\pi q_3 x_3^2}{L_3L_4}}e^{-i 2\pi x_\mu z_\mu}V_{C'}(x)\,,
\end{eqnarray}
keeping in mind $q_3> 0$ and $q_1>1$, and thus, $q_1-r/k>0$.
Substituting Eq. (\ref{def of V}) into the first and second equations in (\ref{the set of eqs to solve 2}) we obtain
\begin{eqnarray}
(\partial_1-i\partial_2)V_{C'}=0\,,\quad (\partial_3-i\partial_4)V_{C'}=0\,.
\end{eqnarray}
We further define the complex coordinates $w_1$ and $w_2$ as $w_{1}\equiv x_1-ix_2$, $w_2\equiv x_3-ix_4$, and thus, the above two equations read
\begin{eqnarray}\label{eqs complex 2 V}
\frac{\partial V_{C'}}{\partial \bar w_1}=0\,,\quad \frac{\partial V_{C'}}{\partial \bar w_2}=0\,,
\end{eqnarray}
and hence, we conclude that $V_{C'}$ is an analytic function of $w_1,w_2$, i.e., $V_{C'}=V_{C'}(w_1,w_2)$. 
We also express the boundary conditions of $\psi_{2C'}$ in (\ref{the set of BCS 2}) in terms of $V_{C'}$ as
\begin{eqnarray}\label{bcs complex 2 V}
\nonumber
V_{C'}(w_1+L_1, w_2)&=&e^{-i\pi r(1-k)/k}e^{\pi(q_1-r/k)\frac{L_1}{L_2}}e^{\frac{2\pi (q_1-r/k)w_1}{L_2}}e^{i 2\pi z_1 L_1}V_{C'-r}(w_1,w_2)\,,\\
\nonumber
V_{C'}(w_1-iL_2, w_2)&=&e^{i\pi\frac{-1-k+2C'}{k}}e^{i 2\pi z_2 L_2}V_{C'}(w_1,w_2)\,,\\
\nonumber
V_{C'}(w_1,w_2+L_3)&=&e^{\frac{\pi q_3L_3}{L_4}}e^{\frac{2\pi q_3 w_2}{L_4}}e^{i 2\pi z_3 L_3}V_{C'}(w_1,w_2)\,,\\
 V_{C'}(w_1,w_2-iL_4)&=&e^{i 2\pi z_4 L_4}V_{C'}(w_1,w_2)\,.
\end{eqnarray}

The most general solution of Eq. (\ref{eqs complex 2 V}) that satisfy the boundary conditions (\ref{bcs complex 2 V}) in the $x_2,x_4$ directions is given by
\begin{eqnarray}\label{exp of VCp}
V_{C'}(w_1,w_2)=e^{-2\pi w_1 z_2-2\pi w_2 z_4}e^{-\frac{\pi w_1 (2C'-1-k)}{k L_2}}\sum_{m,n \in \mathbb Z}g_{C'mn}e^{2\pi\frac{m w_1}{L_2}}e^{2\pi\frac{n w_2}{L_4}}\,,
\end{eqnarray}
for coefficients $g_{C'mn}$. We determine these coefficients in the following by applying the boundary conditions in the $x_1$ and $x_3$ directions.

Applying the boundary condition in the $x_1$ direction, we obtain the difference equation:
\begin{eqnarray}\label{first cond on g}
g_{C'+r,m+q_1,n}=e^{-i \frac{\pi r(1-k)}{k}}e^{\frac{\pi L1}{k L_2}(2C'-1+r+k(-q_1-1-2m))}e^{2\pi (z_2+iz_1)L_1}g_{C',m,n}\,.
\end{eqnarray}
This difference equation must also satisfy the condition $V_{C'}(w_1, w_2) = V_{C' - k}(w_1, w_2)$, meaning that shifting the index $C'$ by $k$ cycles returns us to the same element. This implies the following relation:
\begin{eqnarray}
g_{C'-k,m-1,n}=g_{C',m,n}\,.
\end{eqnarray}
Equivalently, we also must have $V_{C'}(w_1, w_2)=V_{C'+k}(w_1, w_2)$, which yeilds
\begin{eqnarray}\label{consist 1}
g_{C'+k,m+1,n}=g_{C',m,n}\,.
\end{eqnarray}
These above two relations are consistent with Eq. (\ref{first cond on g}), as can be verified by a simple inspection. To find the solution of Eq. (\ref{first cond on g}), we propose the ansatz:
\begin{eqnarray}\label{ansatz}
g_{C'mn}=f(n)e^{-\frac{\pi k L_1}{(q_1k-r)L_2}\left[AC'+Bm+D\right]^2}\,.
\end{eqnarray}
Substituting Eq. (\ref{ansatz}) into (\ref{first cond on g}), we obtain
\begin{eqnarray}
A=-\frac{1}{k}\,,\quad B=1\,,\quad D=\frac{1+k}{2k}+i\frac{r(1-k)L_2}{2k L_1}-L_2(z_2+iz_1)\,.
\end{eqnarray}

The difference equation (\ref{first cond on g}), combined with (\ref{first cond on g}), yields $q_1k - r$ independent solutions. Specifically, there are $\mbox{gcd}(k, r)$ independent $C'$ entries, and each entry has $(q_1k - r)/\mbox{gcd}(k, r)$ independent orbits, such that $m = p + m' (q_1k - r)/\mbox{gcd}(k, r)$, where $p = 0, 1, \dots, (q_1k - r)/\mbox{gcd}(k, r) - 1$, and $m' \in \mathbb{Z}$. The easiest way to understand this is through examples.
 
In the first example, we take $k = 6$ and $r = 5$. Here, all $C'$ entries are related. By repeatedly applying (\ref{first cond on g}) and the condition (\ref{consist 1}), and ignoring the phase factor from (\ref{first cond on g})—as it does not affect the count of independent solutions—we obtain the following. (The arrows indicate the application of the recurrence relation (\ref{first cond on g}), while the equalities follow from the condition (\ref{consist 1}). For simplicity, we suppress the index $n$):
 \begin{eqnarray}\nonumber
&& g_{1,m}\rightarrow g_{6,m+q_1}\rightarrow g_{11,m+2q_1}=g_{5,m+2q_1-1}\rightarrow g_{10,m+3q_1}=g_{4,m+3q_1-2}\\
\nonumber
&& \rightarrow g_{9,m+4q_1-2}=g_{3,m+4q_1-3} \rightarrow g_{8,m+5q_1-3}=g_{2,m+5q_1-4}\\
&& \rightarrow g_{7,m+6q_1-4}=g_{1,m+6q_1-5} \,.
 \end{eqnarray}
 Thus, we conclude $g_{1,m}$ and $g_{1,m+6q_1-5} $ are identified, meaning that there are $6q_1-5$ independent orbits. 
 
 In the second example, we take $k=6,r=4$. In this case, there are $2$ independent $C'$ entries. Applying (\ref{first cond on g}) along with the condition (\ref{consist 1}) repeatedly, we find
 \begin{eqnarray}
 g_{1,m}\rightarrow g_{5,m+q_1} \rightarrow g_{9,m+2q_1}=g_{3,m+2q_1-1}\rightarrow g_{7,m+3q_1-1}=g_{1,m+3q-2}\,,
 \end{eqnarray}
 and
  \begin{eqnarray}
 g_{2,m}\rightarrow g_{6,m+q_1} \rightarrow g_{10,m+2q_1}=g_{4,m+2q_1-1}\rightarrow g_{8,m+3q_1-1}=g_{2,m+3q-2}\,,
 \end{eqnarray}
 Thus, each independent $C'$ entry yields $3q_1-2$ orbits, for a total of $6q_1-4$ solutions.

 Similarly, applying the boundary condition in the $x_3$ direction, we obtain the difference equation:
\begin{eqnarray}
g_{C'mn}=e^{2\pi (z_4+iz_3)L_3}e^{\frac{\pi(q_3-2n)L_3}{L_4}}g_{C'm,n-q_3}\,,
\end{eqnarray}
with a solution given by
\begin{eqnarray}
g_{C'mn}\propto e^{-\frac{\pi L_3}{q_3L_4}\left(n-L_4(z_4+iz_3)\right)^2}\,,
\end{eqnarray}
such that $n=t_4+q_3n'$, $t_4=0,1,\ldots,q_3-1$ and $n'\in \mathbb Z$. Thus, the full solution $\psi_{C'}$ is given by
\begin{eqnarray}\label{general solution of psic'}
\nonumber
\psi_{2C'}(x,z)&=&e^{-\frac{\pi(q_1-r/k)x_1^2}{L_1L_2}}e^{-\frac{\pi q_3 x_3^2}{L_3L_4}}e^{-i 2\pi x_\mu z_\mu}e^{-2\pi w_1 z_2-2\pi w_2 z_4}e^{-\frac{\pi w_1 (2C'-1-k)}{k L_2}}\\
\nonumber
&&\times\sum_{\scriptsize m=t_2+m' (q_1k-r)/\mbox{gcd}(k,r), n=t_4+q_3n', n',m' \in \mathbb Z}{\cal C}^{(t_2,t_4)}_{C'}e^{2\pi\frac{m w_1}{L_2}}e^{2\pi\frac{n w_2}{L_4}} e^{-\frac{\pi L_3}{q_3L_4}\left(n-L_4(z_4+iz_3)\right)^2}\\
\nonumber
&&\times e^{-\frac{\pi k L_1}{(q_1k-r)L_2}\left[-\frac{C'}{k}+m+\frac{1+k}{2k}+i\frac{r(1-k)L_2}{2k L_1}-iL_2(z_2+iz_1)\right]^2}\,.\\
\end{eqnarray}
The analysis above demonstrates that $\psi^{(k)}(x,z)$ solves (\ref{the set of eqs to solve 2}) along with the boundary conditions (\ref{the set of BCS 2}) and admits $q_3\left(q_1 k - r\right)$ independent solutions, each associated with a corresponding ${\cal C}^{(t_2,t_4)}_{C'}$ coefficient. Appendix \ref{Embeddings} provides a systematic method for embedding the independent solutions as column vectors in the $SU(k)$ group.

{\flushleft{\underline{\bf Solution of $\psi_{2C}(x)$}.}} Next, we move to the $\psi_{C}$ solution. In the following, we show that there are $q_1(1+\ell q_3)$ independent nontrivial solutions of $\psi_{C}$; these constitute the $q_1(1+\ell q_3)$ zero modes on $\mathbb T^4$. We proceed as above and define $U_{C}$ via:
\begin{eqnarray}
\psi_{2C}(x)=e^{-\frac{\pi q_1 x_1^2}{L_1 L_2}}e^{-\frac{\pi (1+\ell q_3) x_3^2}{\ell L_3 L_4}}e^{-i2\pi x_\mu z_\mu}U_{C}(w_1,w_2)\,.
\end{eqnarray}
Substituting into (\ref{the set of eqs to solve 2}) we find
\begin{eqnarray}\label{eqs complex U 3}
\frac{\partial U_{C}}{\partial \bar w_1}=0\,,\quad \frac{\partial U_{C}}{\partial \bar w_2}=0\,.
\end{eqnarray}
In terms of the coordinates $w_{1}, w_2$ and the variable $U_{C}$, the boundary conditions (\ref{the set of BCS 2}) read
\begin{eqnarray}\label{BCS for U 3}
\nonumber
U_C(w_1+L_1, w_2)&=&e^{\frac{2\pi q_1 w_1}{L_2}}e^{\frac{\pi q_1 L_1}{L_2}}e^{i 2\pi L_1 z_1}U_C(w_1,w_2)\,,\\
\nonumber
U_C(w_1-iL_2, w_2)&=&e^{i 2\pi L_2 z_2}U_C(w_1,w_2)\,,\\
\nonumber
U_C(w_1,w_2+L_3)&=& e^{\frac{2\pi (1+\ell q_3) w_2}{\ell L_4}}e^{\frac{\pi (1+\ell q_3) L_3}{\ell L_4}}e^{i\frac{\pi (1-\ell)}{\ell}}e^{i 2\pi L_3z_3}U_{C+1}(w_1,w_2)\,,\\
U_C(w_1,w_2-i L_4)&=&e^{i\frac{\pi (2C-1-\ell)}{\ell}}e^{i 2\pi z_4 L_4}U_C(w_1,w_2)\,.
\end{eqnarray}

The solution to (\ref{eqs complex U 3}) that satisfies the second and fourth boundary conditions in (\ref{BCS for U 3}) is given by:
\begin{eqnarray}
U_C(w_1,w_2)=e^{-2\pi (w_1 z_2+w_2z_4)}e^{-\frac{\pi w_2(2C-1-\ell)}{\ell L_4}}\sum_{m,n \in \mathbb Z}d_{Cmn}e^{ \frac{2\pi m w_1}{L_2}} e^{ \frac{2\pi n w_2}{L_4}}\,.
\end{eqnarray}
Applying the boundary condition in the $x_1$ direction gives the difference equation:
\begin{eqnarray}
d_{Cmn}=e^{2\pi L_1(z_2+i z_1)}e^{\frac{\pi q_1 L_1}{L_2}}e^{-\frac{2\pi m L_1}{L_2}}d_{C, m-q_1,n}\,.
\end{eqnarray}
This difference equation admits $q_1$ distinct solutions given by
\begin{eqnarray}\label{first part of dCmn}
d_{Cmn}=f^{(p)}(C,n)e^{-\frac{\pi L_1}{q_1L_2}\left(m-L_2(z_2+iz_1)\right)^2}\,,
\end{eqnarray}
where $m=h+q_1m'$ and $h=0,1,..,q_1-1$, $m'\in \mathbb Z$. Applying the boundary condition in the $x_3$ direction gives the difference equation:
\begin{eqnarray}\label{recurrence relation for d}
d_{Cmn}=e^{2\pi(z_4+iz_3)L_3}e^{i\frac{\pi (1-\ell)}{\ell}}e^{\frac{\pi L_3}{\ell L_4}(2C+\ell(-1-2n+q_3))}d_{C+1,m,n-q_3}\,.
\end{eqnarray}
Demanding that $U_C(w_1,w_2)=U_{C+\ell}(w_1,w_2)$ gives the condition
\begin{eqnarray}\label{d condition}
d_{C+\ell,m,n+1}=d_{Cmn}\,,
\end{eqnarray}
which is consistent with Eq. (\ref{recurrence relation for d}). To solve Eq. (\ref{recurrence relation for d}), we use the ansatz:
\begin{eqnarray}\label{ansatz second}
d_{Cmn}=f(m)e^{-\frac{\pi L_3}{\ell L_4(1+\ell q_3)}\left[A'C+B'n+D'\right]^2}\,,
\end{eqnarray}
and substituting this ansatz into (\ref{recurrence relation for d}), we obtain
\begin{eqnarray}
A'=-1\,,\quad B'=\ell\,,\quad D'=\frac{1+\ell}{2}-i\frac{(1-\ell)L_4}{2 L_3}-\ell L_4(z_4+iz_3)\,.
\end{eqnarray}
Adding the contribution from (\ref{first part of dCmn}), we find that the full solution of $d_{Cmn}$  is given by
\begin{eqnarray}
d_{Cmn}=e^{-\frac{\pi L_3}{\ell L_4(1+\ell q_3)}\left[-C+\ell n+\frac{1+\ell}{2}-i\frac{(1-\ell)L_4}{2 L_3}-\ell L_4(z_4+iz_3)\right]^2}\times e^{-\frac{\pi L_1}{q_1L_2}\left(m-L_2(z_2+iz_1)\right)^2}\,,
\end{eqnarray}
where $n=p_4+(1+q_3\ell)n'$,  $p_4=0,1,\ldots,\ell q_3$, and $n'\in \mathbb Z$. The fact that there are $q_3\ell+1$ solutions to Eq. (\ref{recurrence relation for d}) can be envisaged by recalling Condition (\ref{d condition}), which, after using the recurrence relation (\ref{recurrence relation for d}) $\ell$ times, gives (ignoring the accompanying phases) 
\begin{eqnarray}
d_{C,m,n}\rightarrow d_{C+1,m,n-q_3}\rightarrow d_{C+2,m,n-2q_3}......\rightarrow d_{C+\ell,m,n-\ell q_3}=d_{C,m,n-\ell q_3-1}\,,
\end{eqnarray}
and hence, $n\equiv n-(\ell q_3+1)$, proving that there are $\ell q_3+1$ solutions.  
We conclude that the solutions $\psi_{2C}(x)$ are given by
\begin{eqnarray}\label{solution of psi2c final}
\nonumber
\psi_{2C}(x,z)&=&e^{-\frac{\pi q_1 x_1^2}{L_1 L_2}}e^{-\frac{\pi (1+\ell q_3) x_3^2}{\ell L_3 L_4}}e^{-i2\pi x_\mu z_\mu}e^{-2\pi (w_1 z_2+w_2z_4)}e^{-\frac{\pi w_2(2C-1-\ell)}{\ell L_4}}\\
\nonumber
&&\times\sum_{m=p_2+q_1m', n=p_4+(1+q_3\ell) n', m',n'  \in \mathbb Z}{\cal D}^{(p_2,p_4)}_Ce^{ \frac{2\pi m w_1}{L_2}} \\
\nonumber
&&\times e^{ \frac{2\pi n w_2}{L_4}}e^{-\frac{\pi L_3}{\ell L_4(1+\ell q_3)}\left[-C+\ell n+\frac{1+\ell}{2}-i\frac{(1-\ell)L_4}{2 L_3}-\ell L_4(z_4+iz_3)\right]^2} e^{-\frac{\pi L_1}{q_1L_2}\left(m-L_2(z_2+iz_1)\right)^2}\,.\\
\end{eqnarray}
This analysis shows that there are $q_1(\ell q_3+1)$ independent solutions of $\psi^{(\ell)}_2$: these are solutions of (\ref{the set of eqs to solve 2}) along with the boundary conditions (\ref{the set of BCS 2}). Each independent solution comes with an arbitrary coefficient ${\cal D}^{(p_2,p_4)}_C$. Appendix \ref{Embeddings} provides a systematic method for embedding the independent solutions as column vectors in the $SU(\ell)$ group. 

In conclusion, the undotted fermions $\psi_{2C'}(x)$ and $\psi_{2C}(x)$ in total admit 
$
q_1(\ell q_3+1)+q_3\left(q_1 k-r\right)=q_3\left(Nq_1-r\right)+q_1
$
solutions. Comparing this result with the Dirac index $q_3(Nq_1-r)+q_1$, we find that the undotted fermions saturate the number of the zero modes.

\subsection{Defining the functions $\Sigma_{C'}^{(t_2, t_4)}(x,z)$ and $\Phi_{C}^{(p_2, p_4)}(x,z)$}

In the following, we define new functions $\Sigma_{C'}^{(t_2, t_4)}(x,z)$ and $\Phi_{C}^{(p_2, p_4)}$ that enables us to cast $\psi_{2}^{(k)}$ and $\psi_{2}^{(\ell)}$ as embeddings in $SU(k)$ and $SU(\ell)$, respectively. We shall demonstrate the embedding in Appendix \ref{Embeddings}.

We start by defining the function $\Sigma_{C'}^{(t_2, t_4)}(x,z)$ as
\begin{eqnarray}\label{the definition of Sigma final}
\nonumber
\Sigma_{C'}^{(t_2, t_4)}(x,z)&=&{\cal C}_\Sigma\sum_{\scriptsize m=t_2+m' (q_1k-r)/\mbox{gcd}(k,r), n=t_4+q_3n', n',m' \in \mathbb Z} e^{-i \frac{2\pi n x_4}{L_4}}e^{-\frac{\pi q_3 L_3}{L_4}\left[\frac{x_3}{L_3}-\frac{n-L_4z_4}{q_3} \right]^2}\\
\nonumber
&&\times e^{-i2\pi L_1 z_1\left(\frac{x_1}{L_1}+\frac{2C'-1-k-2mk}{2(q_1k-r)}\right)}  e^{-i 2\pi L_3 z_3\left(\frac{x_3}{L_3}-\frac{n}{q_3}\right)}\\\
\nonumber
&&\times e^{\frac{i\pi x_2 (2C'-1-k-2mk)}{k L_2}}  e^{-\frac{\pi (q_1k-r)  L_1}{kL_2}\left[\frac{x_1}{L_1}+\frac{2C'-1-k-2mk}{2(q_1k-r)}+\frac{k L_2z_2}{q_1k-r}\right]^2}\\
&&\times e^{i\frac{\pi r(1-k)}{k}\left[ \frac{2C'-1-k-2mk}{2(q_1k-r)}\right] }\,,
\end{eqnarray}
 with two superscripts $t_2$ and $t_4$ are taken in the range \begin{equation} t_2=0,1,..., \frac{q_1k-r}{\mbox{gcd}(k,r)}-1\,,\quad t_4=0,1,..,q_3-1\,.\end{equation} We note that the function $\Sigma_{C'}^{(t_2, t_4)}(x,z)$ takes the same structure of $\psi_{2C'}(x,z)$ given in (\ref{general solution of psic'}), leaving aside the coefficients ${\cal C}_{C'}^{t_2,t_4}$. The normalization constant ${\cal C}_\Sigma$ is chosen such that\footnote{Calculating the normalization integral here and in (\ref{normphi}) involves using the summation over Fourier coefficients to extend the range of integration over $x_{1}, x_3$ to the entire real axis, as done in section \ref{sec:normalizability}.}
 \begin{eqnarray}\label{normalization of Sigma for general r}
&&\int_{\mathbb T^4} \sum_{n=0}^{\scriptsize k/\mbox{gcd}(k,r)-1} \Sigma_{i+nr}^{* (t_2+nq_1, t_4)}\Sigma_{i+nr}^{(t_2+nq_1, t_4)}=1,
\end{eqnarray}
 which fixes $ {\cal C}_\Sigma$ as
 \begin{eqnarray}\label{Csigma}
 {\cal C}_\Sigma=\frac{2}{L_2L_4}\sqrt{\frac{q_3(q_1k-r)}{kL_1L_2L_3L_4}}\,.
 \end{eqnarray}
The normalization condition (\ref{normalization of Sigma for general r}) is chosen for later convenience. 
It is straightforward to verify that $\Sigma_{C'}^{(t_2, t_4)}(x,z)$ satisfies the same differential equation as $\psi_{2C'}$:
\begin{eqnarray}
\nonumber
\left(\partial_1-i\partial_2+\frac{2\pi(q_1-r/k)x_1}{L_1L_2}+2\pi(z_2+i z_1)\right)\Sigma_{C'}^{(t_2, t_4)}(x,z)&=&0\,,\\
\left(\partial_3-i\partial_4+\frac{2\pi q_3 x_3}{L_3 L_4}+2\pi(z_4+i z_3)\right)\Sigma_{C'}^{(t_2, t_4)}(x,z)&=&0\,,
\end{eqnarray}
along with the boundary conditions:
\begin{eqnarray}\label{BCS for Sigma full list}
\nonumber
&&\Sigma_{C'}^{(t_2, t_4)}(x+\hat e_1 L_1,z)=\gamma_k^{-r}e^{i\frac{2\pi(r/k- q_1) x_2}{L_2}}\Sigma_{C'-r}^{(t_2-q_1, t_4)}(x,z)\,,\\
\nonumber
&&\Sigma_{C'}^{(t_2, t_4)}(x+\hat e_2 L_2,z)=\gamma_k e^{i\frac{2\pi(C'-1)}{k}}\Sigma_{C'}^{(t_2, t_4)}(x,z)\,, \\
\nonumber
&&\Sigma_{C'}^{(t_2, t_4)}(x+\hat e_3 L_3,z)=e^{-i\frac{2\pi q_3 x_4}{L_4}}\Sigma_{C'}^{(t_2, t_4)}(x,z)\,,\\
&&\Sigma_{C'}^{(t_2, t_4)}(x+\hat e_4 L_4,z)=\Sigma_{C'}^{(t_2, t_4)}(x,z)\,.
\end{eqnarray}
In addition, $\Sigma_{C'}^{(t_2, t_4)}(x,z)$ satisfies the important identities
\begin{eqnarray}\label{peridic relation for Sigma}
\Sigma_{C'+k}^{(t_2, t_4)}(x,z)=\Sigma_{C'}^{(t_2-1, t_4)}(x,z)\,, \quad \Sigma_{C'}^{(t_2, t_4)}(x,z)=\Sigma_{C'}^{(t_2\pm\frac{q_1k-r}{\scriptsize \mbox{gcd}(k,r)}, t_4)}(x,z)\,.
\end{eqnarray}
Notice that the second identity follows trivially from the first one. However, we include it for convenience as it facilitates some subsequent computations.

We also define the functions $\Phi_{C}^{(p_2, p_4)}(x,z)$ as
\begin{eqnarray}\label{final expression of Phi}
\nonumber
 \Phi_{C}^{(p_2, p_4)}(x,z)&=&{\cal C}_{\Phi}\sum_{m=p_2+q_1m', n=p_4+(1+q_3\ell) n', m',n'  \in \mathbb Z}e^{-i \frac{2\pi m x_2}{L_2}}e^{-\frac{\pi q_1 L_1}{L_2}\left[\frac{x_1}{L_1}-\frac{m-L_2z_2}{q_1} \right]^2}\\
\nonumber
&&\times e^{-i 2\pi z_1 L_1\left(\frac{x_1}{L_1}-\frac{m}{q_1}\right)}e^{-i2\pi z_3 L_3\left(\frac{x_3}{L_3}+\frac{2C-1-\ell-2n\ell}{2(1+\ell q_3)}\right)}\\
\nonumber
&&\times e^{\frac{i\pi x_4 (2C-1-\ell-2n\ell )}{\ell L_4}}  e^{-\frac{\pi (1+\ell q_3)  L_3}{\ell L_4}\left[\frac{x_3}{L_3}+\frac{2C-1-\ell-2\ell n}{2(1+\ell q_3)}+\frac{ \ell L_4z_4}{1+\ell q_3}\right]^2}\\
&&\times e^{i\frac{\pi (1-\ell)}{\ell}\left[ \frac{-2C+1+\ell+2n\ell}{2(1+\ell q_3)}\right] }\,.
\end{eqnarray}
The superscripts $p_2,p_4$ are taken in the range
\begin{eqnarray}
p_2=0,1,..,q_1-1\,,\quad p_4=0,1,...,\ell q_3\,.
\end{eqnarray}
The function $\Phi_{C}^{(p_2, p_4)}(x,z)$ takes the same structure of $\psi_{2C}(x,z)$ given in (\ref{solution of psi2c final}). 
The normalization constant ${\cal C}_\Phi$ is chosen such that
\begin{eqnarray}\label{normphi}
\int_{\mathbb T^4}\sum_{C=1}^{\ell} \Phi_{C}^{(p_2, p_4)} \Phi_{C}^{*(p_2, p_4)}=1\,,
\end{eqnarray}
which yields
\begin{eqnarray}\label{Cphi}
{\cal C}_{\Phi}=\frac{2}{L_1L_3}\sqrt{\frac{q_1(1+\ell q_3)}{\ell L_1L_2L_3L_4}}\,.
\end{eqnarray}
It is easy to check that $\Phi_{C}^{(p_2, p_4)}(x,z)$ satisfies the same differential equations as $\psi_{2C}$:
\begin{eqnarray}
\nonumber
\left(\partial_1-i\partial_2+\frac{2\pi q_1x_1}{L_1L_2}+2\pi(z_2+i z_1)\right) \Phi_{C}^{(p_2, p_4)}(x,z)&=&0\,,\\
\left(\partial_3-i\partial_4+\frac{2\pi(1+\ell q_3) x_3}{\ell L_3L_4}+2\pi(z_4+i z_3)\right) \Phi_{C}^{(p_2, p_4)}(x,z)&=&0\,.
\end{eqnarray}
The functions $\Phi_{C}^{(p_2, p_4)}(x,z)$ obey the boundary conditions
\begin{eqnarray}\label{BCS for the Phi final exp}
\nonumber
\Phi_C^{(p_2, p_4)}(x+L_1\hat e_1,z)&=&e^{-i\frac{2\pi q_1 x_2}{L_2}}\Phi_C^{(p_2, p_4)}(x,z)\,,\\
\nonumber
\Phi_C^{(p_2, p_4)}(x+L_2\hat e_2,z)&=&\Phi_C^{(p_2, p_4)}(x,z)\,,\\
\nonumber
\Phi_C^{(p_2, p_4)}(x+L_3\hat e_3,z)&=&e^{i\pi\frac{1-\ell}{\ell}}e^{-i\frac{2\pi (1+\ell q_3)x_4}{\ell L_4}}\Phi_{C+1}^{(p_2, p_4-q_3)}(x,z)\,,\\
\Phi_C^{(p_2, p_4)}(x+L_4\hat e_4,z)&=&e^{i\pi \frac{2C-1-\ell}{\ell}}\Phi_C^{(p_2, p_4)}(x,z)\,.
\end{eqnarray}
In addition, $\Phi_C^{(p_2, p_4)}(x,z)$ satisfies the identity
\begin{eqnarray}\label{periodicity of Phi}
\Phi^{(p_2, p_4)}_{C+\ell}=\Phi^{(p_2, p_4-1)}_C\,.
\end{eqnarray}
%

\subsection{Restoring the internal moduli}
\label{appx:B5}

Up until now, we have been working with the internal moduli switched off. We will now reintroduce them and repeat the calculations. As we will demonstrate, this requires minimal adjustment. With the moduli restored, the $SU(N)$ gauge fields on $\mathbb{T}^4$, described by (\ref{the full abelian bck general r}), now include additional components:
\begin{eqnarray}\label{the full abelian bck general r with moduli C}
\nonumber
\delta A_{1 C'D'}&=&-\delta_{C'D'} 2\pi \ell \phi_{1}^{[C']_r}\,,\quad \delta A_{2 C'D'}=-\delta_{C'D'} 2\pi \ell \phi_{2}^{[C']_r}\\
\delta A_{3 C'D'}&=&-\delta_{C'D'} 2\pi \ell \phi_{3}^{[C']_r}\,,\quad \delta A_{4 C'D'}=-\delta_{C'D'} 2\pi \ell \phi_{4}^{[C']_r}\,,
\end{eqnarray} 
in the $SU(k)$ group and 
\begin{eqnarray}\label{the full abelian bck general r with moduli CP}
\nonumber
\delta A_{1 CD}&=&\delta_{C'D'} 2\pi k \tilde\phi_{1}\,,\quad \delta A_{2 CD}=\delta_{C'D'} 2\pi k \tilde\phi_{2}\\
\delta A_{3 CD}&=&\delta_{C'D'} 2\pi k \tilde\phi_{3}\,,\quad \delta A_{4 CD}=\delta_{C'D'} 2\pi k \tilde \phi_{4}\,,
\end{eqnarray} 
 in the $SU(\ell)$ group, where
 \begin{eqnarray}\label{Rel between Phi and tild Phi}
 \tilde \phi_\mu=\frac{1}{k}\sum_{C'=1}^k \phi_{\mu}^{[C']_r}\,.
 \end{eqnarray}
 The choice of this parametrization of the moduli ensures that $\mbox{tr}[\delta A_\mu]=0$, as required for the unitary group $SU(N)$. We recall that there are only $\phi_\mu^{i}$, where $i=1,2,...,\mbox{gcd}(k,r)$ independent moduli. Therefore, we write (\ref{Rel between Phi and tild Phi}) as
 \begin{eqnarray}\label{contraint on phitild}
  \tilde \phi_\mu=\frac{1}{k}\sum_{C'=1}^k \phi_{\mu}^{[C']_r}=\frac{1}{k}\frac{k}{\mbox{gcd}(k,r)}\sum_{i=1}^{\scriptsize \mbox{gcd}(k,r)}\phi_\mu^i=\frac{1}{\mbox{gcd}(k,r)}\sum_{i=1}^{\scriptsize \mbox{gcd}(k,r)}\phi_\mu^i\,.
 \end{eqnarray}
 
Inspection of the Weyl equations (\ref{set 3}) reveals that one needs to make the substitutions
 \begin{eqnarray}\nonumber
 z_\mu&\rightarrow& z_\mu-\ell \phi_{\mu}^{[C']_r}\, \mbox{in the solutions of}\, \psi_{2C'}\,,\\
  z_\mu&\rightarrow& z_\mu+k \tilde \phi_{\mu}\, \mbox{in the solutions of}\, \psi_{2C}\,.
 \end{eqnarray}
 This, in turn, means that the functions $\Sigma_{C'}^{(t_2,t_4)}$ and $\Phi_{C'}^{(p_2,p_4)}$ defined in (\ref{the definition of Sigma final}) and (\ref{final expression of Phi}) are given by
 \begin{eqnarray}\label{final expression of Sigma moduli}
\nonumber
\Sigma_{C'}^{(t_2,t_4)}(x,z,\phi)&=&{\cal C}_{\Sigma}\sum_{\scriptsize m=t_2+m' (q_1k-r)/\mbox{gcd}(k,r), n=t_4+q_3n', n',m' \in \mathbb Z} e^{-i \frac{2\pi n x_4}{L_4}}e^{-\frac{\pi q_3 L_3}{L_4}\left[\frac{x_3}{L_3}-\frac{n-L_4\left(z_4-\ell \phi_{4}^{[C']_r}\right)}{q_3} \right]^2}\\
\nonumber
&&\times e^{-i2\pi L_1 \left(z_1-\ell \phi_{1}^{[C']_r}\right)\left(\frac{x_1}{L_1}+\frac{2C'-1-k-2mk}{2(q_1k-r)}\right)}  e^{-i 2\pi L_3 \left(z_3-\ell \phi_{3}^{[C']_r}\right)\left(\frac{x_3}{L_3}-\frac{n}{q_3}\right)}\\\
\nonumber
&&\times e^{\frac{i\pi x_2 (2C'-1-k-2mk)}{k L_2}}  e^{-\frac{\pi (q_1k-r)  L_1}{kL_2}\left[\frac{x_1}{L_1}+\frac{2C'-1-k-2mk}{2(q_1k-r)}+\frac{k L_2\left(z_2-\ell \phi_{2}^{[C']_r}\right)}{q_1k-r}\right]^2}\\
&&\times e^{i\frac{\pi r(1-k)}{k}\left[ \frac{2C'-1-k-2mk}{2(q_1k-r)}\right] }\,,
\end{eqnarray}
and
\begin{eqnarray}\label{final expression of Phi moduli}
\nonumber
 \Phi_{C}^{(p_2,p_4)}(x,z,\phi)&=&{\cal C}_{\Phi}\sum_{m=p_2+q_1m', n=p_4+(1+q_3\ell) n', m',n'  \in \mathbb Z}e^{-i \frac{2\pi m x_2}{L_2}}e^{-\frac{\pi q_1 L_1}{L_2}\left[\frac{x_1}{L_1}-\frac{m-L_2\left(z_2+k \tilde \phi_{2}\right)}{q_1} \right]^2}\\
\nonumber
&&\times e^{-i 2\pi \left(z_1+k \tilde \phi_{1}\right) L_1\left(\frac{x_1}{L_1}-\frac{m}{q_1}\right)}e^{-i2\pi \left(z_3 +k \tilde \phi_{3}\right)L_3\left(\frac{x_3}{L_3}+\frac{2C-1-\ell-2n\ell}{2(1+\ell q_3)}\right)}\\
\nonumber
&&\times e^{\frac{i\pi x_4 (2C-1-\ell-2n\ell )}{\ell L_4}}  e^{-\frac{\pi (1+\ell q_3)  L_3}{\ell L_4}\left[\frac{x_3}{L_3}+\frac{2C-1-\ell-2\ell n}{2(1+\ell q_3)}+\frac{ \ell L_4\left(z_4+k \tilde \phi_{4}\right)}{1+\ell q_3}\right]^2}\\
&&\times e^{i\frac{\pi (1-\ell)}{\ell}\left[ \frac{-2C+1+\ell+2n\ell}{2(1+\ell q_3)}\right] }\,,
\end{eqnarray}
where ${\cal C}_{\Sigma}$ and ${\cal C}_{\Phi}$ are given by (\ref{Csigma}, \ref{Cphi}), respectively. 

\subsection{Embedding in $SU(k)$ and $SU(\ell)$}
\label{Embeddings}

In this section, we use the functions $\Sigma_{C'}^{(t_2,t_4)}(x,z,\phi)$ and $\Phi_C^{(p_2,p_4)}(x,z,\phi)$ to write the solutions $\psi_2^{(k)}$ and $\psi_2^{(\ell)}$, given by Eqs. (\ref{general solution of psic'}, \ref{solution of psi2c final}) as column vectors embedded in the defining representations of $SU(k)$ and $SU(\ell)$, respectively. We shall also examine that this embedding satisfies the correct boundary conditions. To avoid unnecessary clutter, we chose to omit the functional dependence on the moduli $\phi_\mu$ in the following discussion.  
We reserve the symbol $\Psi$ to denote the $q_3\left(Nq_1-r\right)+q_1$ independent zero modes of $\psi_2$. We decompose $\Psi$ into an $SU(k)$ and $SU(\ell)$ column vectors  
\begin{eqnarray}\label{general form of fermion zero modes}
\Psi=\Psi_{(k)}\oplus\Psi_{(\ell)}
\end{eqnarray}
meaning that $\Psi_{(k)}$ and $\Psi_{(\ell)}$ are the embeddings inside $SU(k)$ and $SU(\ell)$, respectively.

 Their explicit form is given by  
\begin{eqnarray}\label{The final expressions of PSI in suk and sul}
\Psi_{(k)i}^{[t_2]_{\scriptsize (q_1k-r)/\mbox{gcd}(k,r)},[t_4]_{q_3}}=\left[\begin{array}{c} 0\\.\\.\\ \Sigma_{i}^{(t_2,t_4)}\\0\\.\\.\\\Sigma_{i+r}^{(t_2+q_1,t_4)}\\0\\.\\. \\\Sigma_{i+nr}^{(t_2+nq_1,t_4)}\\0\\...\end{array}\right]\,,\quad \Psi_{(\ell)}^{[p_2]_{q_1},[p_4]_{1+\ell q_3}}=\left[\begin{array}{c} \Phi_1^{(p_2,p_4-q_3)}\\  \Phi_2^{(p_2,p_4-2q_3)}\\.\\.\\  \Phi_{\ell-1}^{(p_2,p_4-(\ell-1)q_3)}\\ \Phi_{\ell}^{(p_2,p_4-\ell q_3)}\end{array}\right]\,.
\end{eqnarray}
$\Psi_{(k)i}^{[t_2]_{\scriptsize (q_1k-r)/\mbox{gcd}(k,r)},[t_4]_{q_3}}$ carries one subscript $i$ denoting $\mbox{gcd}(k,r)$ independent vectors, and two superscripts $t_2$ and $t_4$. Each vector, denoted by $i$, has $\scriptsize \frac{k}{\mbox{gcd}(k,r)}$ nonzero entries $\Sigma_{i+nr}^{(t_2+nq_1,t_4)}$, where $n=0,1,2,..,\frac{k}{\mbox{gcd}(k,r)}-1$. Inspection of (\ref{general solution of psic'}) reveals that the subscript $i$ labels the independent orbits under $C'\rightarrow C'+r$, while $n$ labels the elements in each orbit.  In writing the elements of these vectors, we repeatedly apply (\ref{peridic relation for Sigma}). Also, $\Psi_{(\ell)}^{[p_2]_{q_1},[p_4]_{1+\ell q_3}}$ carries two superscripts $p_2$ and $p_4$. The range of the indices is given by  
\begin{eqnarray}\nonumber
 i&=&1,2,..,\mbox{gcd}(k,r)\,, \quad t_2=0,1,..., \frac{q_1k-r}{\mbox{gcd}(k,r)}-1\,,\quad t_4=0,1,..,q_3-1\,,\\
 n&=&0,1,2,..,\frac{k}{\mbox{gcd}(k,r)}-1\,,\quad p_2=0,1,..,q_1-1\,,\quad p_4=0,1,...,\ell q_3\,.
\end{eqnarray}
In total, there are \begin{equation} q_1(\ell q_3+1)+q_3\left(q_1 k-r\right)=q_3\left(Nq_1-r\right)+q_1\equiv \hat N\end{equation} subscripts and superscripts, denoting the independent $\hat N$ zero modes of $\psi_2$.  

It is trivial to use (\ref{BCS for the Phi final exp}) to show that each $SU(\ell)$ zero mode embedding $\Psi_{(\ell)}^{[p_2]_{q_1},[p_4]_{1+\ell q_3}}$ respects the boundary conditions (\ref{the set of BCS 2}) in the $x_1,x_2,x_4$ directions.  
Under a translation of $x_3$ by $L_3\hat e_3$, the zero mode transforms according to Eq. (\ref{BCS for the Phi final exp}):  
\begin{eqnarray}
\Psi_{(\ell)}^{[p_2]_{q_1},[p_4]_{1+\ell q_3}}(x+\hat e_3 L_3,z)=e^{i\pi\frac{1-\ell}{\ell}}e^{-i\frac{2\pi (1+\ell q_3)x_4}{\ell L_4}}\left[\begin{array}{c} \Phi_2^{(p_2,p_4-2q_3)}(x,z)\\  \Phi_3^{(p_2,p_4-3q_3)}(x,z)\\.\\.\\.\\  \Phi_{\ell}^{(p_2,p_4-\ell q_3)}(x,z)\\ \Phi_{1}^{(p_2,p_4- q_3)}(x,z) \end{array}\right]\,,
\end{eqnarray}
and in the last entry, we used Eq.(\ref{periodicity of Phi}) and shifted the dummy index $n'$ that appears in Eq. (\ref{final expression of Phi}) by unity. Thus, a translation in the $x_3$ direction gives an overall phase accompanied by an $SU(\ell)$ permutation of the components of the zero modes, according to the transition functions (\ref{the set of transition functions for Q equal r over N, general solution}).  

The situation for $\Psi_{(k)i}^{[t_2]_{\scriptsize (q_1k-r)/\mbox{gcd}(k,r)},[t_4]_{q_3}}$ is more involved, and it is best to give explicit constructions for the various cases.  

{\flushleft{\underline {\bf Case I: $r=1$}.}} In this case, there is a single vector $\Psi_{(k)}^{[t_2]_{\scriptsize q_1k-1},[t_4]_{q_3}}$ given by  
\begin{eqnarray}
\Psi_{(k)}^{[t_2]_{q_1k-1},[t_4]_{q_3}}(x,z)=\left[\begin{array}{c} \Sigma_1^{(-t_2+q_1,t_4)}(x,z)\\  \Sigma_2^{(-t_2+2q_1,t_4)}(x,z)\\.\\.\\.\\  \Sigma_{k-1}^{(-t_2+(k-1)q_1,t_4)}(x,z)\\ \Sigma_{k}^{(-t_2+kq_1,t_4)}(x,z) \end{array}\right]\,.
\end{eqnarray}
It is trivial to see that each zero mode respects the boundary conditions (\ref{the set of BCS 2}) in the $x_2,x_3,x_4$ directions.  
Under the $x\rightarrow x+\hat e_1 L_1$, we use (\ref{BCS for Sigma full list}) to find  
\begin{eqnarray}
\Psi_{(k)}^{[t_2]_{q_1k-1},[t_4]_{q_3}}(x+\hat e_1 L_1,z)=\gamma_k^{-1}e^{i\frac{2\pi(1/k-q_1)x_2}{L_2}}\left[\begin{array}{c} \Sigma_k^{(-t_2+kq_1,t_4)}(x,z)\\  \Sigma_1^{(-t_2+q_1,t_4)}(x,z)\\.\\.\\.\\  \Sigma_{k-2}^{(-t_2+(k-2)q_1,t_4)}(x,z)\\ \Sigma_{k-1}^{(-t_2+(k-1)q_1,t_4)}(x,z) \end{array}\right]\,,
\end{eqnarray}
where for the first entry we used the identity $\Sigma_0^{(-t_2,t_4)}=\Sigma_k^{(-t_2+kq_1,t_4)}$. This is the correct boundary condition in the $x_1$ direction.

{\flushleft{\underline{\bf Case II: $r=k$}.}} The $SU(k)$-valued zero modes can be embedded as vectors in the $SU(k)$ group as
\begin{eqnarray}\nonumber
\Psi_{(k)1}^{[t_2]_{q_1-1},[t_4]_{q_3}}(x,z)=\left[\begin{array}{c}\Sigma_{1}^{(t_2,t_4)}(x,z)\\0\\.\\.\\0 \end{array}\right]\,, \Psi_{(k)2}^{[t_2]_{q_1-1},[t_4]_{q_3}}(x,z)=\left[\begin{array}{c}0\\\Sigma_{2}^{(t_2,t_4)}(x,z)\\0\\.\\.\\0 \end{array}\right]\,,\mbox{etc.}\,,\\
\end{eqnarray}  
or in general 
\begin{eqnarray}\label{general form of k eq k zero modes}
\Psi_{(k)C'}^{[t_2]_{q_1-1},[t_4]_{q_3}}(x,z)=\left[\begin{array}{c}0\\.\\.\\\Sigma_{C'}^{(t_2,t_4)}(x,z)\\0\\.\\.\\0 \end{array}\right]\,,
\end{eqnarray}  
$C'=1,2,...,k$, $t_2=0,1,...,q_1-2$, and $t_4=0,1,..,q_3-1$. It is easy to see that such zero modes satisfy the boundary conditions in the $x_2,x_3,x_4$ directions, as can be seen from (\ref{BCS for Sigma full list}). However, along the $x_1$ direction we have
 \begin{eqnarray}
\Psi_{(k)C'}^{[t_2]_{q_1-1},[t_4]_{q_3}}(x+\hat e_1 L_1,z)=\gamma_k^{-k}e^{i\frac{2\pi(1- q_1) x_2}{L_2}}\left[\begin{array}{c}0\\.\\.\\\Sigma_{C'-k}^{(t_2-q_1,t_4)}(x,z)\\0\\.\\.\\0 \end{array}\right]\,.
\end{eqnarray}  
Using the identity 
\begin{eqnarray}
\Sigma_{C'-k}^{(t_2-q_1,t_4)}(x,z)=\Sigma_{C'}^{(t_2,t_4)}(x,z)
\end{eqnarray}
we find
 \begin{eqnarray}\nonumber
\Psi_{(k)C'}^{[t_2]_{q_1-1},[t_4]_{q_3}}(x+\hat e_1 L_1,z)=\gamma_k^{-k}e^{i\frac{2\pi(1- q_1) x_2}{L_2}}\left[\begin{array}{c}0\\.\\.\\\Sigma_{C'}^{(t_2,t_4)}(x,z)\\0\\.\\.\\0 \end{array}\right]=\gamma_k^{-k}e^{i\frac{2\pi(1- q_1) x_2}{L_2}}\Psi_{(k)C'}^{[t_2]_{q_1-1},[t_4]_{q_3}}(x,z)\,,\\
\end{eqnarray}  
showing that $\Psi_{(k)C'}^{[t_2]_{q_1-1},[t_4]_{q_3}}$  respects the appropriate boundary condition.

{\flushleft{\underline{\bf Case III: $1\leq r\leq k$}.}} This is the general case. Here, it is easier to give examples. 

In the first example, we take $k=10, r=4$, with $\mbox{gcd}(k,r)=2$. Then, there are two column vectors given by
\begin{eqnarray}
\Psi_{(k)1}^{[t_2]_{5q_1-2},[t_4]_{q_3}}=\left[ \begin{array}{c} \Sigma_1^{(t_2,t_4)}\\0\\ \Sigma_{3}^{(t_2+3q_1-1,t_4)}\\0\\ \Sigma_{5}^{(t_2+q_1,t_4)}\\0 \\ \Sigma_{7}^{(t_2+4q_1-1,t_4)}\\0 \\\Sigma_{9}^{(t_2+2q_1,t_4)}\\0\end{array}  \right]\,, \quad \Psi_{(k)2}^{[t_2]_{5q_1-2},[t_4]_{q_3}}=\left[ \begin{array}{c} 0\\ \Sigma_2^{(t_2,t_4)}\\0\\ \Sigma_{4}^{(t_2+3q_1-1,t_4)}\\0\\ \Sigma_{6}^{(t_2+q_1,t_4)}\\0 \\ \Sigma_{8}^{(t_2+4q_1-1,t_4)}\\0 \\\Sigma_{10}^{(t_2+2q_1,t_4)}\end{array}  \right]\,.
\end{eqnarray}
Let us explain the rationale behind this construction. Consider, for example, $\Psi_{(k)1}^{[t_2]_{5q_1-2},[t_4]_{q_3}}$. Starting from the element $\Sigma_1^{(t_2,t_4)}$, we add $r=4$ to the subscript $1$ and $q_1$ to the superscript to obtain $\Sigma_{5}^{(t_2+q_1,t_4)}$.  The element  $\Sigma_{9}^{(t_2+2q_1,t_4)}$ is obtained by adding $4$ to the subscript and $q_1$ to the superscript of $\Sigma_{5}^{(t_2+q_1,t_4)}$. Continuing, we find that we should next obtain the term $\Sigma_{13}^{(t_2+3q_1,t_4)}$. Now, however, we use $\Sigma_{C'+k}^{(t_2,t_4)}=\Sigma_{C'}^{(t_2-1,t_4)}$ to  find $\Sigma_3^{(t_2+3q_1-1,t_4)}=\Sigma_{13}^{(t_2+3q_1,t_4)}$. Finally, adding $4$ to the subscript and $q_1$ to the superscript we obtain term $\Sigma_{7}^{(t_2+4q_1-1,t_4)}$. If we try to continue, we obtain the term $\Sigma_{11}^{(t_2+5q_1-1,t_4)}$. Then, using $\Sigma_{C'+k}^{(t_2,t_4)}=\Sigma_{C'}^{(t_2-1,t_4)}$, we find $\Sigma_{11}^{(t_2+5q_1-1,t_4)}=\Sigma_1^{(t_2+5q_1-2,t_4)}$. However, $\Sigma_1^{(t_2+5q_1-2,t_4)}=\Sigma_1^{(t_2,t_4)}$, and thus, everything is consistent. 

In the second example, we take $k=6,r=5$, with $\mbox{gcd}(k,r)=1$. Following the same rationale, we find a single-column vector
\begin{eqnarray}
\Psi_{(k)1}^{[t_2]_{6q_1-5},[t_4]_{q_3}}=\left[\begin{array}{c} \Sigma_1^{(t_2,t_4)}\\ \Sigma_2^{({t_2+5q_1-4,t_4})}\\\Sigma_3^{(t_2+4q_1-3,t_4)}\\ \Sigma_4^{(t_2+3q_1-2,t_4)}\\ \Sigma_5^{(t_2+2q_1-1,t_4)}\\ \Sigma_6^{(t_2+q_1,t_4)} \end{array}\right]\,.
\end{eqnarray}

 We now show that the zero modes $\Psi_{(k)i}^{[t_2]_{\scriptsize (q_1k-r)/\mbox{gcd}(k,r)},[t_4]_{q_3}}(x,z)$ respect the BCS (\ref{the set
 of BCS 2}). First, it is easy to see that translations by $L_\mu$ in the $x_2,x_3,x_4$ satisfy the correct BCS, as is evident from (\ref{BCS for Sigma full list}). A translation by $L_1$ in the $x_1$-direction gives $\Sigma_{C'}^{(t_2,t_4)}(x+\hat e_1 L_1,z)=\gamma_k^{-r}e^{i\frac{2\pi(r/k- q_1) x_2}{L_2}}\Sigma_{C'-r}^{(t_2-q_1,t_4)}(x,z)$, which permutes the bulk entries in the column vectors (\ref{The final expressions of PSI in suk and sul}) by $r$ positions. However, what remains is to show that\\ $\Sigma_{C'}^{(t_2,t_4)}(x+\hat e_1 L_1,z)=\gamma_k^{-r}e^{i\frac{2\pi(r/k- q_1) x_2}{L_2}}\Sigma_{C'-r}^{(t_2-q_1,t_4)}(x,z)$ acts as expected on the elements that live at the top and bottom of (\ref{The final expressions of PSI in suk and sul}). This can be seen using (\ref{peridic relation for Sigma}) and is easily demonstrated in the abovementioned examples. In the example with $k=10,r=4$ we have $\Sigma_{1}^{(t_2,t_4)}(x+\hat e_1 L_1,z)=\gamma_{10}^{-4}e^{i\frac{2\pi(2/5- q_1) x_2}{L_2}}\Sigma_{1-4}^{(t_2-q_1,t_4)}(x,z)=\gamma_{10}^{-4}e^{i\frac{2\pi(2/5- q_1) x_2}{L_2}}\Sigma_{7}^{(t_2+4q_1-1,t_4)}(x,z)$, where in the second equality we used Eqs. (\ref{peridic relation for Sigma}). This demonstrates that a translation in the $x_1$-direction acts as expected by $P_k^{-r}$ permutations. 
 
Restoring the dependence on the moduli, we observe that the previous analysis remains valid, provided we account for the fact $\phi_\mu^{i} \equiv \phi_\mu^{i+nr}$, recalling that the index $i=1,2,...,\mbox{gcd}(k,r)$ labels $\mbox{gcd}(k,r)$ independent moduli.

\subsection{The Nahm dual gauge field on $\hat{\mathbb T}^4$}

\label{sec:nahmdualgaugefield}
Let $\hat N=q_3\left(Nq_1-r\right)+q_1$ be the number of the zero modes of $\psi$. The  $U(\hat N)$ dual gauge field $\hat {\cal A}_\mu(z, \Phi)$ that lives on $\hat{\mathbb T}^4$ is given as a block-diagonal $U(\hat N)$ matrix. Defining 
\begin{eqnarray}
\hat k\equiv q_3(q_1k-r)\,, \quad \hat\ell\equiv q_1(\ell q_3+1)\,,
\end{eqnarray}
we can write $U(\hat N)$ as the direct sum
\begin{eqnarray}\label{composition of U}
U(\hat N)=  U(\hat k)\oplus U(\hat \ell)\,.
\end{eqnarray}
Then, the $U(\hat N)$ dual gauge field $\hat {\cal A}_\mu(z, \phi)$ was already defined in (\ref{explicitnahm1}), which we reproduce below. Its $U(\hat\ell)$ in $U(\hat{k})$ components are:
\ba \label{explicitnahm2}\nonumber
\hat{\cal{A}}_\mu^{(p_2, p_4; p_2', p_4')} &=& - i \int\limits_{\T^4}\sum\limits_{\alpha, B}(\Psi_{\alpha \; B}^{p_2, p_4})^* \; { \partial_{z_\mu}} \Psi_{\alpha \; B}^{p_2', p_4'} \in u(\hat{k}), \nonumber \\
&&~ \; p_2, p_2'= 0,...q_1 -1;\quad p_4, p_4' = 0,..., \ell q_3, \ea
\ba
\hat{\cal{A}}_\mu^{(t_2, i, t_4; t_2', i', t_4')} &=&- i \int\limits_{\T^4} \sum\limits_{\alpha, B'}(\Psi_{\alpha \; B' \; i}^{t_2, t_4})^*  \;{ \partial_{z_\mu}} \Psi_{\alpha \; B' \; i'}^{t_2', t_4'} \in u(\hat{\ell}), \nonumber \\\nonumber
&& t_2, t_2'= 0,..., {k q_1 -r \over \text{gcd}(k,r)}-1;\quad i, i' = 0,...\text{gcd}(k,r)-1;\quad t_4, t_4' = 0,...,  q_3\,.\\ 
\ea
Calculating its components in terms of the normalized zero modes (\ref{The final expressions of PSI in suk and sul}), we find that they are given by the integrals:
\ba
\hat{\cal{A}}_\mu^{(t_2, i, t_4; t_2', i', t_4')} &=& - i \delta^{i,i'} \int_{\mathbb T^4} \sum_{n=0}^{\scriptsize k/\mbox{gcd}(k,r)-1} \Sigma_{i+nr}^{\dagger (t_2+nq_1,t_4)}\partial_{z_\mu}\Sigma_{i+nr}^{(t_2'+nq_1,t_4')} \nonumber \\
\hat{\cal{A}}_\mu^{(p_2, p_4; p_2', p_4')} &=& - i \int_{\mathbb T^4}\sum_{C=1}^{\ell}\Phi_{C}^{*(p_2,p_4-Cq_3)}\partial_{z_\mu}\Phi_{C}^{(p_2',p_4'-Cq_3)}
\ea
By directly calculating the integrals, we find, employing an explicit $\hat{N} \times \hat{N}$ matrix notation:
\begin{eqnarray}\nonumber
\hat {\cal A}_1(z, \phi)&=&2\pi L_1L_2 z_2\left[\begin{array}{cc}\frac{k}{kq_1-r} {\bf 1}_{\hat{k}} &0\\ 0 &\frac{1}{q_1}{\bf 1}_{\hat\ell} \end{array}\right]+\delta\hat{\cal A}_1(\phi)\,,\\\nonumber
\hat {\cal A}_3^{}(z,\phi)&=&2\pi L_3L_4 z_4\left[\begin{array}{cc}\frac{1}{q_3} {\bf 1}_{\hat{k}} &0\\ 0 &\frac{\ell}{1+\ell q_3}{\bf 1}_{\hat\ell} \end{array}\right]+\delta\hat{\cal A}_3(\phi)\,,\\
 \hat {\cal A}_2^{}(z,\phi)&=&\hat {\cal A}_4^{}(z,\phi)=0\,. \label{dual1}
\end{eqnarray}
where the moduli-dependent parts are
 \begin{eqnarray}
 \delta\hat {\cal A}_1(\phi)=2\pi  L_1 L_2\left[ \begin{array}{cc}-\frac{\ell k}{kq_1-r}\left[\begin{array}{ccccc}\phi_2^1 &0&...&\\ 0& \phi_2^2&0&... \\...&...&...&...\\0&...&\phi_2^{\scriptsize \mbox{gcd}(k,r)}&...\\...&...&...&\phi_2^1&...\\....&....&...&....&... \end{array}\right]_{\hat{k}\times \hat{k}} &0 \\0& \frac{k\tilde \phi_2}{q_1}{\bf 1}_{\hat\ell} \end{array}\right]\,, \label{dual2}
 \end{eqnarray}
 and 
 \begin{eqnarray}
 \delta \hat{\cal A}_3(\phi)=2\pi  L_3 L_4\left[ \begin{array}{cc}-\frac{\ell }{q_3}\left[\begin{array}{ccccc}\phi_4^1 &0&...&\\ 0& \phi_4^2&0&... \\...&...&...&...\\0&...&\phi_4^{\scriptsize \mbox{gcd}(k,r)}&...\\...&...&...&\phi_4^1&...\\....&....&...&....&... \end{array}\right]_{\hat{k}\times \hat{k}} &0 \\0& \frac{\ell k\tilde \phi_4}{1+\ell q_3}{\bf 1}_{\hat\ell} \end{array}\right]\,.\label{dual3}
 \end{eqnarray}
 Since there are $\mbox{gcd}(k,r)$ independent moduli (in each spacetime direction), they naturally fit inside the  $\hat{k}\times \hat{k}$ matrix. The gauge fields $\hat{\cal A}_\mu^{}(z)$ do not explicitly depend on $\phi_{1,3}$. However, as we will demonstrate, these fields do appear in the $U(\hat N)$ transition functions on $\hat{\mathbb T}^4$. Through an appropriate gauge transformation, $\phi_{1,3}$ can be removed from the transition functions, only to reemerge in the expressions for the gauge fields. Furthermore, we will show that the Wilson lines exhibit dependence on the moduli across all four dimensions.

We proceed by decomposing the $U(\hat N)$ gauge field into its $SU(\hat N)$ and $U(1)$ components. 
 The dual $U(1)$ gauge field is obtained by tracing over the full $U(\hat N)$ gauge field
 \begin{eqnarray}
\hat a_\mu=\frac{\mbox{tr}[\hat {\cal A}_\mu]}{\hat N}\,,
\end{eqnarray}
and hence
 \begin{eqnarray}\label{expressions of dual abelian field for general r}\nonumber
 \hat a_1&=&\frac{2\pi L_1 L_2 z_2 (q_3N+1)+2\pi k L_1L_2 \tilde \phi_2}{\hat N}\,,\quad  \hat a_3=\frac{2\pi L_3 L_4 z_4(Nq_1-r)+2\pi\ell rL_3L_4\tilde \phi_4}{\hat N}\,,\\
  \hat a_2&=& \hat a_4=0\,,
 \end{eqnarray}
 where we used (\ref{contraint on phitild}). The $SU(\hat N)$ gauge field $\hat A_\mu$ is derived by subtracting the $U(1)$ component from the $U(\hat N)$ gauge field:
\begin{eqnarray}
\hat A_\mu=\hat {\cal A}_\mu-I_{\hat N}\hat a_\mu\,,
\end{eqnarray}
and hence
\begin{eqnarray}\nonumber\label{general SUM fields with moduli}
\hat A_1(z,\phi)&=&L_1L_2 z_2\frac{r}{q_1(kq_1-r)\hat N}\hat \omega+\delta \hat A_1(\phi)\,,\quad \hat A_3(z,\phi)= L_3L_4 z_4\frac{1}{q_3(1+\ell q_3)\hat N}\hat \omega+\delta A_3(\phi)\,,\\
\hat A_2(z,\phi)&=&\hat A_4(z,\phi)=0\,,
\end{eqnarray}
where we define $\hat \omega$ (similar to (\ref{omega})) via
\begin{eqnarray}
\hat \omega=2\pi\left[\begin{array}{cc}\hat\ell\;{\bf 1}_{\hat{k}} &0\\ 0 &-\hat{k}\;{\bf 1}_{\hat\ell} \end{array}\right]\,,
\end{eqnarray}
and the moduli-dependent parts are 
\begin{eqnarray}
\delta\hat A_1(\phi)&=&\delta\hat{\cal A}_1(\phi)-\frac{2\pi k L_1 L_2 \tilde \phi_2}{\hat N} I_{\hat N}\,,\quad
\delta \hat A_3(\phi)=\delta\hat{\cal A}_3(\phi) -\frac{2\pi\ell rL_3L_4\tilde \phi_4}{\hat N}I_{\hat N}\,.
\end{eqnarray}

The $U(\hat N)$ field strength $\hat{\cal F}_{\mu\nu}$ is obtained directly from $\hat {\cal A}_\mu$ via $\hat{\cal F}_{\mu\nu}=\partial_\mu \hat {\cal A}_\nu-\partial_\nu\hat {\cal A}_\mu$, with the nonvanishing components given by
\begin{eqnarray}\nonumber
\hat {\cal F}_{12}(z)&=&-2\pi L_1L_2 \left[\begin{array}{cc}\frac{k}{kq_1-r}\; {\bf 1}_{\hat{k}} &0\\ 0 &\frac{1}{q_1}\; {\bf 1}_{\hat\ell} \end{array}\right]\,,\\
\hat {\cal F}_{34}(z)&=&-2\pi L_3 L_4 \left[\begin{array}{cc}\frac{1}{q_3} \; {\bf 1}_{\hat{k}} &0\\ 0 & \frac{\ell}{1+\ell q_3}\; {\bf 1}_{\hat\ell} \end{array}\right]\,.
\end{eqnarray}
Similarly, the non-vanishing components of the $SU(\hat N)$ field strength are
\begin{eqnarray}\nonumber
\hat F_{12}(z)&=&-2\pi L_1L_2 \left[\begin{array}{cc}\left(\frac{k}{kq_1-r}-\frac{1+Nq_3}{\hat N}\right) {\bf 1}_{\hat k} &0\\ 0 &\left(\frac{1}{q_1}-\frac{1+Nq_3}{\hat N}\right){\bf 1}_{\hat \ell} \end{array}\right]\,,\\
\hat F_{34}(z)&=&-2\pi L_3 L_4 \left[\begin{array}{cc}\left(\frac{1}{q_3}-\frac{Nq_1-r}{\hat N}\right){\bf 1}_{\hat k} &0\\ 0 & \left(\frac{\ell}{1+\ell q_3}-\frac{Nq_1-r}{\hat N}\right){\bf 1}_{\hat \ell} \end{array}\right]\,,
\end{eqnarray}
and the field strength of the dual $U(1)$ gauge field is
\begin{eqnarray}\label{dula U1 fields}
\hat f_{12}=-2\pi L_1 L_2 \frac{(q_3N+1)}{\hat N}\,,\quad \hat f_{34}=-2\pi L_3 L_4 \frac{(Nq_1-r)}{\hat N}\,.
\end{eqnarray}

Using this information, we find that the $SU(\hat N)$ topological charge is given by
\begin{eqnarray}\label{indexappendix1}
Q^{SU(\hat N)}=\frac{1}{4\pi^2}\int_{\hat{\mathbb T}^4}\mbox{tr}\left[\hat F_{12}(z)\hat F_{34}(z) \right]=\frac{r}{\hat N}\,,
\end{eqnarray}
while the $U(1)$ topological charge is
\begin{eqnarray}\label{indexappendix2}
Q^{U(1)}=\frac{1}{4\pi^2}\int_{\hat{\mathbb T}^4}\hat f_{12}(z)\hat f_{34}(z)=\frac{-r}{{{\hat N}}^2}+\frac{N^2q_1q_3-rNq_3+Nq_1}{{\hat N}^2}=\frac{N}{\hat N}-\frac{r}{{\hat N}^2}\,
\end{eqnarray}
From this information, we readily calculate the Dirac index of a fundamental fermion that lives on $\hat {\mathbb T}^4$ in the background of the $U(\hat N)$ gauge field as 
\begin{eqnarray}\label{indexappendix3}
\hat{\cal I}_{\Box}=Q^{SU(\hat N)}+\hat N Q^{U(1)}=\frac{r}{\hat N}+\hat N\left(\frac{N}{\hat N}-\frac{r}{\hat N^2}\right)=N.
\end{eqnarray}
%

\subsection{Dual transition functions on $\hat{\mathbb T}^4$}
\label{appx:transitiondual}

As discussed earlier, the fermion zero modes on the original $\mathbb T^4$ depend on the $U(1)$ moduli $z_\mu$. Let $\hat L_\mu$ denote the dual periods of $\hat{\mathbb T}^4$, defined as \begin{equation}\hat L_\mu \equiv L_\mu^{-1}\,.\end{equation} We introduce the $U(\hat N)$ dual transition functions $\hat\Omega^{\dagger}(z)$, such that when traversing $\hat{\mathbb T}^4$, the fermion zero modes transform as
\begin{eqnarray}\label{how fermions transform under Omega}
\psi(x,z+\hat e_\mu \hat L_\mu) = \psi(x,z)\hat\Omega^{\dagger}(z)^{-i\frac{2\pi x_\mu}{L_\mu}},
\end{eqnarray}
and the $U(\hat N)$ dual gauge field transforms according to
\begin{eqnarray}
\hat {\cal A}_\mu (z+a_\nu \hat L_\nu) = \hat\Omega(z)\hat {\cal A}_\mu(z)\hat\Omega^{\dagger}_\nu(z) -i\hat\Omega_\nu(z)\frac{\partial \hat\Omega_\nu^{\dagger}(z)}{\partial z_\mu}\,.
\end{eqnarray}
Recalling the decomposition of $U(\hat N)$ defined in (\ref{composition of U}), we similarly decompose $\hat \Omega_\mu$ as
 \begin{eqnarray}
 \hat \Omega_\mu=\hat \Omega_{(\hat k)\mu} \oplus \hat \Omega_{(\hat \ell) \mu} =\left[\begin{array}{cc}\hat \Omega_{(\hat k)\mu}  &0\\ 0&\hat \Omega_{(\hat \ell) \mu}\end{array} \right]\,,
 \end{eqnarray}
where $\hat k\equiv q_3(q_1k-r)$ and $\hat \ell\equiv q_1(\ell q_3+1)$.
In the following, we examine the transformation properties of the functions $\Phi^{(p_2,p_4)}_C(x,z,\phi)\) and \(\Sigma^{(t_2,t_4)}_{C'}(x,z,\phi)$, as defined in (\ref{final expression of Sigma moduli}, \ref{final expression of Phi moduli}), to constarct the transition functions $\hat\Omega(z)$.

Under the translations $z\rightarrow z+\hat e_\mu \hat L_\mu$ we find:
\begin{eqnarray}\label{transformation in z on Phi second}\nonumber
\Phi_C^{(p_2,p_4)}(x,z+\hat e_1 \hat L_1, \phi)&=&e^{i\frac{2\pi p_2}{q_1}}\Phi_C^{(p_2,p_4)}(x,z,\phi) e^{-i\frac{2\pi x_1}{L_1}}\,,\\
\nonumber
\Phi_C^{(p_2,p_4)}(x,z+\hat e_2\hat  L_2,\phi)&=&e^{i\frac{2\pi L_1 \left(z_1+k\tilde \phi_1\right) }{q_1}}\Phi_C^{(p_2-1,p_4)}(x,z,\phi)e^{-i\frac{2\pi x_2}{L_2}}\,,\\
\nonumber
\Phi_C^{(p_2,p_4)}(x,z+\hat e_3 \hat L_3, \phi)&=&e^{-i\frac{2\pi (2C-1-\ell -2\ell p_4)}{2(1+\ell q_3)}}\Phi_C^{(p_2,p_4)}(x,z,\phi)e^{-i\frac{2\pi x_3}{L_3}}\,,\\
\Phi_C^{(p_2,p_4)}(x,z+\hat e_4 \hat L_4, \phi)&=& e^{i\frac{2\pi \ell L_3 \left(z_3+k\tilde \phi_3\right) }{1+\ell q_3}}e^{i\frac{\pi (1-\ell)}{(1+\ell q_3)}}\Phi_C^{(p_2,p_4-1)}(x,z,\phi)e^{-i\frac{2\pi x_4}{L_4}}\,,
\end{eqnarray}
and
 \begin{eqnarray}\label{Sigma p trans for general r Phi special values}\nonumber
\Sigma_{i+nr}^{(t_2+nq_1,t_4)}(x,z+\hat e_1 \hat L_1, \phi)&=&e^{-i\frac{2\pi(2i-1-k-2t_2k)}{2(q_1k-r)}} \Sigma_{i+nr}^{(t_2+nq_1,t_4)}(x,z,\phi) e^{-i\frac{2\pi x_1}{L_1}}\,,\\
\nonumber
\Sigma_{i+nr}^{(t_2+nq_1,t_4)}(x,z+\hat e_2 \hat L_2,\phi)&=&e^{i\frac{2\pi k L_1 \left(z_1-\ell \phi_1^{i} \right)  }{(q_1k-r)}}e^{-\frac{i \pi r k (1-k)}{q_1k-r}} \Sigma_{i+nr}^{(t_2+nq_1-1,t_4)}(x,z,\phi)e^{-i\frac{2\pi x_2}{L_2}}\,,\\
\nonumber
\Sigma_{i+nr}^{(t_2+nq_1,t_4)}(x,z+\hat e_3 \hat L_3,\phi)&=&e^{i\frac{2\pi t_4}{q_3}}\Sigma_{i+nr}^{(t_2+nq_1,t_4)}(x,z,\phi)e^{-i\frac{ 2\pi x_3}{L_3}}\,,\\
\Sigma_{i+nr}^{(t_2+nq_1,t_4)}(x,z+\hat e_4 \hat L_4,\phi)&=&e^{i\frac{2\pi L_3 \left(z_3-\ell \phi_3^{i}\right)}{q_3}}\Sigma_{i+nr}^{(t_2+nq_1,t_4-1)}(x,z,\phi)e^{-i\frac{ 2\pi x_4}{L_4}}\,.
\end{eqnarray}
Then, inspection of the structure of the fermion zero modes given by Eqs. (\ref{general form of fermion zero modes}, \ref{The final expressions of PSI in suk and sul}) along with transformation law given by (\ref{how fermions transform under Omega}) reveals the $U(\hat N)$ dual transition functions, which we give in terms of the components:
\begin{eqnarray}\label{transition matrix in the first group}\nonumber
\hat \Omega^{(p_2p_2',p_4p_4')}_{(\hat\ell)1}(z,\phi)&=&e^{-i\frac{2\pi p_2}{q_1}}\delta_{p_2p_2'}\delta_{p_4p_4'}\,,\\\nonumber
\hat \Omega^{ (p_2p_2',p_4p_4')}_{(\hat\ell)2}(z,\phi)&=&e^{-i\frac{2\pi L_1 \left(z_1+k\tilde \phi_1\right)}{q_1}}\delta_{p_2,p_2'+1}\delta_{p_4p_4'}\,,\\\nonumber
\hat \Omega^{ (p_2p_2',p_4p_4')}_{(\hat\ell)3}(z,\phi)&=& e^{-i\frac{2\pi (2\ell p_4 +1+\ell)}{2(1+\ell q_3)}} \delta_{p_2,p_2'} \delta_{p_4,p_4'}\,,\\\nonumber
  \hat\Omega^{(p_2p_2',p_4p_4')}_{(\hat\ell)4}(z,\phi)&=&e^{-i\frac{2\pi \ell L_3 \left(z_3+k\tilde \phi_3\right)}{1+\ell q_3}}e^{-i\frac{\pi(1-\ell)}{1+\ell q_3}}\delta_{p_2,p_2'}\delta_{p_4,p_4'+1}\,,\\
\end{eqnarray}
and
\begin{eqnarray}\label{transition matrix in the second group}
 \nonumber
 \hat \Omega^{  (t_2t_2',t_4t_4')}_{(\hat k)1(ii')}(z,\phi)&=&e^{i\frac{2\pi(2i-1-k-2t_2k)}{2(q_1k-r)}}\delta_{ii'}\delta_{t_2t_2'}\delta_{t_4t_4'}\,,\\\nonumber
 \hat \Omega^{  (t_2t_2',t_4t_4')}_{(\hat k)2(ii')}(z,\phi)&=&e^{-i\frac{2\pi k L_1 \left(z_1-\ell \phi_1^{i} \right)  }{(q_1k-r)}}e^{\frac{i \pi r k (1-k)}{q_1k-r}} \delta_{ii'}\delta_{t_2,t_2'+1}\delta_{t_4t_4'}\,,\\\nonumber
 \hat \Omega^{  (t_2t_2',t_4t_4')}_{(\hat k)3(ii')}(z,\phi)&=&e^{-i \frac{2\pi t_4}{q_3}}\delta_{ii'}\delta_{t_2t_2'}\delta_{t_4t_4'}\,,\\
 \hat \Omega^{ (t_2t_2',t_4t_4')}_{(\hat k)4(ii')}(z,\phi)&=&e^{-i\frac{2\pi L_3 \left(z_3-\ell \phi_3^{i}\right)}{q_3}}\delta_{ii'}\delta_{t_2t_2'}\delta_{t_4,t_4'+1}\,,
 \end{eqnarray}
and we recall the range of indices is
 \begin{eqnarray}\nonumber
 i&=&1,2,..,\mbox{gcd}(k,r)\,, \quad t_2=0,1,..., \frac{q_1k-r}{\mbox{gcd}(k,r)}-1\,,\quad t_4=0,1,..,q_3-1\,,\\
 p_2&=&0,1,..,q_1-1\,,\quad p_4=0,1,...,\ell q_3\,.
\end{eqnarray}

Next, we break the transition functions into its abelian $\hat \omega_\mu$ and nonabelian parts $ \hat \Omega^{SU(\hat N)}$, such that the abelian field and the $SU(\hat N)$ nonabelian fields transform as
\begin{eqnarray} \label{gauge transformation of dual abelian A general}
\hat a_\mu (z+\hat e_\nu \hat L_\nu)=\hat a_\mu\left(z\right)  -i\hat\omega_\nu(z)\frac{\partial \hat\omega_\nu^\dagger(z)}{\partial z_\mu}\,,
\end{eqnarray}
and
 \begin{eqnarray} \label{gauge transformation of dual A general SUM}
\hat A_\mu (z+\hat e_\nu \hat L_\nu)=\hat\Omega^{SU(\hat N)}_\nu(z)\hat A_\mu\left(z\right)\hat\Omega^{\dagger SU(\hat N)}_\nu(z)  -i\hat\Omega^{SU(\hat N)}_\nu(z)\frac{\partial \hat\Omega_\nu^{\dagger SU(\hat N)}(z)}{\partial z_\mu}\,.
\end{eqnarray}

The abelian transition functions are the $\hat N$-th root of the determinants of the $U(\hat N)$ transition functions:
 \begin{eqnarray}\nonumber\label{the abelian transition functions after simplifications}
\hat\omega_1(z,\phi)&=&\left[\mbox{Det}\,\hat\Omega_{(\ell)1}\mbox{Det}\,\hat\Omega_{(k)1}\right]^{1/\hat N} ={\cal T}_1\,,\\\nonumber
\hat\omega_2(z,\phi)&=&\left[\mbox{Det}\,\hat\Omega_{(\ell)2}\mbox{Det}\,\hat\Omega_{(k)2}\right]^{1/\hat N} ={\cal T}_2 e^{-i 2\pi\frac{ L_1\left( z_1(1+Nq_3)+k\tilde \phi_1\right)}{\hat{N}} }\,,\\\nonumber
\hat\omega_3(z,\phi)&=&\left[\mbox{Det}\,\hat\Omega_{(\ell)3}\mbox{Det}\,\hat\Omega_{(k)3}\right]^{1/\hat N} ={\cal T}_3\,,\\
\hat\omega_4(z,\phi)&=&\left[\mbox{Det}\,\hat\Omega_{(\ell)4}\mbox{Det}\,\hat\Omega_{(k)4}\right]^{1/\hat N} ={\cal T}_4e^{-i 2\pi \frac{ L_3\left( z_3(q_1N-r)+r\ell \tilde \phi_3\right)}{\hat{N}} }\,,
\end{eqnarray}
 where ${\cal T}_1,{\cal T}_2,{\cal T}_3,{\cal T}_4$ are $\mathbb C$-valued numbers that are independent of $z_\mu$ and $\phi_\mu$. The $SU(\hat N)$ transition functions are obtained by dividing the $U(\hat N)$ transition functions by $\left(\hat\omega_\mu\right)^{\hat N}$. Their matrix elements are
\begin{eqnarray}\label{transition matrix in the first group for SUM}\nonumber
 \hat\Omega^{SU(\hat N)(p_2p_2',p_4p_4')}_{(\ell)1}(z,\phi)&=&{\cal T}_1^{-\hat N}e^{-i\frac{2\pi p_2}{q_1}}\delta_{p_2p_2'}\delta_{p_4p_4'}\,,\\\nonumber
 \hat\Omega^{SU(\hat N) (p_2p_2',p_4p_4')}_{(\ell)2}(z,\phi)&=&{\cal T}_2^{-\hat N}e^{-i\frac{2\pi L_1 \left(z_1+k\tilde\phi_1\right)}{q_1}}e^{i \frac{2\pi L_1\left( z_1(1+Nq_3)+k\tilde \phi_1\right)}{\hat N} }\delta_{p_2,p_2'+1}\delta_{p_4p_4'}\,,\\\nonumber
\hat \Omega^{SU(\hat N) (p_2p_2',p_4p_4')}_{(\ell)3}(z,\phi)&=&{\cal T}_3^{-\hat N} e^{-i\frac{2\pi (2\ell p_4 +1+\ell)}{2(1+\ell q_3)}} \delta_{p_2,p_2'} \delta_{p_4,p_4'}\,,\\\nonumber
 \hat \Omega^{SU(\hat N)(p_2p_2',p_4p_4')}_{(\ell)4}(z,\phi)&=&{\cal T}_4^{-\hat N}e^{-i\frac{2\pi \ell L_3 \left(z_3+k\tilde\phi_3\right)}{1+\ell q_3}}e^{-i\frac{\pi(1-\ell)}{1+\ell q_3}}e^{i  \frac{2\pi L_3\left( z_3(q_1N-r)+r\ell \tilde \phi_3\right)}{\hat N} }\delta_{p_2,p_2'}\delta_{p_4,p_4'+1}\,.\\
\end{eqnarray}
and 
  \begin{eqnarray}\label{transition matrix in the second group for SUM}
 \nonumber
 \hat\Omega^{SU(\hat N)  (t_2t_2',t_4t_4')}_{(k)1(ii')}(z,\phi)&=&{\cal T}_1^{-\hat N}e^{i\frac{2\pi(2i-1-k-2t_2k)}{2(q_1k-r)}}\delta_{ii'}\delta_{t_2t_2'}\delta_{t_4t_4'}\,,\\\nonumber
 \hat\Omega^{ SU(\hat N) (t_2t_2',t_4t_4')}_{(k)2(ii')}(z,\phi)&=&{\cal T}_2^{-\hat N}e^{-i\frac{2\pi k L_1 \left(z_1-\ell \phi_1^{i} \right)  }{(q_1k-r)}}e^{\frac{i \pi r k (1-k)}{q_1k-r}}e^{i \frac{2\pi L_1\left( z_1(1+Nq_3)+k\tilde \phi_1\right)}{\hat N} } \delta_{ii'}\delta_{t_2,t_2'+1}\delta_{t_4t_4'}\,,\\\nonumber
  \hat\Omega^{SU(\hat N)  (t_2t_2',t_4t_4')}_{(k)3(ii')}(z,\phi)&=&{\cal T}_3^{-\hat N}e^{-i \frac{2\pi t_4}{q_3}}\delta_{ii'}\delta_{t_2t_2'}\delta_{t_4t_4'}\,,\\
   \hat\Omega^{SU(\hat N) (t_2t_2',t_4t_4')}_{(k)4(ii')}(z,\phi)&=&{\cal T}_4^{-\hat N}e^{-i\frac{2\pi L_3 \left(z_3-\ell \phi_3^{i}\right)}{q_3}}e^{i \frac{2\pi L_3\left( z_3(q_1N-r)+r\ell \tilde \phi_3\right)}{\hat N} }\delta_{ii'}\delta_{t_2t_2'}\delta_{t_4,t_4'+1}\,,
 \end{eqnarray}

It can be easily checked that the dual $U(1)$ and $SU(\hat N)$ transition functions satisfy the cocycle conditions
 \begin{eqnarray}\label{cocycle conditions for both 1}
 \nonumber
 \hat \Omega^{SU(\hat N)}_{\mu}(z+\hat e_\nu \hat L_\nu)\hat \Omega^{SU(\hat N)}_\nu(z)&=&e^{i\frac{2\pi \hat n_{\mu\nu}}{\hat N}} \hat \Omega^{SU(\hat N)}_\nu(z+\hat e_\mu \hat L_\mu)\hat \Omega^{SU(\hat N)}_\mu (z)\,,\\
  \hat\omega^{}_{\mu}(z+\hat e_\nu \hat L_\nu)\hat \omega^{}_\nu(z)&=&e^{-i\frac{2\pi \hat n_{\mu\nu}}{\hat N}} \hat\omega^{}_\nu(z+\hat e_\mu \hat L_\mu)\hat \omega^{}_\mu (z)\,,
 \end{eqnarray}
 with  
 \begin{eqnarray}\label{dualtwists}
 \hat n_{12}=- \hat n_{21}=1+Nq_3\,,\quad \hat n_{34}=-\hat n_{43}=q_1N-r\,.
 \end{eqnarray}
Also, it can be easily checked that the dual abelian gauge field, given in (\ref{expressions of dual abelian field for general r}), as well as the $SU(\hat N)$  field, given in (\ref{general SUM fields with moduli}),  respect the transformation laws (\ref{gauge transformation of dual abelian A general}) and (\ref{gauge transformation of dual A general SUM}).

\subsection{Canonical form of the dual gauge fields and transition functions}
\label{appx:transitiondualcanonical}

In the previous analysis, it was noted that the moduli $\phi_\mu$ are distributed between the gauge fields and transition functions. Specifically, the gauge fields exhibit dependence on $\phi_2$ and $\phi_4$, while the transition functions are influenced by $\phi_1$ and $\phi_3$. However, it would be preferable to adopt a gauge where the moduli dependence resides entirely in the gauge fields, leaving the transition functions independent of them.
 
To this end, we apply a $U(\hat N)$ gauge transformation by using the gauge function $g(z)$:
 \begin{eqnarray}\nonumber
 g(z)=\left[\begin{array}{cc}\left[\begin{array}{ccccc}Y_1 &0&...&\\ 0& Y_2&0&... \\...&...&...&...\\0&...&Y_{\rm{gcd}(k,r)}&...\\...&...&...&Y_1&...\\....&....&...&....&... \end{array}\right]_{\hat k \times \hat k}   &0 \\ 0 &  e^{\frac{i 2\pi k L_1\tilde\phi_1 z_2L_2}{q_1}}e^{i\frac{2\pi \ell k L_3\tilde \phi_3 z_4L_4}{1+\ell q_3}}\bm 1_{\hat \ell} \end{array} \right]\,,\\
 \end{eqnarray}
  where
 \begin{eqnarray}
 Y_i\equiv e^{-\frac{i 2\pi \ell k L_1 \phi_1^i z_2 L_2}{kq_1-r} -\frac{i 2\pi \ell L_3\phi_3^i z_4L_4}{q_3}} 
 \end{eqnarray}
 to remove the dependence on $\phi_{1}^i$ and $\phi_{3}^i$ from the transition functions.
Under this gauge transformation, the $U(\hat N)$ transition functions and gauge fields transform as
 \begin{eqnarray}\nonumber
 \hat \Omega'^{}_\mu(z)&=&g(z_\mu=L_\mu^{-1})\hat \Omega^{}_\mu(z) g^{-1}(z=0)\,,\\
  {\cal A}'^{}_\mu(z)&=&g(z)\left({\cal A}^{}_\mu-i\partial_\mu^z\right)g^{-1}(z)\,.
 \end{eqnarray}

A direct application of this gauge transformation gives the new $\phi$-independent $U(\hat N)$ transition functions:
\begin{eqnarray}\label{k space primed TF}
\nonumber
\hat\Omega'^{  (t_2t_2',t_4t_4')}_{1(ii')}(z)&=&e^{i\frac{2\pi(2i-1-k-2t_2k)}{2(q_1k-r)}}\delta_{ii'}\delta_{t_2t_2'}\delta_{t_4t_4'}\,,\\\nonumber
\hat\Omega'^{  (t_2t_2',t_4t_4')}_{2(ii')}(z)&=&e^{-i\frac{2\pi k L_1 z_1  }{(q_1k-r)}}e^{\frac{i \pi r k (1-k)}{q_1k-r}} \delta_{ii'}\delta_{t_2,t_2'+1}\delta_{t_4t_4'}\,,\\\nonumber
\hat\Omega'^{  (t_2t_2',t_4t_4')}_{3(ii')}(z)&=&e^{-i \frac{2\pi t_4}{q_3}}\delta_{ii'}\delta_{t_2t_2'}\delta_{t_4t_4'}\,,\\
\hat\Omega'^{ (t_2t_2',t_4t_4')}_{4(ii')}(z)&=&e^{-i\frac{2\pi L_3 z_3}{q_3}}\delta_{ii'}\delta_{t_2t_2'}\delta_{t_4,t_4'+1}\,,
\end{eqnarray}
and
\begin{eqnarray}\label{l space primed TF}
\nonumber
\hat\Omega'^{(p_2p_2',p_4p_4') }_{1}(z)&=&e^{-i\frac{2\pi p_2}{q_1}}\delta_{p_2p_2'}\delta_{p_4p_4'}\,,\\\nonumber
\hat \Omega'^{(p_2p_2',p_4p_4') }_{2}(z)&=&e^{-i\frac{2\pi L_1 z_1}{q_1}}\delta_{p_2,p_2'+1}\delta_{p_4p_4'}\,,\\\nonumber
\hat\Omega'^{ (p_2p_2',p_4p_4') }_{3}(z)&=&e^{-i\frac{2\pi (2\ell p_4 +1+\ell)}{2(1+\ell q_3)}}\delta_{p_2,p_2'}\delta_{p_4,p_4'}\,,\\
\hat\Omega'^{(p_2p_2',p_4p_4') }_{4}(z)&=& e^{-i\frac{2\pi \ell L_3 z_3}{1+\ell q_3}}e^{-i\frac{\pi(1-\ell)}{1+\ell q_3}}\delta_{p_2,p_2'}\delta_{p_4,p_4'+1}\,,
\end{eqnarray}
and recall that the range of indices is
\begin{eqnarray}\nonumber
i&=&1,2,..,\mbox{gcd}(k,r)\,, \quad t_2=0,1,..., \frac{q_1k-r}{\mbox{gcd}(k,r)}-1\,,\quad t_4=0,1,..,q_3-1\,,\\
p_2&=&0,1,..,q_1-1\,,\quad p_4=0,1,...,\ell q_3\,.
\end{eqnarray}
It is also instructive to quote the $SU(\hat N )$ transition functions:
\begin{eqnarray}\label{transition matrix in the first group for SUM P}
\nonumber
\hat \Omega'^{SU(\hat N)(p_2p_2',p_4p_4')}_{1}(z)&=&{\cal T}_1^{-\hat N}e^{-i\frac{2\pi p_2}{q_1}}\delta_{p_2p_2'}\delta_{p_4p_4'}\,,\\\nonumber
\hat \Omega'^{SU(\hat N) (p_2p_2',p_4p_4')}_{2}(z)&=&{\cal T}_2^{-\hat N}e^{-i\frac{2\pi L_1 z_1}{q_1}}e^{i \frac{2\pi L_1z_1(1+Nq_3)}{\hat N} }\delta_{p_2,p_2'+1}\delta_{p_4p_4'}'\,,\\\nonumber
\hat \Omega'^{SU(\hat N) (p_2p_2',p_4p_4')}_{3}(z)&=&{\cal T}_3^{-\hat N} e^{-i\frac{2\pi (2\ell p_4 +1+\ell)}{2(1+\ell q_3)}} \delta_{p_2,p_2'} \delta_{p_4,p_4'}\,,\\\nonumber
\hat  \Omega'^{SU(\hat N)(p_2p_2',p_4p_4')}_{4}(z)&=&{\cal T}_4^{-\hat N}e^{-i\frac{2\pi \ell L_3 z_3}{1+\ell q_3}}e^{-i\frac{\pi(1-\ell)}{1+\ell q_3}}e^{i  \frac{2\pi L_3 z_3(q_1N-r)}{\hat N} }\delta_{p_2,p_2'}\delta_{p_4,p_4'+1}\,.\\
\end{eqnarray}
and 
\begin{eqnarray}\label{transition matrix in the second group for SUM P}
\nonumber
\hat\Omega'^{SU(\hat N)  (t_2t_2',t_4t_4')}_{(ii')1}(z)&=&{\cal T}_1^{-\hat N}e^{i\frac{2\pi(2i-1-k-2t_2k)}{2(q_1k-r)}}\delta_{ii'}\delta_{t_2t_2'}\delta_{t_4t_4'}\,,\\\nonumber
\hat\Omega'^{ SU(\hat N) (t_2t_2',t_4t_4')}_{(ii')2}(z)&=&{\cal T}_2^{-\hat N}e^{-i\frac{2\pi k L_1 z_1  }{(q_1k-r)}}e^{\frac{i \pi r k (1-k)}{q_1k-r}}e^{i \frac{2\pi L_1 z_1(1+Nq_3)}{\hat N} } \delta_{ii'}\delta_{t_2,t_2'+1}\delta_{t_4t_4'}\,,\\\nonumber
\hat\Omega'^{SU(\hat N)  (t_2t_2',t_4t_4')}_{(ii')3}(z)&=&{\cal T}_3^{-\hat N}e^{-i \frac{2\pi t_4}{q_3}}\delta_{ii'}\delta_{t_2t_2'}\delta_{t_4t_4'}\,,\\
\hat\Omega'^{SU(\hat N) (t_2t_2',t_4t_4')}_{(ii')4}(z)&=&{\cal T}_4^{-\hat N}e^{-i\frac{2\pi L_3 z_3}{q_3}}e^{i \frac{2\pi L_3z_3(q_1N-r)}{\hat N} }\delta_{ii'}\delta_{t_2t_2'}\delta_{t_4,t_4'+1}\,.
\end{eqnarray}
As the index notation shows, the transition functions on $\hat\T^4$ are also composed of clock and shift matrices, similar to (\ref{the set of transition functions for Q equal r over N, general solution}).

In addition, the components of the gauge-transformed $U(\hat N)$ gauge field are given by
\begin{eqnarray}\nonumber
\hat A'^{ (t_2it_4,t_2'i't_4')}_1(z, \phi) &=&\left\{ \frac{2\pi L_1L_2 kz_2}{kq_1-r}-\frac{2\pi L_1L_2\ell k \phi_2^i}{kq_1-r} \right\}\delta_{t_2t_2'}\delta_{ii'}\delta_{t_4t_4'}\,,\\\nonumber
\hat {\cal A}'^{ (t_2it_4,t_2'i't_4')}_2(z, \phi) &=&\frac{2\pi L_1L_2 \ell k \phi_1^i}{q_1k-r}\delta_{t_2t_2'}\delta_{ii'}\delta_{t_4t_4'}\,,\\\nonumber
\hat {\cal A}'^{ (t_2it_4,t_2'i't_4')}_3(z, \phi) &=&\left\{\frac{2\pi L_3L_4   z_4}{q_3}-\frac{2\pi  L_3L_4 \ell \phi_4^i  }{q_3}\right\}\delta_{t_2t_2'}\delta_{ii'}\delta_{t_4t_4'}\,,\\
\hat {\cal A}'^{(t_2it_4,t_2'i't_4')}_4(z, \phi) &=&\frac{2\pi L_3L_4 \ell \phi_3^i}{q_3}\delta_{t_2t_2'}\delta_{ii'}\delta_{t_4t_4'}\,, \label{Uhatkfield}
\end{eqnarray}
and
\begin{eqnarray}\nonumber
\hat {\cal A}'^{(p_2p_2',p_4p_4')}_1(z, \phi) &=&\left\{\frac{2\pi L_1L_2   z_2}{q_1}+\frac{2\pi L_1L_2 k \tilde \phi_2}{q_1}\right\}\delta_{p_2p_2'}\delta_{p_4p_4'}\,,\\
\nonumber
\hat {\cal A}'^{ (p_2p_2',p_4p_4')}_2(z, \phi) &=&-\frac{2\pi L_1L_2 k\tilde\phi_1}{q_1}\delta_{p_2p_2'}\delta_{p_4p_4'}\,,\\\nonumber
\hat {\cal A}'^{ (p_2p_2',p_4p_4')}_3(z, \phi) &=&\left\{\frac{2\pi L_3L_4  \ell z_4}{(1+\ell q_3)}+\frac{2\pi L_3L_4\ell k \tilde \phi_4}{1+\ell q_3} \right\}\delta_{p_2p_2'}\delta_{p_4p_4'}\,,\\
\hat {\cal A}'^{ (p_2p_2',p_4p_4')}_4(z, \phi) &=&-\frac{2\pi L_3L_4 \ell k \tilde \phi_3}{1+\ell q_3}\delta_{p_2p_2'}\delta_{p_4p_4'}\,.\label{Uhatellfield}
\end{eqnarray} 

Recalling the abelian gauge field is given by
 \begin{eqnarray}
\hat a'^{}_\mu=\frac{\mbox{tr}[\hat {\cal A}'^{}_\mu]}{\hat N}\,,
\end{eqnarray}
 we find 
  \begin{eqnarray}\nonumber
\hat a'^{}_1(z, \phi) &=&\frac{2\pi L_1 L_2 z_2 (q_3N+1)+2\pi  L_1L_2 k \tilde \phi_2}{\hat N}\,,\\\nonumber
\hat a'^{}_2 (z, \phi) &=&-\frac{2\pi L_1L_2 k\tilde \phi_1}{\hat N}\,,\\\nonumber
\hat a'^{}_3(z,\phi)&=&\frac{2\pi L_3 L_4 z_4(Nq_1-r)+2\pi L_3L_4 r\ell \tilde \phi_4}{\hat N}\,,\\
\hat a'^{}_4(z,\phi)&=&-\frac{2\pi L_3L_4 r \ell \tilde \phi_3}{\hat N}\,.
 \end{eqnarray}
In addition, the $SU(\hat N)$ gauge field is $\hat A'_\mu=\hat {\cal A}'_\mu-I_{\hat N}\hat a'_\mu$, with components given by
\begin{eqnarray}\label{the primed gauge field}\nonumber
\hat A'^{ (t_2it_4,t_2'i't_4')}_1(z, \phi) &=&\left\{ \frac{2\pi L_1L_2 kz_2}{kq_1-r}-\frac{2\pi L_1L_2\ell k \phi_2^i}{kq_1-r}- \frac{2\pi L_1 L_2 z_2 (q_3N+1)+2\pi  L_1L_2 k \tilde \phi_2}{\hat N} \right\}\delta_{t_2t_2'}\delta_{ii'}\delta_{t_4t_4'}\,,\\\nonumber
\hat A'^{ (t_2it_4,t_2'i't_4')}_2(z, \phi) &=&\left\{\frac{2\pi L_1L_2 \ell k \phi_1^i}{q_1k-r}+\frac{2\pi L_1L_2 k\tilde \phi_1}{\hat N}\right\}\delta_{t_2t_2'}\delta_{ii'}\delta_{t_4t_4'}\,,\\\nonumber
\hat A'^{ (t_2it_4,t_2'i't_4')}_3(z, \phi) &=&\left\{\frac{2\pi L_3L_4   z_4}{q_3}-\frac{2\pi  L_3L_4 \ell \phi_4^i  }{q_3}-\frac{2\pi L_3 L_4 z_4(Nq_1-r)+2\pi L_3L_4 r\ell \tilde \phi_4}{\hat{N}}\right\}\delta_{t_2t_2'}\delta_{ii'}\delta_{t_4t_4'}\,,\\
\hat A'^{ (t_2it_4,t_2'i't_4')}_4(z, \phi) &=&\left\{\frac{2\pi L_3L_4 \ell \phi_3^i}{q_3}+\frac{2\pi L_3L_4 r \ell \tilde \phi_3}{\hat{N}}\right\}\delta_{t_2t_2'}\delta_{ii'}\delta_{t_4t_4'}\,,
\end{eqnarray}
and
\begin{eqnarray}\label{the primed gauge field2}\nonumber
\hat A'^{(p_2p_2',p_4p_4')}_1(z, \phi) &=&\left\{\frac{2\pi L_1L_2   z_2}{q_1}+\frac{2\pi L_1L_2 k \tilde \phi_2}{q_1}-\frac{2\pi L_1 L_2 z_2 (q_3N+1)+2\pi  L_1L_2 k \tilde \phi_2}{\hat N}\right\}\delta_{p_2p_2'}\delta_{p_4p_4'}\,,\\
\nonumber
\hat A'^{(p_2p_2',p_4p_4')}_2(z, \phi) &=&\left\{-\frac{2\pi L_1L_2 k\tilde\phi_1}{q_1}+\frac{2\pi L_1L_2 k\tilde \phi_1}{\hat N}\right\}\delta_{p_2p_2'}\delta_{p_4p_4'}\,,\\\nonumber
\hat A'^{ (p_2p_2',p_4p_4')}_3(z, \phi) &=&\left\{\frac{2\pi L_3L_4  \ell z_4}{(1+\ell q_3)}+\frac{2\pi L_3L_4\ell k \tilde \phi_4}{1+\ell q_3}-\frac{2\pi L_3 L_4 z_4(Nq_1-r)+2\pi L_3L_4 r\ell \tilde \phi_4}{\hat N} \right\}\delta_{p_2p_2'}\delta_{p_4p_4'}\,.\\\nonumber
\hat A'^{(p_2p_2',p_4p_4')}_4(z, \phi) &=&\left\{-\frac{2\pi L_3L_4 \ell k \tilde \phi_3}{1+\ell q_3}+\frac{2\pi L_3L_4 r \ell \tilde \phi_3}{\hat N}\right\}\delta_{p_2p_2'}\delta_{p_4p_4'}\,.\\
\end{eqnarray}

\section{Moduli space on $\hat{\mathbb T}^4$ and the shape of the $SU(\hat N)$ moduli}
\label{appx:moduli}

This section discusses the dual gauge field zero modes (moduli) on the dual $\hat{\mathbb{T}}^4$.  First, inspection of the dual transition functions (\ref{k space primed TF}, \ref{l space primed TF}), we argue that there is a maximum number of $\mbox{gcd}(k,r)$ holonomies; this is the dimension of the subspace spanned by the index $i$. Basically, the group $U(\hat N )$ is decomposed as $U(\hat N)\supset U(1)\times SU(q_3(kq_1-r))\times SU(q_1(1+\ell q_3))$. The holonomies take the form $\hat \phi_\mu^a H^{a,SU(q_3(kq_1-r))}_\mu$ in $SU(q_3(kq_1-r))$ and $\hat \phi_\mu^A H^{A,SU(q_3(kq_1-r))}_\mu$ in $SU(q_1(1+\ell q_3))$ subgroups, where $H^a$ and $H^A$ are the corresponding Cartan generators. There exists no Cartan generator that commutes with the transition functions  (\ref{l space primed TF}) since they contain the shift matrix in the $2$ and $4$ directions. However, there are $\mbox{gcd}(k,r)$ Cartan generators $H^A$ of  $SU(q_3(kq_1-r))$ that commute with the transition functions  (\ref{k space primed TF}), since, as said above, the transition functions contain diagonal matrices in the space spanned by the index $i$. When $k=r$, there are $\mbox{gcd}(k,r)=r$ holonomies in each direction, coinciding with the number of the bosonic zero modes, as expected from the index theorem.  In this case, the dual holonomies are mapped to $\phi_\mu^i$ on the original $\mathbb T^4$. 


In this section, we calculate Wilson's lines on $\hat{\mathbb T}^4$ and use them to gain insights into the structure of the moduli space on the $SU(\hat N)$ dual gauge group. 
The dual $SU(\hat N)$ and $U(1)$ lines are given by
\begin{eqnarray}\nonumber 
\hat W_\mu^{SU(\hat  N)}(z,\phi)&=&\mbox{tr}\left[\exp\left[i\int_z^{z+\hat L_\mu^{}} dz'_\mu \hat A_\mu^{}(z', \phi) \right]\hat \Omega_\mu^{SU(\hat N )}(z,\phi) \right]\,,\\
\hat W_\mu^{U(1)}(z,\phi)&=&\exp\left[i\int_z^{z+\hat L_\mu} dz'_\mu \hat a_\mu^{}(z',\phi)\right]\hat \omega^{}(z,\phi)\,.
\end{eqnarray}
 Using the abelian gauge field and the transition functions given by Eqs. (\ref{expressions of dual abelian field for general r}, \ref{the abelian transition functions after simplifications}), we readily find
  \begin{eqnarray}\nonumber
 \hat W_1^{U(1)}(z,\phi)&\propto&\exp\left[\frac{i2\pi L_2 \left( z_2(1+q_3N)+k \tilde \phi_2\right)}{\hat N} \right]\,,\\\nonumber
  \hat W_2^{U(1)}(z,\phi)&\propto&\exp\left[\frac{-i2\pi L_1 \left( z_1(1+q_3N)+k\tilde \phi_1\right)}{\hat N} \right]\,,\\\nonumber
 \hat W_3^{U(1)}(z,\phi)&\propto&\exp\left[\frac{i2\pi L_4 \left( z_4(Nq_1-r)+\ell r\tilde \phi_4\right)}{\hat N} \right]\,,\\
  \hat W_4^{U(1)}(z,\phi)&\propto&\exp\left[\frac{-i 2\pi L_3\left( z_3(Nq_1-r)+\ell r \tilde \phi_3\right)}{\hat N} \right]\,,\label{dualU1wilson}
 \end{eqnarray}
 and the proportionality constants are not essential in what follows. 
 
 On the other hand, examination of the transition functions $\hat \Omega_{2,4}^{SU(\hat N )}(z,\phi)$ reveals that their traces must vanish. This is because these transition functions are off-diagonal matrices.  Moreover, taking the trace over the transition functions in the \( x_1 \) and \( x_3 \) directions yields zero, as a result of summing over clock phases; for example, $\sum_{j=1}^h e^{i\frac{2\pi j}{h}} = 0$. We conclude that
 \begin{eqnarray}
  \hat W_\mu^{SU(\hat N)}(z,\phi)=0\,,
 \end{eqnarray}
 for $\mu=1,2,3,4$. Yet, higher powers of $  \hat W_\mu^{SU(\hat N)}(z,\phi)$ do not generally vanish. The key idea is that, while the traces of the $p \times p$ clock and shift matrices are zero, the traces of their  $p$th powers are non-zero. For example, the $SU(\hat N)$ transition functions (\ref{transition matrix in the first group for SUM}, \ref{transition matrix in the second group for SUM}) in the $z_2$-direction is given by the $\scriptsize\frac{(q_1k-r)}{\mbox{gcd}(k,r)}$ shift matrices, and thus, its $\scriptsize\frac{(q_1k-r)}{\mbox{gcd}(k,r)}$th power has a nonzero trace. Similar observations apply to the transition functions in the $z_{1,3,4}$ directions. 
 
 Straightforward calculations yield the following two sets of nonvanishing Wilson's lines:\footnote{\label{footnote:assume}The constant phases appearing in front of the r.h.s. are not essential in what follows and are omitted. In addition, in writing (\ref{nonvainishing powers set1}, \ref{nonvainishing powers set2}) we make the simplifying assumption that  none of the ratios $r/q_1$, $q_3/(1 + \ell q_3)$, $(1 + \ell q_3)/q_3$, and $q_1/(q_1 k - r)$  are integer.}
 \begin{eqnarray}\label{nonvainishing powers set1}\nonumber
  \left(\hat W_1^{SU(\hat N)}\right)^{\scriptsize(q_1k-r)} (z,\phi)&\propto& \sum_{i=1}^{\scriptsize \mbox{gcd}(k,r)}e^{i\frac{2\pi L_2}{\hat N}\left(r(1+\ell q_3)z_2-k(kq_1-r)\tilde\phi_2-\ell k \hat N \phi_2^i\right)}\,,\\\nonumber
\left(\hat W_2^{SU(\hat N)}\right)^{\scriptsize \frac{(q_1k-r)}{\mbox{gcd}(k,r)}}(z,\phi)&\propto& \sum_{i=1}^{\scriptsize \mbox{gcd}(k,r)} e^{\scriptsize \frac{-i2\pi k L_1 \left(z_1-\ell \phi_1^{i} \right)}{\mbox{gcd}(k,r)}}  e^{\scriptsize i \frac{2\pi L_1\left( z_1(1+Nq_3)+k\tilde \phi_1\right)(q_1k-r)}{\hat N \mbox{gcd}(k,r)}}\,, \\ 
  \left(\hat W_3^{SU(\hat N)}\right)^{q_3} (z,\phi)&\propto&  \sum_{i=1}^{\scriptsize \mbox{gcd}(k,r)}e^{i\frac{2\pi L_4}{\hat N}\left(q_1 z_4-\ell r q_3 \tilde \phi_4-\ell \hat N \phi_4^i\right)}\,,\\
  \left(\hat W_4^{SU(\hat N)}\right)^{q_3} (z,\phi)&\propto&  \sum_{i=1}^{\scriptsize \mbox{gcd}(k,r)}  e^{-i2\pi L_3 \left(z_3-\ell \phi_3^{i}\right)}e^{i \frac{2\pi L_3 q_3\left( z_3(q_1N-r)+r\ell \tilde \phi_3\right)}{\hat N} }\,, \nonumber
 \end{eqnarray}
 and 
  \begin{eqnarray}\label{nonvainishing powers set2}\nonumber
       \left(\hat W_1^{SU(\hat N)}\right)^{q_1}(z,\phi)&\propto& q_1(1+\ell q_3)e^{i \frac{2\pi L_2}{\hat N}\left(-rq_3z_2+k(\hat N-q_1)\tilde \phi_2\right)}\,,\\
   \left(\hat W_2^{SU(\hat N)}\right)^{q_1}(z,\phi)&\propto& q_1(1+\ell q_3)e^{i\frac{2\pi L_1}{\hat N}\left(-(z_1+k\tilde \phi_1)\hat N+q_1z_1(1+N q_3)+k\tilde \phi_1 q_1\right)}\,,\\\nonumber
           \left(\hat W_3^{SU(\hat N)}\right)^{1+\ell q_3}(z,\phi)&\propto& q_1(1+\ell q_3)e^{-i\frac{2\pi L_4}{\hat N}\left((kq_1-r)z_4-\ell k \hat N\tilde \phi_4 + \ell r (1+\ell q_3)\tilde\phi_4\right)}\,. \\
     \left(\hat W_4^{SU(\hat N)}\right)^{1+\ell q_3}(z,\phi)&\propto& q_1(1+\ell q_3) e^{\frac{i 2\pi L_3}{\hat N}\left(- \ell \hat N\left(z_3+k\tilde \phi_3\right)+(1+\ell q_3)\left( z_3(q_1N-r)+r\ell \tilde \phi_3\right)\right) }\,, \nonumber \end{eqnarray}
The original $SU(N)$ Yang-Mills theory on $\mathbb{T}^4$ possesses a $\mathbb{Z}_N^{(1)}$ 1-form symmetry, while the $SU(\hat{N})$ theory on the dual $\hat\T^4$ has a $\mathbb{Z}_{\hat{N}}^{(1)}$ symmetry. The Wilson lines in the two theories carry charges under their respective symmetries. However, since we are not within the thermodynamic limit, the expectation value of these Wilson lines must vanish identically; symmetry must be preserved, i.e., they cannot exhibit spontaneous breaking at a finite volume. This procedure was utilized in \cite{Anber:2022qsz,Anber:2024mco} to determine the shape and volume of the bosonic moduli space on the original $\mathbb T^4$. Here, we use the same logic to determine the shape and moduli of the dual gauge theory $SU(\hat N)$. To this end, as reviewed in \cite{Anber:2022qsz,Anber:2024mco} by considering the theory on a twisted four torus ($\T^4$ or $\hat\T^4$) we must have for every non-vanishing Wilson line
 \begin{eqnarray}\label{vanishing condition}
 \left\langle\left( \hat W_\nu^{SU(\hat N)}\right)^{{\cal P}}\right\rangle&=&\int_{\Gamma(\phi)} \left[\prod_{\mu=1}^4\prod_{i=1}^{\scriptsize \mbox{gcd}(k,r)}d\phi_\mu^i\right] \left(\hat W_\nu^{SU(\hat N)}(z,\phi)\right)^{\cal P}=0\,.
 \end{eqnarray}
The expectation value is computed in the background of a pure $SU(\hat{N})$ gauge bundle, with the moduli space parameterized by the moduli of the original $SU(N)$ theory. Here, ${\cal P}$ represents an appropriate power, as defined in (\ref{nonvainishing powers set1}, \ref{nonvainishing powers set2}), ensuring that the Wilson lines do not vanish. Additionally, $\Gamma(\phi)$ denotes the moduli space. An equivalent representation of the dual theory moduli space in terms of dual moduli $\hat \phi_i$ will also be considered below and will work as a check on our construction. 

This analysis, as we describe below, can also be phrased in terms of the images under the global $\Z_{\hat{N}}^{(1)}$ form symmetry of the fractional instanton solution. 

To simplify the treatment and streamline our analysis, we will focus on the specific case where $r = k = 1$, and thus, $\hat N=q_3\left(Nq_1-1\right)+q_1$. Here, we need to remember that $\phi_\mu^i=\tilde\phi_\mu=\phi_\mu$. Then, the nonvanishing Wilson's lines are, recalling that $\ell = N-1$, for the $z_1$ and $z_2$ directions on $\hat\T^4$:
\begin{eqnarray}\label{r1k1duallines} \left(\hat W_1^{SU(\hat N)}\right)^{\scriptsize(q_1-1)}&\propto& e^{i  {2\pi (\ell q_3 + 1) \over \hat N} ({z_2 \over \hat L_2} - (Nq_1-1)\phi_2L_2)}\,,\quad
 \left(\hat W_1^{SU(\hat N)}\right)^{\scriptsize q_1}\propto  e^{-i\frac{2\pi  q_3}{\hat N} ({z_2 \over \hat L_2} - (Nq_1-1)\phi_2L_2)}\,, \nonumber \\
  \left(\hat W_2^{SU(\hat N)}\right)^{\scriptsize (q_1-1)}&\propto& e^{-i\frac{2\pi  (1+\ell q_3)}{\hat N}({z_1 \over \hat L_1} - (N q_1 -1) L_1 \phi_1)}\,,\quad
  \left(\hat W_2^{SU(\hat N)}\right)^{\scriptsize q_1}\propto e^{i\frac{2\pi q_3}{\hat N}({z_1 \over \hat L_1} - (N q_1 -1) L_1 \phi_1)}\,.\nonumber \\
\end{eqnarray}
Next, we observe that in the presence of a twist $\hat n_{12} = N q_3 +1$, as per (\ref{dualtwists}), a shift $z_2 \rightarrow z_2 + \hat L_2$ ($z_1 \rightarrow z_1 + \hat L_1$) performs a 
center symmetry transformation in the $z_1$ ($z_2$) direction, multiplying $W_1^p$ by a phase $e^{-i {2 \pi \over \hat{N}} p \hat{n}_{12}}$ ($W_2^p$ by an opposite phase). These two shifts are equivalent to shifts of $\phi_2 L_2$ ($\phi_1 L_1$) by $- {1 \over N q_1 -1}$. 

Since center symmetry is a global symmetry, the $\hat{N}$ images of the instanton under $\Z_{\hat{N}}^{(1)}$ in each direction are different solutions. To account for these images  in a path integral calculation, the range of the moduli $\phi_{1,2}$ has to be extended  $\hat N$ times, to the range
\begin{eqnarray}\label{the correct range condition 1}
 \phi_{1,2}\in \hat N \left[0,\frac{1}{ (Nq_1-1) L_{1,2}}\right)~.
 \end{eqnarray}
As is also clear from the form of the Wilson loops, the appropriate integral (\ref{vanishing condition}) also vanishes when the region of integration is taken as (\ref{the correct range condition 1}).

For the Wilson loops winding in $z_3$ and $z_4$,
\begin{eqnarray}\label{r1k1duallines2}
 \left(\hat W_3^{SU(\hat N)}\right)^{\scriptsize q_3}&\propto& e^{i\frac{2\pi  q_1}{\hat N}({z_4 \over \hat L_4} - (N q_3 + 1)\ell L_4 \phi_4)}\,,\quad
 \left(\hat W_3^{SU(\hat N)}\right)^{\scriptsize (1+\ell q_3)}\propto e^{-i\frac{2\pi  (q_1-1)}{\hat N}({z_4 \over \hat L_4} - (N q_3 + 1)\ell L_4 \phi_4)}\,\nonumber\\
  \left(\hat W_4^{SU(\hat N)}\right)^{\scriptsize  q_3}&\propto& e^{-i\frac{2\pi  q_1}{\hat N}({z_3 \over \hat L_3} - (N q_3 + 1)\ell L_3 \phi_3)}\,,\quad
  \left(\hat W_4^{SU(\hat N)}\right)^{\scriptsize (1+\ell q_3)}\propto e^{i\frac{2\pi   (q_1-1) }{\hat N}({z_3 \over \hat L_3} - (N q_3 + 1)\ell L_3 \phi_3)}\,,\nonumber \\ 
\end{eqnarray}
the analysis is similar (recalling (\ref{dualtwists}) that the $3-4$ plane twist on $\hat\T^4$ is $n_{34} = N q_1 -1$)  and leads to the conclusion that including all $\Z_{\hat{N}}^{(1)}$ center symmetry images of the instanton  leads to
the vanishing of (\ref{vanishing condition}),   and implies that the moduli lives in the fundamental intervals\footnote{Provided the same assumptions as in footnote \ref{footnote:assume} are made.}
\begin{eqnarray}\label{the correct range condition}
\Gamma(\phi)\equiv \left\{\begin{array}{c}
 \phi_{1,2}\in \hat N \left[0,\frac{1}{ (Nq_1-1) L_{1,2}}\right)\,\\ \\ \phi_{3,4}\in \hat N \left[0,\frac{1}{(N-1) (Nq_3+1) L_{3,4}}\right)\end{array}\right.\,.
\end{eqnarray}
We recall that an identical consideration in the $SU(N)$ theory on $\T^4$ found that the corresponding ranges in the $\T^4$ theory are $[0,1)$ and $[0,1/(N-1))$, for $\phi_{1(2)} L_{1(2)}$ and  $\phi_{3(4)} L_{3(4)}$, respectively.

We proceed by computing the metric on the moduli space:
\begin{eqnarray}
 {\cal U}_{\alpha\beta}^{SU(\hat N)}=\frac{2}{g^2}\int_{\hat{\mathbb T}^4}\mbox{tr}_{\hat N}\left[\frac{\partial \hat A'^{}_\mu}{\partial \phi_\alpha}\frac{\partial \hat A'^{}_\mu}{\partial \phi_\beta}\right]\,,\quad \alpha,\beta=1,2,3,4\,,
\end{eqnarray}
and we used the $SU(\hat N)$ gauge fields given by (\ref{the primed gauge field}, \ref{the primed gauge field2}) to perfom the computations and $g$ is the coupling of the $SU(\hat N)$ gauge theory on $\hat\T^4$. We remind the reader that this is the gauge field in the primed frame, obtained after a gauge transformation that shifts the moduli dependence from the transition functions entirely onto the gauge field. Direct computations give the nonvanishing metric components 
\begin{eqnarray}\nonumber
  {\cal U}_{11}^{SU(\hat N)}&=& {\cal U}_{22}^{SU(\hat N)}=V_{\hat{\mathbb T}^4}\frac{8\pi^2 L_1^2L_2^2  q_3(-1+Nq_1)^2(1+\ell q_3)}{g^2q_1(q_1-1)\hat N}\,,\\
 {\cal U}_{33}^{SU(\hat N)}&=&  {\cal U}_{44}^{SU(\hat N)}=V_{\hat{\mathbb T}^4}\frac{8\pi^2 L_3^2L_4^2 \ell^2 q_1(1+Nq_3)^2(-1+ q_1)}{g^2q_3(1+\ell q_3)\hat N}\,,
\end{eqnarray}
where $V_{\hat{\mathbb T}^4}=\prod_{\mu=1}^4\hat L_\mu=\prod_{\mu=1}^4 L_\mu^{-1}$ is the volume of $\hat{\mathbb T}^4$, and recall that $\ell=N-1$.
Thus, we find
\begin{eqnarray}
\sqrt{\mbox{Det}\, {\cal U}_{\alpha\beta}^{SU(\hat N)}}=\frac{64\pi^4  (N-1)^2}{g^4}\left(\frac{1}{\hat N}-N\right)^2\,.
\end{eqnarray}
After this point, the computations of the measure on the moduli space become a standard procedure, and the reader is referred to the literature for details. The upshot is that the measure of the bosonic moduli space is given by
\begin{eqnarray}\nonumber
d\mu^{SU(\hat N)}&=&\sqrt{\mbox{Det}\,  {\cal U}_{\alpha\beta}^{SU(\hat N)}}\prod_{\mu=1}^4\frac{d\phi_\mu L_\mu}{\sqrt{2\pi}}=\frac{16\pi^2}{g^4}(N-1)^2 \left(\frac{1}{\hat N}-N\right)^2 \prod_{\mu=1}^4d\phi_\mu L_\mu~.
\end{eqnarray}
Finally, the volume of the moduli space $\Gamma(\phi)$ is obtained by integrating the measure over the range given in (\ref{the correct range condition}):
\begin{eqnarray}\label{final result sun using phi}
\mu^{SU(\hat N)}=\int_{\Gamma(\phi)}d\mu^{SU(\hat N)}=\frac{16\pi^2 \hat N^2}{g^4}\,,
\end{eqnarray}

\bibliography{Nahm2.bib}
  \bibliographystyle{JHEP}

\end{document}